\documentclass[10pt,aps,prd,twocolumn,superscriptaddress,
showpacs,preprintnumbers,nofootinbib,notitlepage]{revtex4-1}

\usepackage{amsmath,amssymb,hyperref,subfigure,xspace}
\usepackage{mciteplus,color,mathtools,graphicx}
\usepackage{braket,bbm,slashed}
\usepackage{multirow}
\usepackage[english]{babel}
\usepackage{mathrsfs}
\usepackage{enumerate}
\usepackage[utf8]{inputenc}

\usepackage{lineno}

\usepackage{xspace}
\usepackage{slashed}
\newcommand{\cf}{{cf.}\xspace}

\newcommand{\sw}{$S$-}

\renewcommand{\P}{\mathbf{P}}

\newcommand{\p}{\mathbf{p}}
\newcommand{\q}{\mathbf{q}}

\newcommand{\0}{\mathbf{0}}
\newcommand{\1}{\mathbbm{1}}
\newcommand{\2}{\mathbf{2}}
\newcommand{\3}{\mathbf{3}}

\newcommand{\x}{\mathbf{x}}

\newcommand{\A}{\mathcal{A}}
\newcommand{\B}{\mathcal{B}}
\newcommand{\C}{\mathcal{C}}
\newcommand{\D}{\mathcal{D}}
\newcommand{\E}{\mathcal{E}}
\newcommand{\F}{\mathcal{F}}

\newcommand{\Kc}{\mathcal{K}}
\newcommand{\R}{\mathcal{R}}
\newcommand{\N}{\mathcal{N}}

\newcommand{\wt}[1]{\widetilde{ #1 }}
\newcommand{\wh}[1]{\widehat{ #1 }}

\newcommand{\ov}[1]{\overline{ #1 }}
\newcommand{\bs}[1]{\boldsymbol{ #1 }}
\newcommand{\sss}[1]{\scriptscriptstyle{ #1 }}

\DeclareMathOperator{\im}{Im}

\newcommand{\eg}{{e.g.}\xspace}
\newcommand{\ie}{{i.e.}\xspace}
\newcommand{\etal}{ {et al.}\xspace}
\newcommand{\nn}{\nonumber}

\newcommand{\mev}{\ensuremath{{\mathrm{\,Me\kern -0.1em V}}}\xspace}
\newcommand{\gev}{\ensuremath{{\mathrm{\,Ge\kern -0.1em V}}}\xspace}
\newcommand{\tev}{\ensuremath{{\mathrm{\,Te\kern -0.1em V}}}\xspace}

\usepackage{mfirstuc} 
\newcommand{\addReviewer}[2]{
  \expandafter\newcommand\csname #1\endcsname[1]{{\bf \color{#2} \capitalisewords{#1}:\,##1}}
  \expandafter\newcommand\csname #1cor\endcsname[2]{{\color{#2} \capitalisewords{#1}:\,\st{##1}{\bf ##2}}}
  \expandafter\newcommand\csname #1color\endcsname{#2}
}

\usepackage{tikz, marginnote} 
\newcommand{\checkedby}[1]{
\ifdefined\CROSSCHECKS
  \marginnote{
    \begin{tikzpicture}
      \foreach \x [count=\xi] in {#1} {
         \node[shape=circle,inner sep=0mm,
         minimum size=2mm,
         fill=\csname \x color\endcsname] at (\xi*3mm,0) {};
       }
    \end{tikzpicture}
  }
\else
\fi
}

\usepackage{soul,color}
\definecolor{chromeyellow}{rgb}{1.0, 0.65, 0.0}
\definecolor{DodgeBlue}{rgb}{0.118, 0.565,1.000}
\definecolor{asparagus}{rgb}{0.53, 0.66, 0.42}
\definecolor{cadmiumgreen}{rgb}{0.0, 0.42, 0.24}

\addReviewer{adam}{red}
\addReviewer{ale}{chromeyellow}
\addReviewer{vincent}{brown}
\addReviewer{misha}{cadmiumgreen}
\addReviewer{jannes}{blue}
\addReviewer{cesar}{magenta}
\addReviewer{tomasz}{cyan}
\addReviewer{miguel}{DodgeBlue}
\addReviewer{andrew}{purple}
\addReviewer{nathan}{orange}
\addReviewer{arkaitz}{asparagus}

  \hypersetup{
      pdfnewwindow=true,      
      citecolor=blue,        
      filecolor=blue,      
      urlcolor=blue,        
     colorlinks=true,    
     linkcolor=blue     
  }

\newcommand{\bonn}{Universit\"at Bonn,
Helmholtz-Institut f\"ur Strahlen- und Kernphysik,
53115 Bonn, Germany}
\newcommand{\ceem}{Center for  Exploration  of  Energy  and  Matter,
Indiana  University,
Bloomington,  IN  47403,  USA}
\newcommand{\ghent}{Department of Physics and Astronomy,
Ghent University, Belgium}
\newcommand{\icn}{Instituto de Ciencias Nucleares,
Universidad Nacional Aut\'onoma de M\'exico,
Ciudad de M\'exico 04510, Mexico}
\newcommand{\indiana}{Physics  Department,
Indiana  University,
Bloomington,  IN  47405,  USA}
\newcommand{\jlab}{Theory Center,
Thomas  Jefferson  National  Accelerator  Facility,
Newport  News,  VA  23606,  USA}
\newcommand{\csu}{California State University, Bakersfield, CA 93311, USA}
\newcommand{\ectstar}{European Centre for Theoretical Studies in Nuclear Physics and Related
Areas (ECT$^*$) and Fondazione Bruno Kessler,
I-38123 Villazzano (TN), Italy}

\newcommand{\jpac}{Joint Physics Analysis Center}

\begin{document}

\title{Phenomenology of Relativistic $\3\to\3$ Reaction Amplitudes\\
within the Isobar Approximation}

\author{A.~Jackura}\email{ajackura@iu.edu}\affiliation{\ceem}\affiliation{\indiana}
\author{C.~Fern\'andez-Ram\'{\i}rez}\affiliation{\icn}
\author{V.~Mathieu}\affiliation{\jlab}
\author{M.~Mikhasenko}\affiliation{\bonn}
\author{J.~Nys}\affiliation{\ghent}
\author{A.~Pilloni}\affiliation{\jlab}\affiliation{\ectstar}
\author{K. Salda\~na}\affiliation{\ceem}\affiliation{\csu}
\author{N.~Sherrill}\affiliation{\ceem}\affiliation{\indiana}
\author{A.~P.~Szczepaniak}\affiliation{\ceem}\affiliation{\indiana}\affiliation{\jlab}

\collaboration{\jpac}
\noaffiliation
\preprint{JLAB-THY-18-2817}
\begin{abstract}
Further progress in hadron spectroscopy necessitates the phenomenological description of three particle reactions. We consider the isobar approximation, where the connected part of the $\3\to\3$ amplitude is first expressed as a sum over initial and final pairs, and then expanded into a truncated partial wave series. The resulting 
unitarity equation is automatically fulfilled by the $B$-matrix solution, which is an integral equation for the partial wave amplitudes, analogous to the $K$-matrix parameterization used 
 to describe $\2\to\2$ amplitudes. 
We study the one particle exchange and how its analytic structure impacts rescattering solutions such as the triangle diagram. The analytic structure is compared to other parameterizations discussed in the literature. We briefly discuss the 
analogies with recent formalisms for extracting $\3\to\3$ scattering amplitudes in lattice QCD.
\end{abstract}
\date{\today}
\maketitle

\section{Introduction}\label{sec:Introduction}
Modern high-energy experiments
 are accumulating high quality data on three-hadron final states, 
 that are  expected 
 to be the
 main decay channels of several poorly known or missing resonances. 
 These include, for example,  the enigmatic $a_1$, $\omega_2$,  and the exotic  $\pi_1$ resonances that can be studied in peripheral production at 
 COMPASS, GlueX, and CLAS12~\cite{Adolph:2014rpp,Adolph:2015pws,Adolph:2015tqa,Akhunzyanov:2018pnr,Bookwalter:2011cu,Ghoul:2015ifw,AlGhoul:2017nbp}.  
 In addition to conventional hadrons,  many of the exotic $XYZ$ and pentaquark states observed in the heavy quarkonium sector~\cite{Esposito:2016noz,Lebed:2016hpi,Olsen:2017bmm}, are also found in three particle final states.
 
 Many of these newly observed or anticipated states lie close to  thresholds of their decay products. For example, the mass difference between
   the  $X(3872)$~\cite{Choi:2003ue} and the $D^{0}\bar{D}^0 \pi^0$ threshold is only $6 \text{ MeV}$.  The proximity of the three particle threshold together with
   the possibility of long-range interactions mediated by a single pion exchange can significantly influence the $X(3872)$ line-shape~\cite{Braaten:2010mg} and  one needs to carefully 
    analyze the role of pion exchange and whether it is able to bind $D^{*0}$ and $\bar D^{0}$~\cite{Thomas:2008ja,Baru:2011rs,Kalashnikova:2012qf,Guo:2017jvc}.
  In the light meson sector, the recently observed $a_1(1420)$~\cite{Adolph:2015pws} is yet another candidate for a state  not expected in the quark model that can be  influenced
    by meson exchange interactions and thresholds ~\cite{Basdevant:2015wma, Ketzer:2015tqa}.
  
    On the theory side, lattice QCD has made substantial progress in extracting the resonance spectrum from simulations of
   $\2\to\2$  reactions~\cite{Wilson:2014cna,Lang:2015sba,Dudek:2016cru,Briceno:2016mjc,Moir:2016srx,Briceno:2017qmb,Briceno:2017max,Woss:2018irj,Brett:2018jqw}, and recently, the formalism
for $\3\to\3$ scattering has been developed~\cite{Hansen:2014eka,Hansen:2015zga,Hansen:2016ync,Briceno:2017tce,Briceno:2018mlh,Mai:2017bge,Mai:2018djl,Polejaeva:2012ut,Hammer:2017kms,Hammer:2017uqm,Doring:2018xxx}.  Analysis and interpretation of both  experimental data and lattice simulations require input in the form of reaction amplitudes that can be analytically continued into the complex energy plane.
For example, in partial waves, resonances appear as pole singularities, 
while particle exchanges lead to logarithmic branch points.  
   Fortunately, reaction amplitudes are constrained by unitarity, which can be used to determine 
    the discontinuities of partial waves in the 
    near threshold region.   

   The problem of
     constraining $\3\to\3$ reactions from the $S$-matrix principles of unitarity and analyticity
      has been studied previously in Refs.~\cite{Fleming:1964zz,Holman:1965,Aitchison:1966lpz,Grisaru:1966,Ascoli:1975mn,Mai:2017vot}. In this  paper we extend these earlier works and clarify some of the results. Moreover, we
       present the $\3 \to \3$ reaction amplitudes in a way 
       that can be directly translated to the finite volume.
       
           Our description relies on the isobar approximation, where the amplitude is constructed as a sum of truncated partial wave expansions.
  This provides a good description of three-particle final states in the resonance region, where analyses of Dalitz plots indicate that they are dominated by
        intermediate two-body resonances. For example, the decay of the $a_1(1260)$ resonance into
three pions occurs primarily via a decay to the $\rho\pi$ intermediate state
         with the subsequent decay of  $\rho$ to two pions~\cite{Adolph:2015tqa,Akhunzyanov:2018pnr}. The isobar approximation can be regarded as an effective way to incorporate the relevant singularities in all Mandelstam variables, and will be discussed in detail later. 

 The rest of the paper is organized
 as follows.  In Sec.~\ref{sec:Kin_Inv_Amp} we define the $\3\to\3$ amplitude for three spinless
 particles and discuss the relevant  kinematics.  In Sec.~\ref{sec:Isobar_Model} we introduce the isobar approximation and investigate the consequences of unitarity. We explain the  difference between 
 isobar and the partial wave amplitudes
 , 
 which are often confused. 
 In short, we use the isobar representation to
  describe the $\3\to\3$ amplitude,
  $\A = \sum \A_{kj}$, where the indices  $k$ and $j$ label the spectator particle in the final and initial state, respectively. We refer to the $ \A_{kj}$'s  as  isobar-spectator amplitudes, since  they can be pictured as  scattering of a quasi-particle, the isobar, and a stable spectator.
  The latter are expanded in partial waves of the three-particle  system.
  Unitarity constrains the  $\3\to\3$ amplitudes on the real energy axis,
  which results in specific relations involving the imaginary parts of the partial-wave-projected  isobar-spectator amplitudes.
Unitarity alone does not uniquely specify  partial wave amplitudes,  as evident, for example, in the $K$-matrix parametrization of 
 $\2\to\2$ scattering amplitudes~\cite{Castillejo:1955ed,Gribov:2009zz}. In 
 Sec.~\ref{sec:B-Matrix} we
  discuss a specific parameterization for the isobar-spectator amplitudes which satisfies the three-body and two-body unitarity.  It is given as a solution of a set of linear integral equations that involve, among others, the one particle exchange (OPE) as a driving term. We call this the  $B$-matrix parameterization and it satisfies, \begin{equation}
\A_{kj} = \B_{kj} + \B_{kn} \tau_{n} \A_{nj}, \label{1}
\end{equation}
where $\B$ is the driving term that contains the OPE, $\tau$ is a known function of the phase space and of the $\2\to\2$ 
amplitudes. The product formally represents an integration over the intermediate isobar mass. In contrast to Ref.~\cite{Mai:2017vot}, we
restrict the domain of the integrals to physical values of energies. This enables us to use the experimentally accessible  subchannel amplitudes and we also  discuss the consequences of this restriction. We derive Eq.~(\ref{1}) for isobars with arbitrary spin $s$, and for any value of the isobar-spectator orbital angular momentum $\ell$.
 
The $B$-matrix parameterization can be 
analytically continued to the complex energy plane and in Sec.~\ref{sec:AnalyticProperties} we discuss aspects of its analytic properties. Specifically, the one particle exchange  process has some unique features, as it contains a kinematic singularity due to the exchange of a real particle, which can be isolated from the full $\3\to\3$ scattering amplitude.
In addition, we also study the triangle amplitude that emerges from the $B$-matrix parameterization, and the relation to the Bethe-Salpeter solution. We summarize our results in Sec.~\ref{sec:Conclusion}.

\section{Kinematics, Invariants, \& Amplitudes}\label{sec:Kin_Inv_Amp}
\begin{figure}[b!]
    \centering
    \includegraphics[ width=.7\columnwidth]{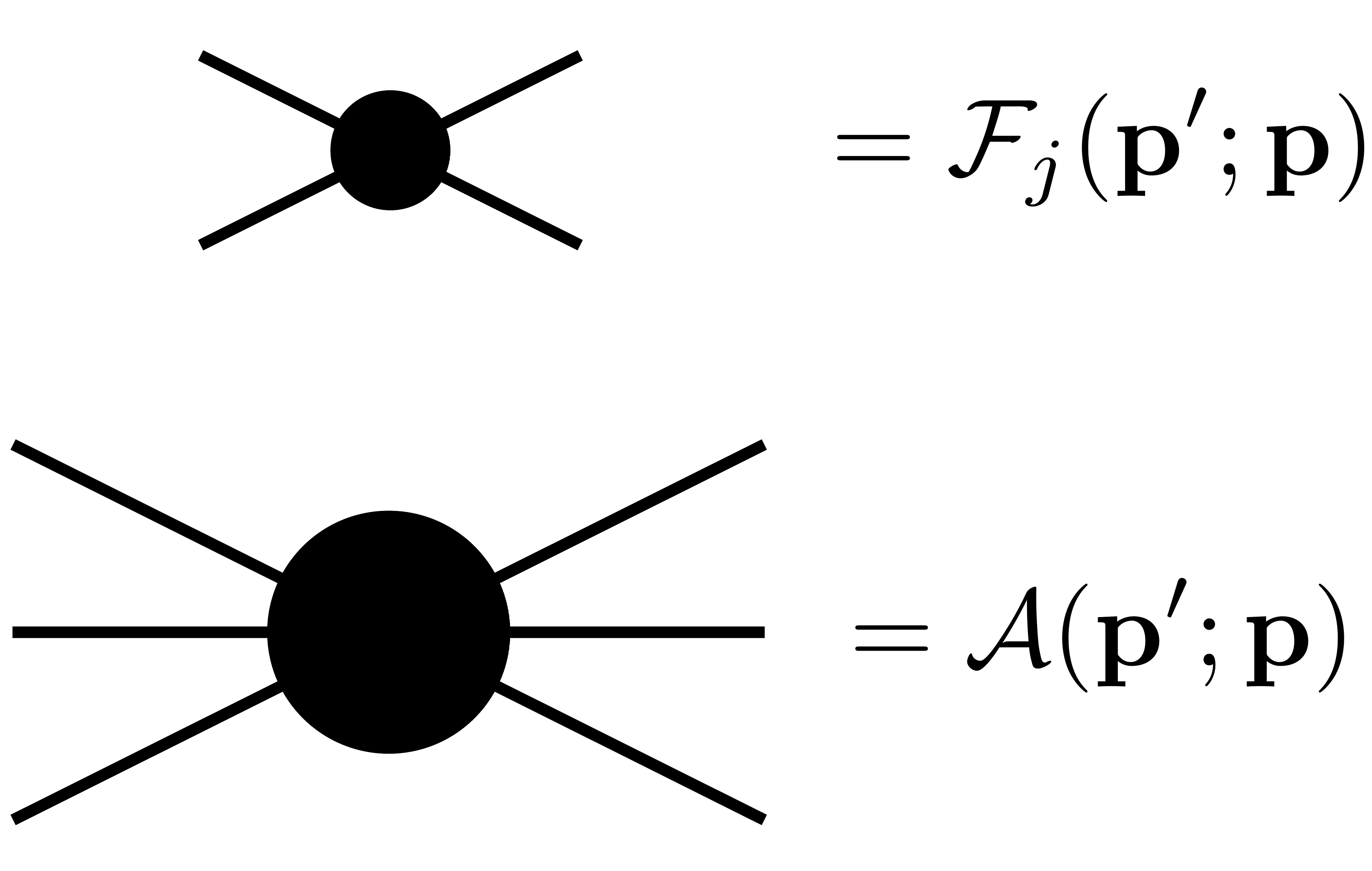}
    \caption{Diagrammatic representation of the disconnected $\2\to\2$ amplitude of Eq.~\eqref{eq:2to2d_Amp} (black disk with four external legs) and connected $\3\to\3$ amplitude of Eq.~\eqref{eq:3to3c_Amp} (black disk with six external legs).}
    \label{fig:3to3_Symbols}
\end{figure}
%
\begin{figure}[b!]
    \centering
    \includegraphics[ width=1.0\columnwidth]{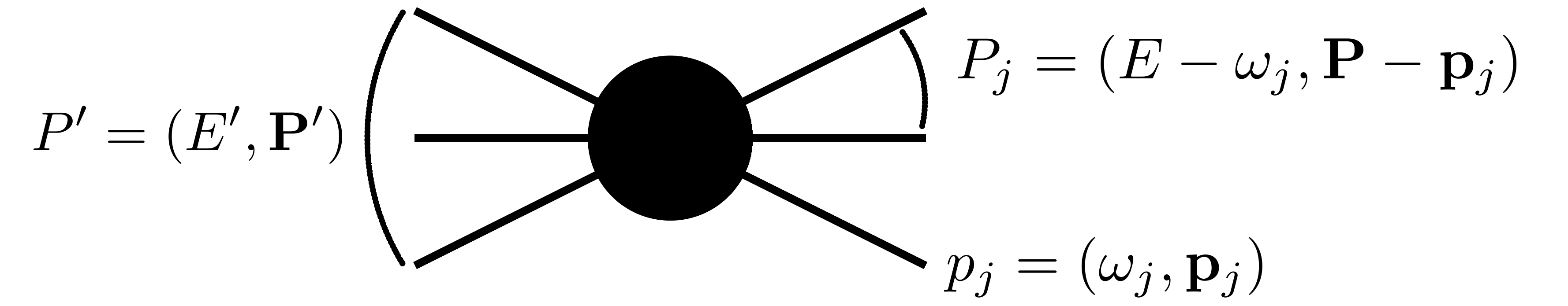}
    \caption{Momenta labels for three particles. The left side denotes the final state particles, while the right is the initial state.
    }
    \label{fig:3to3_Kinematics}
\end{figure}
We consider elastic scattering of three distinguishable, spinless particle, \eg $D\pi \bar D$, $K\pi \bar K$, or
$\pi^+\pi^-\pi^0$. The particles have mass $m_j$
, where $j = 1, 2, $ or $3$ labels the individual particles.
A single particle state, with four-momentum $p_j = (\omega_j,\p_j)$, where $\omega_j = \sqrt{m_j^2 + \lvert  \p_j\rvert^2}$ is the energy and $\p_j$ is the three-momentum, is denoted $\ket{\p_j}$ and has relativistic normalization
$\braket{\p_k'|\p_j} = (2\pi)^{3}\,2\omega_j \delta^{(3)}(\p_k'-\p_j)\delta_{kj}$.
We are interested in the $S$-matrix element of the elastic $\3\to\3$ scattering process. We can decompose the $S$-matrix as $S = \1 +  i T$. 
The $T$-matrix contains two terms, $T = T_d + T_c$, where the disconnected part, $T_d$, involves interactions of two particles at the time
 with the third one being a spectator, while the
connected part, $T_c$, contains interactions of all three particles. The disconnected part can always be identified kinematically by the spectator momentum conserving delta function \cite{Itzykson:1980rh}.
The disconnected part is written as $T_d = \sum_{j}\1_{j} \otimes T^{(j)}$, where $\1_{j}$ is the identity operator in the single particle space of the spectator, $j$ and $T^{(j)}$ describes $\2 \to \2$ scattering between the other two particles. 
The amplitudes associated with the matrix elements of scattering operators $T_d$ and $T_c$ are defined as ${\F}$ and ${\A}$, respectively.
Specifically, the  connected amplitude  $\A$ is given by
\begin{equation}\label{eq:3to3c_Amp}
\bra{\p'}T_c\ket{\p} = (2\pi)^4\delta^{(4)}(P' - P) \A(\p';\p),
\end{equation}
where $\ket{\p} \equiv \ket{\p_1\p_2\p_3}$ and $\ket{\p'} \equiv \ket{\p_1'\p_2'\p_3'}$ denote the initial and final states of the three particles, and $P = p_1 + p_2 + p_3$ and $P' = p_1' + p_2' + p_3'$ are the initial and final total four-momenta, respectively, as illustrated in Fig.~\ref{fig:3to3_Symbols}.
Time-Reversal symmetry implies that the amplitude is symmetric in the initial-final state variables, $\A(\p';\p) = \A(\p;\p')$.
The chosen normalization implies that the amplitude $\A(\p';\p)$ has mass dimension $-2$. The disconnected amplitudes $\F_{j}$ are defined by
\begin{equation}\label{eq:2to2d_Amp}
\begin{split}
\bra{\p'} T_d \ket{\p}  & = (2\pi)^4\delta^{(4)}(P' - P)
 \\ &  \times
 \sum_{j=1}^{3} (2\pi)^{3}\,2\omega_j \delta^{(3)}(\p_j'-\p_j) \F_{j}(\p';\p),
\end{split}
\end{equation}
where the delta function enforces that the spectator $j$ does not interact. We remark that the $\F_{j}$ is the genuine $\2\to\2$ scattering amplitude, as required by the LSZ construction~\cite{Itzykson:1980rh}. We also define $P_{j} \equiv  P - p_{j}$ and $P_{j}' \equiv P' - p_{j}'$ as  the initial and final total four-momenta of the interacting pair recoiling against spectator $j$, \cf Fig.~\ref{fig:3to3_Kinematics}.

In this paper we adopt the so-called {\it{spectator notation}} or {\it{odd-one-out notation}} \cite{Giebink:1985zz}, where the $\2\to \2$ amplitudes associated with the spectator $j$ are labeled by the spectator index.  The spectator notation requires additional information specifying the first particle in the two-particle system. There are two conventions which are useful for our discussions: the two-pair convention, and the cyclic convention.
The two-pair convention is more practical when
 interaction in one of the three pairs is negligible.
   An example of such a system is $\pi^+\pi^+\pi^-$, where the $\pi^+\pi^+$ system interacts weakly. In this case it is  convenient to choose the noninteracting system as, say, particles $(13)$ and designate
    particle $2$ as the second particle for both the interacting sub-systems.
     Therefore, the spectator index  $j=1$ and $j=3$ uniquely identifies the two orderings in the pairs to be $(32)$ and $(12)$, respectively.
  If the interactions in all three subchannels are important,
   one can define the ordering through cyclical permutation, \ie the spectator label $j=1,2,3$ corresponds to ordering of the two particles subsystems as $(23)$, $(31)$, and $(12)$, respectively. For simplicity, in the following we assume only two relevant subchannels, and use the former convention. Generalization to the latter case is straightforward.
   The type of applications we have in mind are systems like $M\bar M \pi$ elastic scattering, where $M$ is an open-flavor meson, such as $K$, $D$, and $B$. The interacting pairs will be assumed in the $M\pi$ and $\bar{M}\pi$ channels  only, and pion being designated as particle $j=2$.

The $\3\to\3$ amplitude depends on eight independent kinematic variables. The choice of variables largely depends on the kinematical range of interest, \eg the low vs high total energy region. Here  we are interested in the low-energy region and use the following redundant set of Mandelstam variables,
\begin{subequations}
\begin{align}
s & = (p_1 + p_2 + p_3)^2 = (p_1'+p_2'+p_3')^2, \\
t_{jk} & = (p_j - p_k')^2 = (P_{j} - P_{k}')^2, \\
u_{jk} & = (P_{j} - p_k')^2 = ((P - p_j) - p_k')^2, \\
\sigma_{j} & = P_{j}^2 = (P - p_{j})^2, \\
\sigma_{k}' & = P_{k}'^2 = (P - p_{k}')^2.
\end{align}
\end{subequations}
where  $s$, $\sigma_j$, and $\sigma_k'$ are the invariant mass squares of the total three particle system, the initial pair, and the final pair, respectively.
The  transferred momenta,
  $t_{jk}$ and $u_{jk}$, are between the initial and final spectators and the initial pair and final spectator, respectively.
 The Mandelstam invariants are related by energy-momentum conservation,
 \begin{subequations}
\begin{align}
s + t_{jk} + u_{jk} & = \sigma_{j} + \sigma_{k}' + m_j^2 + m_k^2, \label{eq:NRG_const} \\
\sum_{j=1}^{3}\sigma_{j} & = s + \sum_{j=1}^{3}m_{j}^2, \label{eq:subNRG_const1} \\
\sum_{k=1}^{3}\sigma_{k}' & = s + \sum_{k=1}^{3}m_{k}^2. \label{eq:subNRG_const2}
\end{align}
\end{subequations}
In the physical region of the $\3\to\3$ reaction,
  $s$ can take any value above the three particle threshold, $s \ge s_{\mathrm{th}} = (m_1 + m_2 + m_3)^2$, while the subchannel invariant masses $\sigma_{j}$ and $\sigma_{k}'$ are bounded by ${\sigma_j^{(\mathrm{th})}} \le \sigma_{j} \le (\sqrt{s} - m_j)^2$ and ${\sigma_k^{(\mathrm{th})}} \le \sigma_{k}' \le (\sqrt{s} - m_k)^2$, where ${\sigma_j^{(\mathrm{th})}}$ are the sub-energy thresholds, \eg ${\sigma_1^{(\mathrm{th})}} = (m_{2} + m_{3})^2$.
 We will need the relations between the invariants and
    energies and scattering angles, in three reference frames.
     The  frames of interest will be the overall center-of-momentum frame
      (CMF) and the isobar rest frame (IRF). There are two IRFs corresponding to the initial and final states: the initial IRF$_j$,
      labeled by the spectator $j$, and the final IRF$'_k$, labeled with spectator $k$ and a prime.

\begin{figure*}[t!]
    \centering
    \includegraphics[trim={0 5cm 0 7cm},clip, width=.8\textwidth]{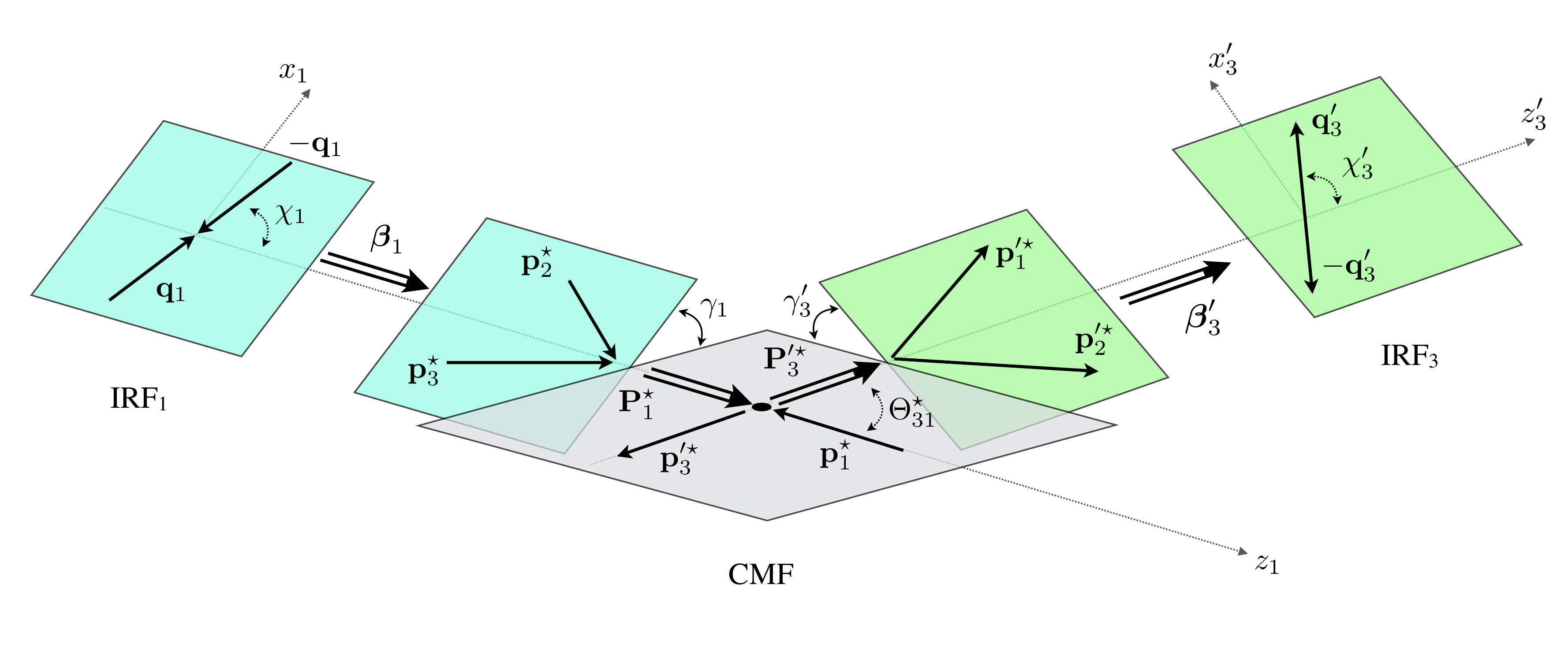}
    \caption{Connection between the three reference frames for the $(32)1 \to (12)3$ system. The total reaction plane in the CMF is shown in gray, and the initial and final IRFs are shown in blue and green, respectively. The Lorentz boosts $\bs{\beta}_1$ and $\bs{\beta}_3'$ indicate the transformations between the three frames. }
    \label{fig:3bodykins}
\end{figure*}
To distinguish momenta in the CMF we denote them by a $\star$, \ie  $\P^{\star} = \P'^{\star} = \0$. In the CMF
 the scattering angle, $\Theta_{kj}^{\star}$, is defined as the angle between the initial and final state spectator momenta, $\cos\Theta_{kj}^{\star} \equiv \wh{\p}_k'^{\star} \cdot \wh{\p}_j^{\star}$, where $\wh{\p}_j^{\star}$ and $\wh{\p}_k'^{\star}$ denote the CMF orientations of the initial and final spectators, respectively.
 The kinematic variables in the other frames,
 IRF$_j$ ($\P_j=\0$) and IRF$'_k$ ($\P_k'=\0$)
 are obtained from the CMF
by a Lorentz boost 
 in the direction opposite
 to momentum of the corresponding spectator. The momentum of the first particle in the pair is denoted by $\q_j$,  and $\q_k'$ in IRF$_j$ and IRF$'_k$, respectively.
  Orientation of these momenta are given by solid angles, $\wh{\q}_j=(\gamma_j,\chi_j)$ and
   $\wh{\q}_k' = (\gamma_k',\chi_k')$, respectively. Here, $\gamma_j$ and $\gamma_k'$ are the azimuthal angles between the decay plane of the isobar and the isobar-spectator scattering plane and $\chi_j$ and $\chi_k'$ are helicity angles,
   \cf Fig.~\ref{fig:3bodykins} for the specific scheme $(23)1 \to (12)3$. The relations between all relevant kinematical variables and the Mandelstam invariants are given in Appendix \ref{sec:app_A}. 
 In the following, we will use the set $(\wh{\q}'_k, \sigma_k, s, t_{jk}, \sigma_j, \wh{\q}_j)$ to describe the isobar-spectator amplitude.

\subsection{Unitarity Relations}\label{sec:UnitarityRelations}
\begin{figure}[t!]
    \centering
    \includegraphics[width=.8\columnwidth]{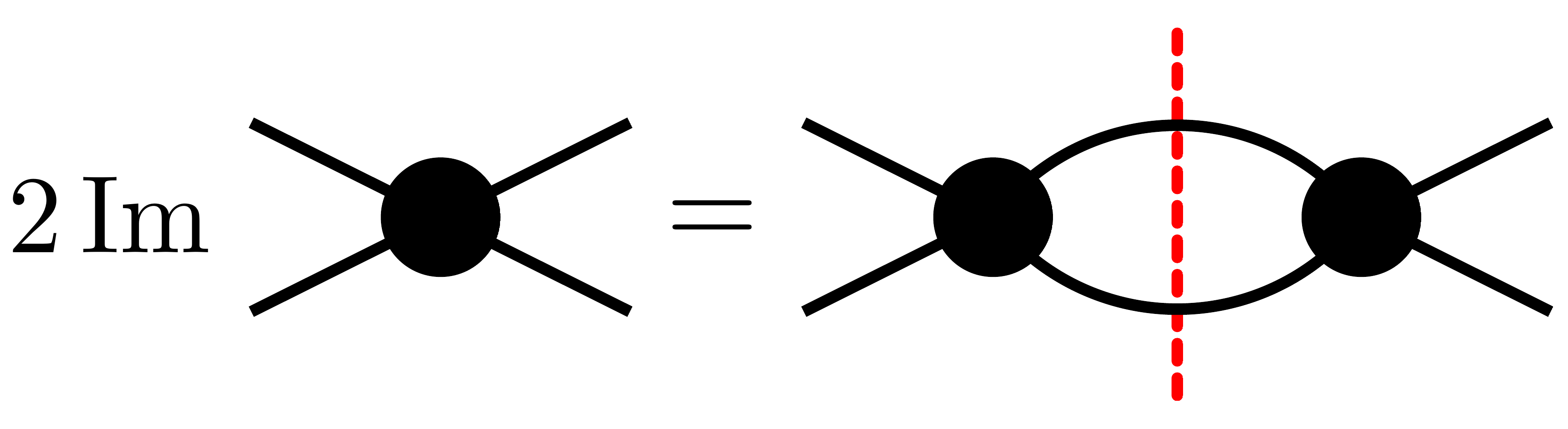}
    \caption{Diagrammatic representation for the $\2\to\2$ disconnected amplitude unitarity relation in Eq.~\eqref{eq:2to2discUnit}. The red vertical dashed line indicates the intermediate particles are put on-shell. 
    }
    \label{fig:2to2_disc_unitarity}
\end{figure}

\begin{figure*}[t!]
    \centering
    \includegraphics[ width=0.7\textwidth]{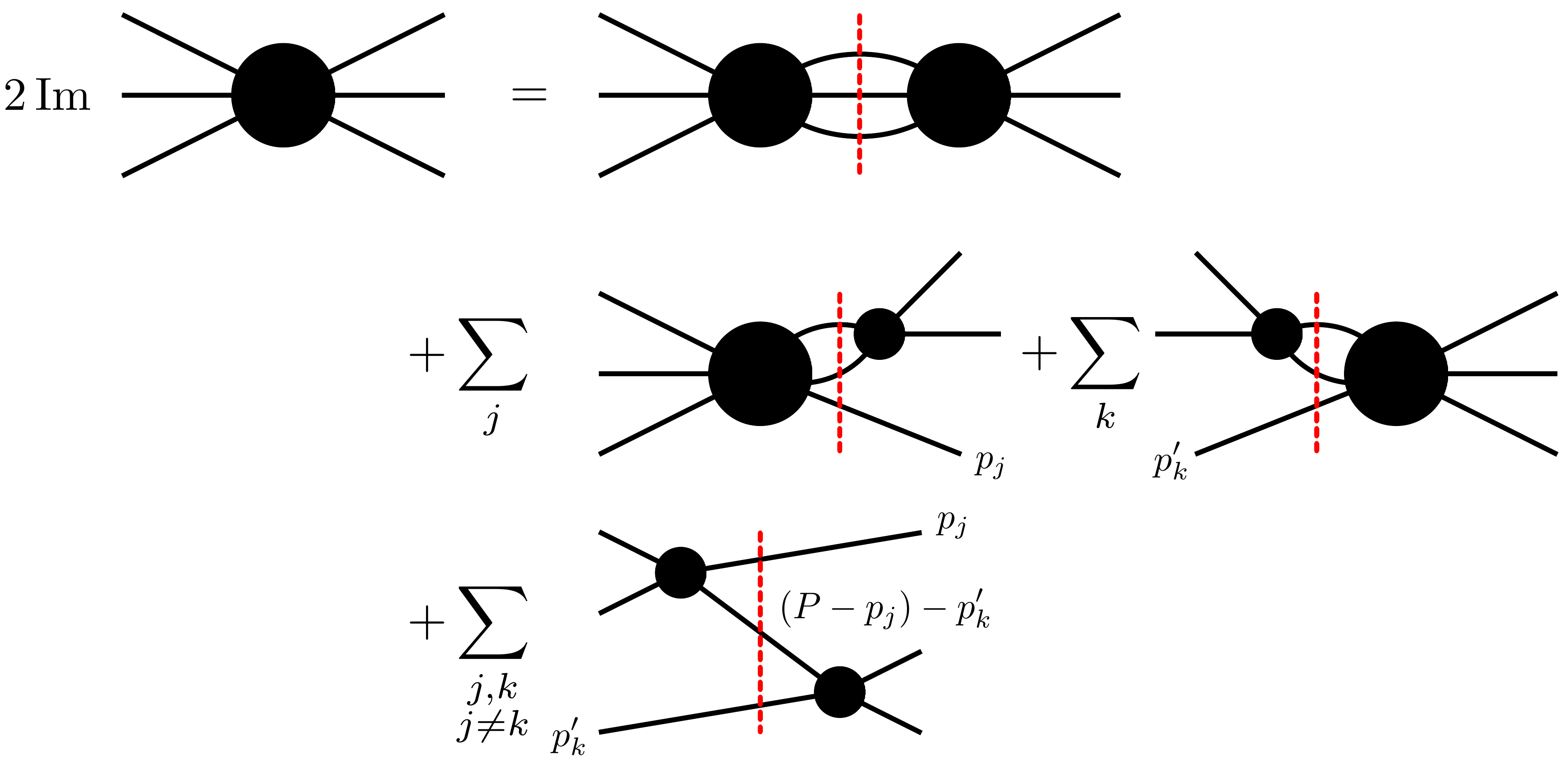}
    \caption{Diagrammatic representation for the $\3\to\3$ connected amplitude unitarity relation in Eq.~\eqref{eq:3to3connUnit}. The red vertical dashed line indicates the intermediate particles are put on-shell.}
    \label{fig:3to3_conn_unitarity}
\end{figure*}

We consider elastic unitarity in the physical region
 of the $\3 \to \3$ reaction below inelastic thresholds. 
  It yields two relations~\cite{Eden:1966dnq}, one for the disconnected $\2\to\2$ amplitude $\F_j$ and one for the connected $\3\to\3$ amplitude $\A$. For $\F_{j}$, one finds
   \begin{equation}\label{eq:2to2discUnit}
\im{\F_{j}(\p';\p)} = \rho_2(\sigma_j) \int d\wh{\q}_j'' \, \F_{j}^{*}(\p'';\p') \F_{j}(\p'';\p)
\end{equation}
where
\begin{equation}\label{eq:2bodyPS}
\rho_2(\sigma_j) = \frac{1}{64\pi^2} \frac{2\lvert \q_j \rvert}{\sqrt{\sigma_j}}
\end{equation}
is the phase space for the two particle system, and $\q''_j$ is the intermediate state relative momentum. 
The IRFs are defined with their $z$-axes defined along the opposite direction of the spectator and their $x$-axes defined by their azimuthal angles w.r.t. the total CMF plane spanned by the initial and final spectator momenta, \cf Fig.~\ref{fig:3bodykins}.
 Note that from energy-momentum conservation, $ \lvert \q_j \rvert =  \lvert \q_j' \rvert =  \lvert \q_j'' \rvert$. Figure~\ref{fig:2to2_disc_unitarity} is a diagrammatic representations of Eq.~\eqref{eq:2to2discUnit}.
Elastic unitarity yields the following condition for the
connected $\3\to\3$ amplitude,
\begin{widetext}
\begin{equation}\label{eq:3to3connUnit} \
\begin{split}
 \im{\A(\p';\p)} & = \frac{1}{2(2\pi)^{5}} \int \frac{d^{3}\p_1''}{2\omega_1''} \frac{d^{3}\p_2''}{2\omega_2''} \frac{d^{3}\p_3''}{2\omega_3''} \, \delta^{(4)}(P'' - P) \,\A^{*}(\p'';\p') \A(\p'';\p) \\
\
&  + \sum_{k}\, 
\rho_2(\sigma_k')
\int d\wh{\q}_k''  \,\F_k^{*}(\p'';\p') \A(\p'';\p)\rvert_{\p''_k = \p_k'} \Theta(\sigma_k' - \sigma_{k}^{(\mathrm{th})}) \\
\
&  + \sum_{j}\, 
\rho_2(\sigma_j)
\int d\wh{\q}_j'' \,\A^{*}(\p'';\p') \rvert_{\p''_j = \p_j} \F_{j}(\p'';\p)\Theta(\sigma_j - \sigma_{j}^{(\mathrm{th})})  \\
\
&  + \sum_{\substack{j,k \\ j\ne k}} \pi \, \delta(u_{jk} - \mu_{jk}^2)\,\F_{k}^{*}(\p'';\p')\rvert_{\p''_j = \p_j} \F_{j}(\p'';\p) \rvert_{\p''_k = \p_k'},
\end{split}
\end{equation}
\end{widetext}
where $\mu_{jk}$ is the mass of the exchanged particle that is neither $j$ nor $k$, \eg if $j=1$, and $k=3$, then the exchanged mass is $\mu_{13} = m_2$. Note that the evaluations $\p_k'' = \p_k'$ in the second and fourth lines enforce that $\sigma_k' = \sigma_k''$, and similarly in lines three and four, $\p_j'' = \p_j$ implies that $\sigma_j'' = \sigma_j$.
Figure~\ref{fig:3to3_conn_unitarity} is a diagrammatic representation of Eq.~\eqref{eq:3to3connUnit} and its derivation 
is given in Appendix~\ref{sec:app_B}.

The implications of unitarity for the 
$\F_j$ are summarized below. The unitarity relation for the connected,  $\A$ amplitude is more complicated. The first term in Eq.~\eqref{eq:3to3connUnit} is analogous to the $\2\to\2$ case, in the sense that it is given by the product of the same connected amplitude  $\A$. 
The next two terms originate from the contribution to $S^{\dag}S$ given by the product of 
$T_c$ and $T_d$, and represents the situation when only two of the three particles rescatter. 
The last term is the contribution to the imaginary part of the connected amplitude from the product of two disconnect amplitudes and reflects the real one particle exchange process.  Since the unitarity relation deals with physical, on-shell amplitudes, this last contribution is non-vanishing only when the exchanged  particle is on-shell, where it is singular and proportional to $\delta(u_{jk} - \mu_{jk}^2)$.

The implications of unitarity  for the analytic properties of the $\2\to\2$ amplitude are well known \cite{Gribov:2009zz}. In the physical region the partial wave expansion
\begin{equation}\label{eq:2to2PW}
\F_{j}(\p';\p) = \sum_{s_j=0}^{\infty} \N_{s_j}^{2}
f_{s_j}(\sigma_j) P_{s_j}(\wh{\q}'_{j}  \cdot \wh{\q}_j ),
\end{equation}
converges and reduces the integral relation given by Eq.~\eqref{eq:2to2discUnit} to a countable set of algebraic ones.
Here $s_j$ is the angular momentum of the  two-particle system $j$ defined in the IRF$_j$-frame, $\N_{s_j}^2 = ( 2s_j + 1) / 4\pi$ is a normalization constant, $f_{s_j}(\sigma_j)$ is the partial wave amplitude, and
$P_{s_j}(\wh{\q}'_{j}  \cdot \wh{\q}_j )$
 is the Legendre polynomial describing the rotation dependence in terms of the cosine of the $\2\to\2$ scattering angle.  
The unitarity relation is diagonalized to the partial wave unitarity relation,
\begin{equation}\label{eq:2to2UnitPW}
\im f_{s_j}(\sigma_j) = \rho_2(\sigma_j) \lvert f_{s_j} (\sigma_j)\rvert^{2}  \Theta(\sigma_j - \sigma_j^{\sss (\mathrm{th})}).
\end{equation}
This equation is automatically satisfied by
\begin{equation}\label{eq:AmpKmat}
f_{s_{j}}^{-1}(\sigma_{j}) = K_{s_j}^{-1}(\sigma_j) - \frac{1}{\pi} \int_{\sigma_{j}^{(\mathrm{th})}}^{\infty} d\wh{\sigma} \, \frac{\rho_2(\wh{\sigma})}{\wh{\sigma} - \sigma_j}
\end{equation}
where the $K$-matrix is a real function along the unitarity cut.

The $\3\to\3$ amplitude in the physical region can be expanded in  partial waves in any of the $(12)$, $(13)$, $(23)$ subsystems. We refer to a subchannel of choice, \eg $(12)$
 as the direct channel and to the others
  as the cross channels. Since each term in the partial wave expansion is analytic in the angular variables, and therefore in the $(13)$ and $(23)$ invariant masses, singularities in the latter variables can happen only when the series  diverges.
  In contrast to the $\2\to\2$ case, the unitarity equations for each partial wave would not decouple, and would contain an infinite number of terms. Since in practice one must truncate the series, the amplitude would be regular in the $(13)$ and $(23)$ invariant masses, and the information about the cross channels dynamics would be lost. Instead, we will represent $\3\to\3$ amplitude in an isobar approximation, where only a finite number of terms in the direct and cross channels are included.
  
\section{The Isobar Representation}\label{sec:Isobar_Model}
To be concrete, the partial wave expansion of the connected $\3\to\3$ amplitude reads
\begin{equation}\label{eq:PW_expansion}
\begin{split}
\A(\p';\p) & = \sum_{J} \sum_{\ell_k',s_k'} \sum_{\ell_j,s_j} \mathcal{M}_{\ell_k' s_k'; \ell_j s_j}^{J}(\sigma_k',s,\sigma_j) \\
& \times \sum_{M}Z_{\ell_k' s_k'}^{JM\,*}(\wh{\P}_k'^{\star},\wh{\q}'_{k}) Z_{\ell_j s_j}^{J M}(\wh{\P}_j^{\star},\wh{\q}_j) ,
\end{split}
\end{equation}
where we project the amplitude onto the chosen $j$ and $k$ initial and final channels. Here $s_j$ ($s_k'$) is the angular momentum of the initial (final) pair, $\ell_j$ ($\ell_k'$) is the angular momentum between the pair and the spectator, $J$ and $M$ are the total angular momentum of the three particles and its projection, and $\mathcal{M}_{\ell_k' s_k';\ell_j s_j}^{J}$ is the partial wave amplitude. The angles $\wh{\P}_j^{\star}$ and $\wh{\P}_k'^{\star}$ are the orientations of the initial and final pair, which are related to the CMF scattering angle via $\cos{\Theta_{kj}^{\star}}=\wh{\P}_k'^{\star}\cdot\wh{\P}_j^{\star}$. The functions $Z_{\ell s}^{JM}$
contain the rotational dependence of the amplitude $\A$, which are defined as
\begin{equation}\label{eq:Zfcn}
Z_{\ell s}^{JM} (\wh{\P},\wh{\q}) = \N_{\ell} \N_{s} \sum_{\lambda=-s}^{s} \braket{J \lambda | \ell 0 s \lambda} \D_{M \lambda}^{(J)}(\wh{\P}) \D_{\lambda 0}^{(s)}(\wh{\q}).
\end{equation}
The $Z$-functions contain all the angular dependence, and they fulfill 
the orthonormality condition
\begin{equation}\label{eq:ZfcnOrtho}
\int d\wh{\P} \int d\wh{\q} \, Z_{\ell' s'}^{J'M'  *}(\wh{\P},\wh{\q}) Z_{\ell s}^{JM}(\wh{\P},\wh{\q}) = \delta_{JJ'}\delta_{MM'}\delta_{\ell\ell'}\delta_{ss'}.
\end{equation}
More details are in Appendix~\ref{sec:app_C}.

We next discuss the relation between partial wave expansion, isobar representation, and finally the isobar approximation.  
The partial wave expansion given by  Eq.~\eqref{eq:PW_expansion} is in principle an exact representation of the amplitude in the physical region of  $\3 \to \3$ scattering. However, 
unlike the analogous expansion in  $\2 \to \2$ scattering, the partial wave expansion cannot be used in practice in the $\3 \to \3$ case. 

In practice, one needs to restrict the series to a finite number of partial waves. In the physical region of 
 $\2 \to \2$ scattering, the 
 low-energy behavior of the partial waves is determined by barrier factors due to the finite range of interactions. This suppresses the strength of higher partial waves at threshold, provided the latter are regular in the cross channel Mandelstam variables. 
Cross channel exchanges generate singularities that spoil the convergence of the partial wave series. However, in the $\2 \to \2$ kinematics, these singularities do not overlap with the 
 direct channel physical region.
Therefore, the partial wave series can be safely truncated in a finite domain of 
 CMF energies above the two particle threshold.   

   This is not the case, for example,  when one of the particles can decay to the other three, and similarly it is never the case for $\3\to\3$ scattering. 
If we consider indistinguishable particles, explicit Bose symmetry is lost for the $\mathcal{M}_{\ell_k' s_k';\ell_j s_j}^{J}$ partial waves, since the partial wave expansion in the initial and final states singles out specific two-body channels. The symmetry is only recovered upon resummation.
      The isobar representation, in principle, takes care of this problem. One writes the connected $\3\to\3$ amplitude  as a redundant sum of  expansions in all the initial and final pairs to make the symmetry explicit. Bose symmetry is thus preserved upon truncation.
      
\begin{figure}[t]
    \centering
    \includegraphics[width=.85\columnwidth]{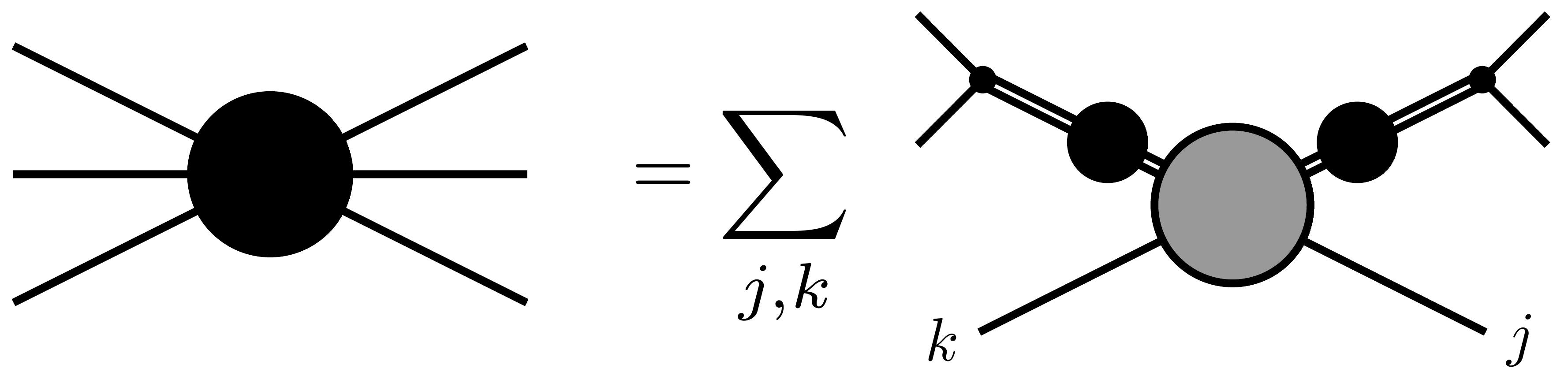}
    \caption{Diagrammatic representation of the isobar approximation amplitude in Eq.~\eqref{eq:Isobar_Expand}. The double lines with the black disk represents the isobar amplitude $f_{s_j}(\sigma_j)$, while the gray disk represents the isobar-spectator amplitude $\A_{kj}(\p';\p)$.
    }
    \label{fig:3to3_Isobar_Expand}
\end{figure}

As discussed above, one can manage only a finite number of terms in the sums over the subchannel spins.  
Therefore one reduces the isobar 
representation 
\begin{equation}\label{eq:Isobar_Expand}
\A(\p';\p) = \sum_{j,k} \A_{kj}(\p';\p),
\end{equation}
to the isobar approximation, 
by representing the connected $\3\to\3$ amplitude as a sum over a finite number of isobar-spectator amplitudes,
\begin{equation}\label{eq:PWIS_expansion}
\begin{split}
\A_{kj}(\p';\p) & = \sum_{J} \sum_{\ell_k',s_k'}^{\text{max}'} \sum_{\ell_j,s_j}^{\text{max}} \A_{\ell_k' s_k'; \ell_j s_j}^{J}(\sigma_k',s,\sigma_j) \\
& \times \sum_{M}Z_{\ell_k' s_k'}^{JM\,*}(\wh{\P}_k'^{\star},\wh{\q}'_{k}) Z_{\ell_j s_j}^{J M}(\wh{\P}_j^{\star},\wh{\q}_j) ,
\end{split}
\end{equation}
as shown in Fig.~\ref{fig:3to3_Isobar_Expand}. The truncation is reflected by ``max" in the sums. We  projected the isobar-spectator amplitudes onto the total angular momentum $J$ of the three particle system. In the following, we refer to  
 $\A_{\ell_k' s_k'; \ell_j s_j}^{J}(\sigma_k',s,\sigma_j)$
as the partial wave isobar spectator (PWIS) amplitudes. 
We emphasize that, while 
truncation in $s_k'$ and $s_j$ cannot be avoided in practice,  unitarity is diagonal in the total angular momentum. Amplitudes for each $J$ are thus independent and can in principle be resummed. 

We also stress that the PWIS amplitudes $\A_{\ell_k' s_k'; \ell_j s_j}^{J}$ are not the genuine $\3\to\3$ partial wave amplitudes $\mathcal{M}_{\ell_k' s_k'; \ell_j s_j}^{J}$ in Eq.~\eqref{eq:PW_expansion}:
\begin{equation}\label{eq:3to3_recover}
\begin{split}
\mathcal{M}_{\ell_k' s_k' ; \ell_j s_j}^{J}(\sigma_k',s,\sigma_j) & = \A_{\ell_k' s_k' ; \ell_j s_j}^{J}(\sigma_k',s,\sigma_j)  \\
&+ X_{\ell_k' s_k' ; \ell_j s_j}^{J}(\sigma_k',s,\sigma_j),
\end{split}
\end{equation}
where $X_{\ell_k' s_k' ; \ell_j s_j}^{J}$ contains all the cross channel terms which recouple to the direct channel amplitude, 
\begin{equation}\label{eq:cross_channel}
\begin{split}
& X_{\ell_k' s_k' ; \ell_j s_j}^{J}(\sigma_k',s,\sigma_j) \\
 & = \sum_{\substack{\, a\ne j,\\ b\ne k} }  \sum_{\ell_b' , s_b'} \sum_{\ell_a, s_a} \int d\wh{\P}_k'^{\star} \int d\wh{\q}_k' \, \int d\wh{\P}_j^{\star} \int d\wh{\q}_j \,   \\
& \times Z_{\ell_k' s_k' }^{J M }(\wh{\P}_k'^{\star},\wh{\q}_k') Z_{\ell_b' s_b' }^{JM\,*}(\wh{\P}_b'^{\star},\wh{\q}_b') \\
& \times  Z_{\ell_j s_j}^{JM\,*}(\wh{\P}_j^{\star},\wh{\q}_j) Z_{\ell_a s_a}^{JM}(\wh{\P}_a^{\star},\wh{\q}_a) \\
& \times \A_{\ell_b' s_b' ; \ell_a s_a}^{J}(\sigma_b',s,\sigma_a) .
\end{split}
\end{equation}
The kinematic relations given in Appendix \ref{sec:app_A} can be used to write the cross channel variables in terms of the direct channel variables. 

Often in the literature, Bose symmetry is considered as a motivation for Eq.~\eqref{eq:Isobar_Expand}. However, this is completely independent: the representation can be applied to the distinguishable particle case (in this case the various $\A_{kj}(\p';\p)$ contain different physics and have different functional forms), and Bose symmetry can be imposed to the expansion in Eq.~\eqref{eq:PW_expansion} without requiring an explicit sum over channels. For example we consider the $\pi^+\pi^-\pi^0 \to \pi^+\pi^-\pi^0$ process in the isoscalar vector channel, where the $\omega$ is observed. Thinking in isospin basis, where the three pions are indistinguishable, and in the charge basis, where they are distinguishable, leads to the same form of the amplitude, showing that Bose symmetry plays no role in defining the representation.

Isobars parameterize the $\2\to\2$ dynamics in a given subchannel and angular momentum state. Contrary to the $\2\to\2$ partial waves, they have only right hand singularities constrained by unitarity. In the $N/D$ formalism, the isobars can be identified with the denominator function, where the left hand cuts are removed via a dispersive integral~\cite{Bjorken:1960zz}. In the following, we will ignore all left hand singularities of the $\2\to\2$ amplitudes, and identify their partial waves with the isobars.  Although we do not need to assume any resonant content for the isobars (\eg we could use an isobar to describe the $\pi^+\pi^+$ dynamics), it is a popular picture to think of them as a quasi-particle, and to identify the invariant mass and angular momentum of the pair with the isobar mass and spin. Isobars are customarily labeled with the name of the dominant resonance, if any. 
Isobars can be parameterized as in Eq.~\eqref{eq:AmpKmat}. 
For example, the $a_1(1260)$ decays into three pions dominantly in the $\rho\pi$ and $\sigma\pi$ channels~\cite{Tanabashi:2018oca}. If one chooses to perform a truncated partial wave expansion of the $\3\to\3$ amplitude in only the $\rho\pi \to a_{1}(1260) \to \rho\pi$ channel, rescattering effects between the $\rho\pi \to \sigma\pi$ isobars are ignored. The isobar approximation corrects this by including amplitudes for $\sigma\pi \to a_{1}(1260) \to \rho\pi$, $\rho\pi \to a_{1}(1260) \to \sigma\pi$, and $\sigma\pi \to a_{1}(1260) \to \sigma\pi$. 

The approximation is expected to be valid at low values of energy, where a finite number of singularities dominate the amplitude. Moving to higher energies, the left hand cuts controlling the crossed $\2\to\mathbf{4}$ processes will become relevant, and the behavior of the amplitude will be controlled by analyticity in angular momentum, rather than direct-channel unitarity.

Since the isobar approximation includes the cross channel effects in the summation, the isobar-spectator amplitudes contain only normal threshold singularities determined by unitarity. Therefore, the analytic structure of each isobar-spectator amplitude 
in the energy variables, $s$, $\sigma_j$, and $\sigma_k'$, are determined by unitarity.

The problem of convergence in $J$ is more severe. The $\3 \to \3$ amplitude contains an OPE process  (see the last diagram in Fig.~\ref{fig:3to3_conn_unitarity}), which can go 
  on-shell in the direct channel,  and results in an interaction of infinite range. In this case the cross channel singularities overlap with the physical region and project onto an infinite number of partial waves. The analytic properties of the projected amplitude 
   are highly nontrivial. We discuss them in detail 
    detail in Section~\ref{sec:AnalyticProperties}. 
However, since the main goal in this and similar studies of three particle scattering is to identify the spectrum, ultimately one needs to deal with amplitudes of well defined total angular momentum $J$. In other words,  these amplitudes diagonalize unitarity, which is the basis for analytic continuation and identification of complex singularities as resonance poles. For this reason, in the following we will not address the problem of convergence in $J$.

\subsection*{Unitarity Relations}\label{sec:IM_unitarity}
It is advantageous to introduce an amputated PWIS amplitude $\wt{\A}_{\ell_k' s_k';\ell_j s_j}^{J}$, in which the isobar amplitudes are factorized,
\begin{equation}\label{eq:amputation}
\A_{\ell_k' s_k'; \ell_j s_j}^{J}= f_{s_k'}(\sigma_k') \wt{\A}_{\ell_k' s_k'; \ell_j s_j}^{J}(\sigma_k',s,\sigma_j) f_{s_j}(\sigma_j).
\end{equation}
The amputation reduces the number of terms in the isobar-spectator unitarity relation by making use of subchannel unitarity in Eq.~\eqref{eq:2to2UnitPW}. However, the amputated PWIS amplitudes still have a non-trivial dependence on the subchannel energies due to rescattering effects. As shown in detail in Appendix \ref{sec:app_C}, combining Eqs.~\eqref{eq:3to3connUnit}, \eqref{eq:2to2UnitPW}, \eqref{eq:Zfcn}, \eqref{eq:ZfcnOrtho},  \eqref{eq:PWIS_expansion}, and \eqref{eq:amputation}  results in the amputated PWIS unitarity relation
\begin{widetext}
\begin{equation}\label{eq:PWIS_unitarity}
\begin{split}
 \im \, & \wt{\A}^{J}_{ \ell_k' s_k'  ; \ell_j s_j }(\sigma_k',s,\sigma_j) \\
 & = \frac{1}{\pi(32\pi^2)^2} \sum_{n} \sum_{\ell_n'' , s_n''} \int_{\sigma_{n}^{(\mathrm{th})}}^{(\sqrt{s} - m_n)^2} d\sigma_n'' \, \frac{\lvert \q_n'' \rvert \lvert \p_n''^{\star} \rvert }{ \sqrt{\sigma_n''} \sqrt{s}} \lvert f_{s_n''}(\sigma_n'') \rvert^2 \, \wt{\A}^{J\,*}_{\ell_n'' s_n'' ; \ell_k' s_k' }(\sigma_n'',s,\sigma_k') \wt{\A}^{J}_{\ell_n'' s_n'' ; \ell_j s_j }(\sigma_n'',s,\sigma_j) \Theta(s - s_{\mathrm{th}})  \\
\
&  + \frac{1}{2\pi s(32\pi^2)^2} \sum_{\substack{n,r \\ n \ne r}} \sum_{\ell_n'' , s_n''}\sum_{\ell_r'' , s_r''} \int_{\sigma_{n}^{(\mathrm{th})}}^{(\sqrt{s} - m_n)^2} d\sigma_n'' \,\int_{\sigma_r^{(-)}(\sigma_n'')}^{\sigma_r^{(+)}(\sigma_n'')} d\sigma_r'' \,   f_{s_r''}^{*}(\sigma_r'')f_{s_n''}(\sigma_n'')   \\
\
& \qquad\qquad \times \C_{\ell_n'' s_n'' ; \ell_r'' s_r''}^{J}(\sigma_n'',s,\sigma_r'')\,  \wt{\A}^{J\,*}_{\ell_r'' s_r''; \ell_k' s_k' }(\sigma_r'',s,\sigma_k') \wt{\A}^{J}_{\ell_n'' s_n'' ;  \ell_j s_j }(\sigma_n'',s,\sigma_j) \Theta(s - s_{\mathrm{th}}) \\
\
&  +  \frac{1}{64 \pi^2 \sqrt{s}}  \frac{1}{\lvert \p_k^{\prime\,\star} \rvert}  \sum_{\substack{r \ne k}} \sum_{\ell_r'', s_r''}\, \int_{\sigma_r^{(\mathrm{th})}}^{(\sqrt{s}-m_r)^2} d\sigma_r'' \,  \C_{\ell_k' s_k' ; \ell_r'' s_r''}^{J}(\sigma_k',s,\sigma_r'') \, f_{s_r''}(\sigma_r'') \, \wt{\A}_{\ell_r'' s_r''  ; \ell_j s_j }^{J}(\sigma_r'',s,\sigma_j) \Theta(\sigma_k' - \sigma_{k}^{(\mathrm{th})}) \\
\
&  + \frac{ 1 }{64 \pi^2 \sqrt{s}} \frac{1}{\lvert \p_j^{\star} \rvert}  \sum_{\substack{n \ne j}} \sum_{\ell_n'' , s_n''} \,  \int_{\sigma_n^{(\mathrm{th})}}^{(\sqrt{s}-m_n)^2} d\sigma_n'' \,   \C_{\ell_n'' s_n'' ; \ell_j s_j}^{J}(\sigma_n'',s,\sigma_j) \, f_{s_n''}^{*}(\sigma_n'') \, \wt{\A}_{ \ell_k' s_k'; \ell_n'' s_n''}^{J\,*}(\sigma_k',s,\sigma_n'') \Theta(\sigma_j - \sigma_{j}^{(\mathrm{th})})  \\
\
&   + \frac{\pi}{2\lvert \p_j^{\star} \rvert \lvert \p_k'^{\star} \rvert}  \, \C_{\ell_k' s_k' ; \ell_j s_j}^{J}(\sigma_k',s,\sigma_j) 
(1 - \delta_{jk}) \Theta(1 - \cos^2\theta_{kj}^{\star})  ,
\end{split}
\end{equation}
where $\C_{\ell_n'' s_n'' ; \ell_r'' s_r''}^{J}(\sigma_n'',s,\sigma_r'') $ is a purely kinematical recoupling coefficient between different intermediate state isobars, 
\begin{equation}\label{eq:recoupling_coef}
\begin{split}
\C_{\ell_k s_k ; \ell_j s_j}^{J}(\sigma_k,s,\sigma_j) & =  2\pi \, \N_{s_j} \N_{s_k} \N_{\ell_j}\N_{\ell_k} \N_{J}^{-2} \sum_{\lambda_k,\lambda_k} \braket{J \lambda_k | \ell_k 0 s_k \lambda_k} \braket{J \lambda_j | \ell_j 0 s_j \lambda_j}\,\\
& \times d_{\lambda_k 0}^{(s_k)}(\cos\chi_k
)\,d_{\lambda_k \lambda_j}^{(J)}(\cos\theta_{kj}^{\star}
) \,d_{\lambda_j 0}^{(s_j)}(\cos\chi_j
) .
\end{split}
\end{equation}
The recoupling coefficients relate two different orientations of three particles in the same frame ~\cite{Ascoli:1974sp,Ascoli:1975mn,Giebink:1985zz}. Appendix 
\ref{sec:app_C} contains details on the derivation of the recoupling coefficients from the rotational matrices in Eq.~\eqref{eq:Zfcn}. The helicity angles and the CMF angle between particles $j$ and $k$, $\theta_{kj}^{\star}$, are functions of the invariants (\cf Appendix \ref{sec:app_A}).
\end{widetext}

\begin{figure*}[t!]
    \centering
    \includegraphics[ width=0.8\textwidth]{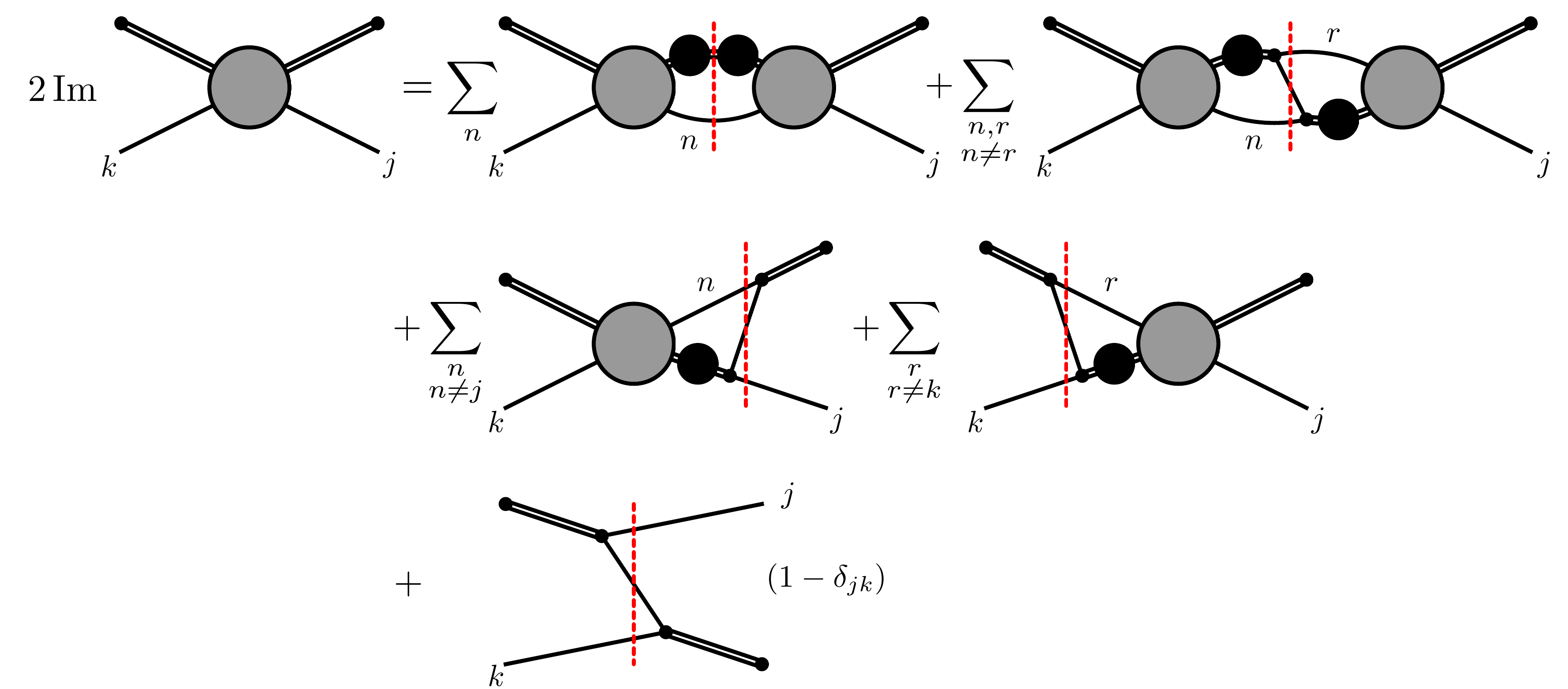}
    \caption{Diagrammatic representation for the amputated PWIS unitarity relation in Eq.~\eqref{eq:PWIS_unitarity}. The black disks in the internal legs represent the isobars, which are amputated from the external legs, see Eq.~\eqref{eq:amputation}. The cuts across the OPE in the intermediate states yield recoupling coefficients. 
     }
    \label{fig:PWIS_unitarity}
\end{figure*}

The second term contains two integrals over the Dalitz region of the three-particles in the intermediate state, where the physical region is bounded by $\sigma_n^{(\mathrm{th})} \le \sigma_n'' \le (\sqrt{s} - m_n)^2$ and $\sigma_r^{(-)} \le \sigma_r'' \le \sigma_r^{(+)}$, where $\sigma_r^{\,(\pm)}$ is a function of $\sigma_n''$ and gives the physical boundary $\cos\chi_{n}'' = \pm 1$, \eg for $n = 1$ and $r=3$, 
\begin{equation}
\begin{split}
\sigma_3^{(\pm)}(\sigma_1'') & = m_1^2 + m_2^2 - \frac{1}{2\sigma_1''} (\sigma_1'' - s + m_1^2)(\sigma_1'' + m_2^2 - m_3^2) \\
&  \pm \frac{1}{2\sigma_1''} \lambda^{1/2}(s,\sigma_1'',m_1^2)\lambda^{1/2}(\sigma_1'',m_2^2,m_3^2).
\end{split}
\end{equation}
Eq.~\eqref{eq:PWIS_unitarity} is illustrated in Fig.~\ref{fig:PWIS_unitarity}. Appendix \ref{sec:app_C} contains a sketch of the derivation of the amputated PWIS unitarity relations.
The first term of Eq.~\eqref{eq:PWIS_unitarity} involves the direct propagation of an isobar in the intermediate state, whereas the second, third, and fourth term involve the exchange of a particle between isobars. 
The rescattering between isobars modifies the line shape of the isobar amplitudes~\cite{Niecknig:2012sj,Danilkin:2014cra}. The final term is the contribution from the OPE process, which gives and additional imaginary part to the amplitude in the physical region.
At this stage we have not factored out the threshold factors from partial waves. This is straightforward to implement, however we do not do it
here as we consider angular momenta in $S$-wave in further sections.

\section{The B-Matrix Parameterization}\label{sec:B-Matrix}
\begin{figure}[t!]
    \centering
    \includegraphics[width=0.95\columnwidth]{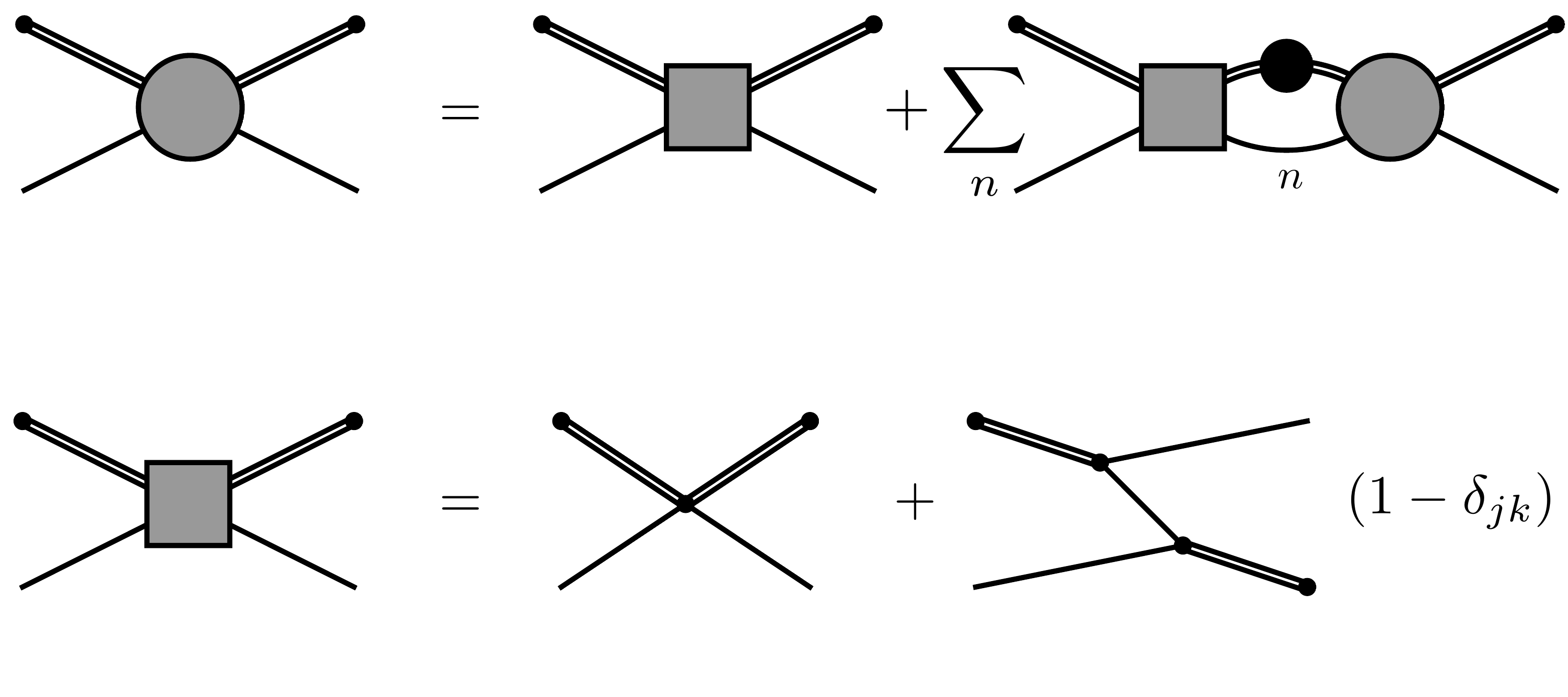}
    \put(-120,55){{(a)}}
    \put(-120,-5){{(b)}}
    \caption{(a) Diagrammatic representation of the  $B$-matrix parameterization in Eq.~\eqref{eq:BMatrixParam}. The gray disk represent the amputated PWIS, and the gray box  the $B$-matrix. (b) The $B$-matrix is composed of  a short-range real $\R$ amplitude, and the OPE $\E$, see Eq.~\eqref{eq:Bmatrix}. 
     }
    \label{fig:B-Matrix_param}
\end{figure}

Motivated by $S$-matrix theory, we present a parameterization for the PWIS amplitudes that satisfies real axis unitarity given by Eq.~\eqref{eq:PWIS_unitarity}. In the $\2\to\2$ case, the $K$-matrix, $f^{-1} = K^{-1} - i\rho_2$, is an example of a parameterization satisfying unitarity.
For the $\3\to\3$ case, we present the $B$-matrix parameterization for the PWIS amplitudes.  The $B$-matrix parameterization is a linear integral equation for the amputated PWIS amplitudes that satisfy the unitarity relations Eq.~\eqref{eq:PWIS_unitarity}:
\begin{equation}\label{eq:BMatrixParam}
\begin{split}
& \wt{\A}_{\ell_k' s_k' ; \ell_j s_j}^{J}(\sigma_k',s,\sigma_j)  = \wt{\B}_{\ell_k' s_k' ; \ell_j s_j}^{J}(\sigma_k',s,\sigma_j) \\
& + \sum_{n} \sum_{\ell_n'',s_n''} \int_{\sigma_n^{(\mathrm{th})}}^{(\sqrt{s} - m_n)^2} d\sigma_n'' \,  \wt{\B}_{\ell_k' s_k' ; \ell_n'' s_n''}^{J}(\sigma_k',s,\sigma_n'') \\
& \qquad\qquad \times  \tau_n(s,\sigma_n'') \wt{\A}_{\ell_n'' s_n'' ; \ell_j s_j}^{J}(\sigma_n'',s,\sigma_j),
\end{split}
\end{equation}
where the $B$-matrix $\wt{\B}_{\ell_k' s_k' ; \ell_j s_j}^{J}$ contains two terms,
\begin{equation}\label{eq:Bmatrix}
\wt{\B}_{\ell_k' s_k' ; \ell_j s_j}^{J} = \wt{\R}_{\ell_k' s_k' ; \ell_j s_j}^{J} + \wt{\E}_{\ell_k' s_k' ; \ell_j s_j}^{J}.
\end{equation}
The function $\wt{\E}_{\ell_k' s_k' ; \ell_j s_j}^{J}$ is the amputated partial wave OPE amplitude, $\wt{\R}_{\ell_k' s_k' ; \ell_j s_j}^{J}$ is a real function that represents the short-distance three-body interactions unconstrained by unitarity, and $\tau_n$ is the product of the isobar-spectator phase space between and of the isobar amplitude
\begin{equation}\label{eq:tau}
\tau_{n}(s,\sigma_n) = \rho_{3}(s,\sigma_n) f_{s_n}(\sigma_n),
\end{equation}
with 
\begin{equation}\label{eq:3bodyPhaseSpace}
\rho_{3}(s,\sigma_n) = \frac{1}{64\pi^3 } \frac{2\lvert \p_n^{\star} \rvert }{\sqrt{s}}.
\end{equation}
The parameterization is diagrammatically represented in Fig.~\ref{fig:B-Matrix_param}. The OPE amplitude is defined as
\begin{equation}\label{eq:OPE}
\begin{split}
\E_{kj}(\p';\p) & = \F_k(\p';\p)  \frac{1}{ \mu_{jk}^2 - u_{jk} -i\epsilon } \F_j(\p';\p),
\end{split}
\end{equation}
where we note that the OPE only contributes to off-diagonal amplitudes, \ie $j\ne k$. In principle, the OPE could contain a regular function of the energy in addition to the pole term, however unitarity only constrains the pole, and we assume all other real functions to be absorbed by $\R$.
The amputated partial wave projected OPE amplitude $\wt{\E}_{\ell_k' s_k' ; \ell_j s_j}^{J}$ can be constructed using Eqs.~\eqref{eq:PWIS_expansion} and \eqref{eq:amputation}.
The $\R$ represents the freedom of short-distance physics for the scattering of three particles, and can be any real function. In an effective field theory approach, it represents a low order polynomial of contact interactions. For simplicity, in the following we assume the latter for $\R$.
Appendix \ref{sec:app_E} illustrates how the $B$-matrix parameterizations satisfies the amputated PWIS unitarity relations.
Aspects of its analytical properties are examined in Sec.~\ref{sec:AnalyticProperties}.

The $B$-matrix parameterization in Eq.~\eqref{eq:BMatrixParam} differs from Mai et al.~\cite{Mai:2017vot} in the lower limit of the integral: the latter is derived using Lippman-Schwinger equations with a relativistic potential model, and includes contributions from the unphysical subthreshold region, $\sigma_n < \sigma_n^{(\text{th})}$. Obviously, both parameterization have the same imaginary part in the physical region, since both satisfy unitarity. 

\begin{figure*}
    \centering
    \includegraphics[ width=1.\textwidth]{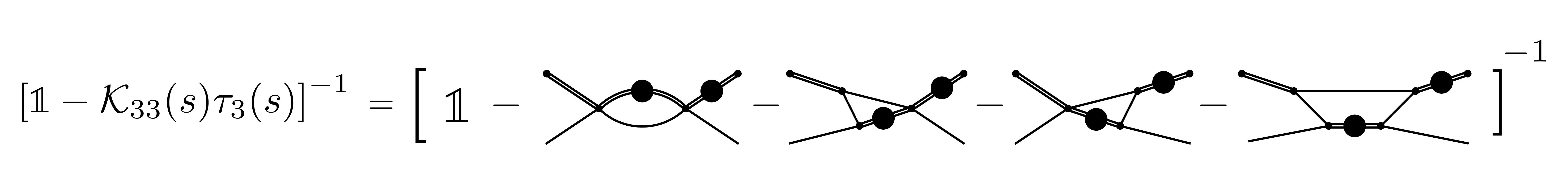}
    \put(-330,0){{(a)}}
    \put(-240,0){{(b)}}
    \put(-150,0){{(c)}}
    \put(-80,0){{(d)}}
    \caption{The denominator of the $B$-matrix parameterization contains four primitive diagrams associated with the rescattering of the $B$-matrix: (a) bubble diagram, (b) and (c) triangle diagrams, and (d) box diagram.}
    \label{fig:denom}
\end{figure*}

For notational simplicity, let $\wt{\A}_{kj}(s) \equiv \wt{\A}_{\ell_k' , s_k' ; \ell_j, s_j }^{J }(\sigma_k',s,\sigma_j)$, so that the amplitudes are matrices in the isobar sub-energies and angular momenta, which are indicated by the spectator indices. Equation~\eqref{eq:BMatrixParam} is then a matrix relation with the integrations over intermediate isobars formally represented as matrix multiplications. Recalling that we work with the convention that isobars exists only in the (12) and (32) channels, we write the $B$-matrix parameterization as the set of coupled equations
\begin{subequations}\label{eq:Bmat_coupled_12}
\begin{align}
\wt{\A}_{13}(s) & = \wt{\B}_{13}(s) + \wt{\B}_{13}(s)\tau_{3}(s) \wt{\A}_{33}(s) ,\label{eq:Bmat_coupled_1} \\
\wt{\A}_{33}(s) & = \wt{\B}_{31}(s)\tau_{1}(s)\wt{\A}_{13}(s) \label{eq:Bmat_coupled_2} ,
\end{align}
\end{subequations}
with the other two amplitudes given by a similar set of equations,
\begin{subequations}\label{eq:Bmat_coupled_34}
\begin{align}
\wt{\A}_{31}(s) & = \wt{\B}_{31}(s) + \wt{\B}_{31}(s)\tau_{1}(s) \wt{\A}_{11}(s) ,\label{eq:Bmat_coupled_3} \\
\wt{\A}_{11}(s) & = \wt{\B}_{13}(s)\tau_{3}(s)\wt{\A}_{31}(s) \label{eq:Bmat_coupled_4} .
\end{align}
\end{subequations}
The Eqs.~\eqref{eq:Bmat_coupled_12} can be combined into one integral equation for $\wt{\A}_{13}$,
\begin{equation}\label{eq:BMat_integral1}
\wt{\A}_{13}(s) = \wt{\B}_{13}(s) + \Kc_{11}(s) \tau_1(s) \wt{\A}_{13}(s),
\end{equation}
where the kernel $\Kc_{11}$ is
\begin{equation}\label{eq:kernel}
\Kc_{11}(s) = \wt{\B}_{13}(s) \tau_{3}(s) \wt{\B}_{31}(s).
\end{equation}
Similarly, Eqs.~\eqref{eq:Bmat_coupled_34} give
\begin{equation}\label{eq:BMat_integral2}
\wt{\A}_{31}(s) = \wt{\B}_{31}(s) + \Kc_{33}(s) \tau_3(s) \wt{\A}_{31}(s),
\end{equation}
where the kernel $\Kc_{33}$ is given by exchanging the $1 \leftrightarrow 3$ indices in Eq.~\eqref{eq:kernel}. Eqs.~\eqref{eq:BMat_integral1} and \eqref{eq:BMat_integral2} can be formally inverted to yield the solutions,

\begin{subequations}\label{eq:Bmat1234}
\begin{align}
\wt{\A}_{13}(s) & = \left[ \mathbbm{1} - \Kc_{11}(s) \tau_1(s) \right]^{-1} \wt{\B}_{13}(s), \label{eq:Bmat1} \\
\wt{\A}_{33}(s) & = \left[ \mathbbm{1} - \Kc_{33}(s) \tau_{3}(s) \right]^{-1} \Kc_{33}(s), \label{eq:Bmat2} \\
\wt{\A}_{31}(s) & = \left[ \mathbbm{1} - \Kc_{33}(s) \tau_3(s) \right]^{-1} \wt{\B}_{31}(s), \label{eq:Bmat3} \\
\wt{\A}_{11}(s) & = \left[ \mathbbm{1} - \Kc_{11}(s) \tau_{1}(s) \right]^{-1} \Kc_{11}(s), \label{eq:Bmat4}
\end{align}
\end{subequations}
Several terms can be identified in the kernels, $\Kc_{kj}(s) = \mathcal{G}_{kj}(s) + \mathcal{H}_{kj}(s) + \mathcal{T}_{kj}^{(1)}(s) + \mathcal{T}_{kj}^{(2)}(s)$, where $\mathcal{G}$ is a bubble diagram, $\mathcal{H}$ is a box diagram, and the $\mathcal{T}$'s are triangle diagrams, generated by integrals over OPE and contact terms in Eq.~\eqref{eq:Bmatrix}.  Explicitly,
\begin{subequations}\label{eq:kernels_all}
\begin{align}
\mathcal{G}_{kj}(s) & = \sum_{n}\wt{\R}_{kn}(s)\tau_{n}(s) \wt{\R}_{nj}(s), \label{eq:kernel_bubble} \\
\mathcal{T}_{kj}^{(1)}(s) & = \sum_{n}\wt{\E}_{kn}(s) \tau_{n}(s) \wt{\R}_{nj}(s), \label{eq:kernel_tri1}\\
\mathcal{T}_{kj}^{(2)}(s) & = \sum_{n}\wt{\R}_{kn}(s) \tau_{n}(s) \wt{\E}_{nj}(s), \label{eq:kernel_tri2} \\
\mathcal{H}_{kj}(s) & = \sum_{n}\wt{\E}_{kn}(s) \tau_n(s) \wt{\E}_{nj}(s). \label{eq:kernel_box}
\end{align}
\end{subequations}
These diagrams occur in the denominators of the amplitudes in Eqs.~\eqref{eq:Bmat1234}, \cf Fig.~\ref{fig:denom}.
They differ to the Feynman diagrams obtained in a perturbative QFT since the integrations are only over the physical region, changing the analytic structure below threshold (see Sec.~\ref{sec:AnalyticProperties}).

The solutions can be interpreted as an infinite series of exchange and bubble diagrams. For example, expanding the solution for $\wt{\A}_{13}$,
\begin{equation}
\begin{split}
\wt{\A}_{13}(s) & = \wt{\B}_{13}(s) + \Kc_{11}(s)\tau_1(s) \wt{\B}_{13}(s) \\
&  + \Kc_{11}(s)\tau_1(s)\Kc_{11}(s)\tau_1(s)\wt{\B}_{13}(s) + \cdots.
\end{split}
\end{equation}
The first term is the OPE and contact interaction, the second term is a ladder diagram with three exchanges, and various combinations of bubbles and OPE, and so on. The unitarization of bubble diagrams has been considered in quasi-two-body models \cite{Basdevant:1978tx,Bhandari:1982ck,Jackura:2016llm, Mikhasenko:2017jtg}. In these models it is easy to show how additional cuts appear in the unphysical sheets due to the isobar decay.

Three-body resonances manifest as poles in the complex $s$-plane of the scattering amplitude.
Rearranging the constituents of the kernel relates the two denominators
\begin{equation}\label{eq:denom_shift}
\begin{split}
 \wt{\B}_{13}(s)\tau_{3}(s) & \left[\mathbbm{1}  - \Kc_{33}(s)\tau_3(s)\right]^{-1}  \\
& \,\, = \left[\mathbbm{1} - \Kc_{11}(s)\tau_1(s)\right]^{-1} \wt{\B}_{13}(s) \tau_3(s)  .
\end{split}
\end{equation}
Thus, we can write the full $\3\to\3$ amplitude in terms of a single Fredholm determinant. The determinants are independent of the external isobar energies, and the intermediate integrations will modify the phase space factors to incorporate rescattering effects. Resonance poles can be determined by solving
\begin{equation}\label{eq:det}
\det{\left[ \mathbbm{1} - \Kc_{11}(s) \tau_{1}(s) \right]} = 0.
\end{equation}
The $B$-matrix solutions are real-boundary values of analytic functions in the complex $s$-plane. The physical amplitudes are defined by $s \to s + i\epsilon$, $\sigma_j \to \sigma_j + i\epsilon$, and $\sigma_k' \to \sigma_k' + i\epsilon$. Aspects of its analytic properties are discussed in the following section.

 \subsection*{Relation to  the finite volume formalism }
In finite volume studies for lattice QCD, 
 substantial progress has been made to understand the connection between discrete energy levels and properties of hadron scattering 
 amplitudes ~\cite{Hansen:2014eka,Hansen:2015zga,Hansen:2016ync,Briceno:2017tce,Briceno:2018mlh,Polejaeva:2012ut,Mai:2017bge,Hammer:2017uqm,Hammer:2017kms,Doring:2018xxx,Mai:2018djl}. In the case of 
  $\2\to\2$ scattering the two-particle finite volume spectrum constrains the values of the infinite volume partial wave amplitudes via the L\"uscher quantization condition~\cite{Luscher:1986pf}.
The multi-variable  nature of  $\3\to\3$ scattering amplitudes makes the derivation of the finite volume quantization condition much more complicated and different groups have approached the problem from a different angle. For example, in Refs.~\cite{Hansen:2014eka,Hansen:2015zga,Hansen:2016ync,Briceno:2017tce,Briceno:2018mlh}
 the authors introduce amplitudes labeled by subchannel spins and the spectator 3-momenta. 
 Furthermore,  ladder diagrams generated by OPE are considered independently from other interactions. This implies that partial wave projection to total spin, which is necessary if one is interested in extracting properties of three-body resonances, would be performed after resummation of the OPE ladder.   On the other hand, in Refs.~\cite{Mai:2017bge,Mai:2018djl}, the quantization conditions are derived starting from a set of amplitudes projected onto the total and subchannel spins from the start, \ie before OPE resummation, in a spirit close to our work. In Refs.~\cite{Polejaeva:2012ut,Hammer:2017uqm,Hammer:2017kms} the quantization conditions are derived in a nonrelativistic EFT framework, and the direct comparison with our $S$-matrix approach is more complicated.
It is more interesting to discuss the differences with Refs.~\cite{Hansen:2014eka,Hansen:2015zga,Hansen:2016ync,Briceno:2017tce}. Since we do not aim to address the subtleties of the finite volume here, we compare with the 
 infinite volume equations derived in there on the basis of the finite volume formalism. For simplicity 
   we ignore coupling to the $\2$-body channel.
In Refs.~\cite{Hansen:2014eka,Hansen:2015zga,Hansen:2016ync,Briceno:2017tce}, the $\3 \to \3$ connected amplitude is denoted by  $\mathcal{M}_{33}(\vec k,\vec k')$ (see Eq. (112) in Ref.~\cite{Briceno:2017tce}). It  contains the resummed OPE ladder 
and the amputated amplitude $\mathcal{T}_{33}(\vec k,\vec k')$ that is generated by the kernel 
$\mathcal{K}_{\text{df},33}(\vec k,\vec k')$, which is analogous to our driving term  $\wt{\R}_{\ell_k' s_k' ; \ell_j s_j}^{J}(\sigma_k',s,\sigma_j)$. 
Both the OPE ladder and the amplitude 
 $\mathcal{T}_{33}(\vec k,\vec k')$ are solutions of linear integral equations (see Eqs. (87) and (106) in Ref.~\cite{Briceno:2017tce}), 
  which are analogous to our Eq.~\eqref{eq:BMatrixParam}.

 To further illustrate the connection between our amplitudes and those of Ref.~\cite{Hansen:2014eka,Briceno:2017tce} we shall consider the case of three identical particles in $S$-waves. The phase space $\rho_2$ (Eq.~\eqref{eq:2bodyPS}) of the two particle subsystem gains a factor $1/2!$ to account for their identical nature. The resulting unitarity relations have a similar form as in Eq.~\eqref{eq:PWIS_unitarity}. In matrix notation, one finds
\begin{equation}\label{imd}
\begin{split}
\im{\wt{\A}} & = \im{\wt{\E}} + \im{\wt{\E}} \rho_3 f \wt{\A} + \wt{\A}^{*} f^{*} \rho_3 \im{\wt{\E}} \\
& + \wt{\A}^* \rho_3 \im f \wt{\A} + \wt{\A}^* f^* \rho_3 \im\wt{\E} \rho_3 f \wt{\A},
\end{split}
\end{equation}
where the matrices are in the $\sigma'$, $\sigma$ space with $f$ and $\rho_3$ diagonal matrices. The $S$-wave projection of the OPE is given by the symmetric matrix $\wt{\E}$, and is found by the inverse relation of Eq.~\eqref{eq:PWIS_expansion} on the OPE amplitude in Eq.~\eqref{eq:OPE}. It is straightforward to show that the $B$-matrix parameterization (\cf Eq.~\eqref{eq:BMatrixParam}), $\wt{\A} = \wt{\B} + \wt{\B} \rho_3 f \wt{\A}$, \ie
\begin{equation}
\wh{\A} = [1 - (\wt{\E}  + \wt{\R}) \rho_3 f ]^{-1} (\wt{\E} + \wt{\R})  \label{alat}
\end{equation}
satisfies the unitarity relation, Eq.~\eqref{imd}. 
Moreover, after simple manipulations, Eq.~\eqref{alat} can be rewritten as 
\begin{equation}  
 \A = f \wt{\A} f 
 = \mathcal{D} + 
\mathcal{L} [1 - 
\wt{R} \rho_3 \mathcal{L}  ]^{-1} \wt{R}
 \mathcal{L}^{\top} 
\end{equation}
where the $\mathcal{D}$ amplitude is the ladder sum of OPE, given by 
\begin{equation} 
\mathcal{D} = f \wt{\E}  f  +  f \wt{\E}\rho_3  \mathcal{D}
\end{equation} 
and $\mathcal{L} \equiv f + \mathcal{D} \rho_3$.   Finally we introduce the amplitude 
 $\mathcal{T}$ satisfying  
\begin{equation} 
\mathcal{T} = \wt{\R}  + 
\wt{\R} \rho_3 \mathcal{L}
 \mathcal{T},
\end{equation} 
and obtain an expression closely resembling that in Refs.~\cite{Hansen:2014eka,Briceno:2017tce},
\begin{equation} 
\A = \mathcal{D}  + \mathcal{L} \mathcal{T}\mathcal{L}^{\top}.\label{af}
\end{equation}
The difference between Eq.~\eqref{af} and the corresponding expression for ${\cal M}_{33}^{(u,u)}$ in Refs.~\cite{Hansen:2014eka,Briceno:2017tce} is in the definition of $\mathcal{L}$. In our notation, the $\mathcal{L}$ of Refs.~\cite{Hansen:2014eka,Briceno:2017tce} contains an additional $1/3$ constant, and the $f$ and $\mathcal{D}$ matrices contain an extra factor of $\rho_2$.

  Although these analogies should be verified with care, two main differences appear. One is in the  treatment of the OPE dynamics, which in Ref.~\cite{Briceno:2017tce}  is resummed before projection onto the total spin and in our case the projection is done first. It is likely that these approaches will ultimately prove to be equivalent, since in practical applications only a finite number of partial waves in total spin or spectator momentum components can be kept.
The other difference is in the $\mathcal{L}$ function, which could possibly be related to a discrepancy in the definitions of $\wt{\R}$ and $\mathcal{K}_{\textrm{df},33}$. It would be interesting to see if our approach and the corresponding equation of Refs.~\cite{Hansen:2014eka,Briceno:2017tce} provide the same result, and to determine the origin of the difference.

\section{Aspects of Analytic Properties}\label{sec:AnalyticProperties}
In this section, we examine the singularities of the OPE amplitude and the triangle diagram from the $B$-matrix parameterization. We numerically evaluate an amplitude where all external particles have unit mass ($m_1 = m_2 = m_3 = 1$) and coupling. In these  studies, the units are arbitrary. For simplicity, we consider  \sw waves only, \ie $J(\ell' s')_{k} (\ell s)_{j} = 0(00)_{k}(00)_{j}$. Generalizing to nonzero angular momenta does not change the analytic properties.

\subsection{One Particle Exchange}\label{sec:ope}
As seen in Eqs.~\eqref{eq:kernels_all}, the building block for the $B$-matrix kernels is the OPE amplitude.
Projecting Eq.~\eqref{eq:OPE} using Eq.~\eqref{eq:PWIS_expansion} gives the $S$-wave OPE amplitude,
\begin{equation}\label{eq:OPE_Swave}
\begin{split}
\wt{\E}^S_{kj}(\sigma_k',s,\sigma_j) & = \frac{1}{4 \lvert \p_k'^{\star}\rvert \lvert \p_j^{\star}\rvert}  \log{ \left( \frac{z_{kj} - 1 }{z_{kj} + 1} \right) }.
\end{split}
\end{equation}
where $z_{kj}$ is given as 
\begin{equation}\label{eq:z_mu}
\begin{split}
z_{kj}  = \frac{2s(\sigma_{j} + m_{k}^2 - \mu_{jk}^2) - (s + \sigma_{j} - m_j^2) (s + m_k^2 - \sigma_{k}' )  }{ \lambda^{1/2}(s,\sigma_{j},m_j^2)\lambda^{1/2}(s,\sigma_{k}',m_k^2) },
\end{split}
\end{equation}
where $\lambda(a,b,c) = a^2 + b^2 + c^2 - 2(ab+bc+ca)$ is the K\"all\'en triangle function. 
Eq.~\eqref{eq:z_mu}.
We investigate the OPE as a function of $s$ for fixed real $\sigma_j$ and $\sigma_k'$. The imaginary part of the OPE is
\begin{equation}\label{eq:OPE_Swave_imag}
\im {\wt{\E}^S_{kj}(\sigma_k',s,\sigma_j) } = \frac{\pi }{4\lvert \p_j^{\star} \rvert \lvert \p_k'^{\star} \rvert}  \Theta(1 - \lvert z_{kj} \rvert^2),
\end{equation}
which is given by the unitarity relations in Eq.~\eqref{eq:PWIS_unitarity}. 
The OPE has four branch points in $s$, one at zero, one at infinity, and two which we label $s_{kj}^{(\pm)}$,
\begin{equation}\label{eq:spm}
\begin{split}
s_{kj}^{(\pm)} & = \frac{1}{2\mu_{jk}^2} \bigg[ (m_k^2 - \sigma_j)(m_j^2 - \sigma_k') - \mu_{jk}^4 \\
&  + \mu_{jk}^2(m_k^2 + m_j^2 + \sigma_j + \sigma_k') \\
&  \pm \lambda^{1/2}(\mu_{jk}^2,m_k^2,\sigma_j)  \lambda^{1/2}(\mu_{jk}^2,m_j^2,\sigma_k') \bigg],
\end{split}
\end{equation}
which depend on the isobar invariant masses. The momenta in the denominator do not contribute additional branch points, because the logarithm vanishes and cancel the singularity, as expected from the \sw-wave threshold behavior. The $s_{kj}^{(\pm)}$ branch points are in general complex. There are then two branch cuts: one where $s\in (-\infty,0]$, called the virtual particle exchange (VPE) cut, and one connecting $s_{kj}^{(-)}$ to $s_{kj}^{(+)}$, called the real particle exchange (RPE) cut. The VPE cut is associated with the exchange of a virtual particle, generating long-range forces. Historically, the RPE cut is associated with the exchange of a  real particle between isobars, \ie when it is kinematically allowed for an isobar to decay. This corresponds to when the RPE branch points lie on the real axis above the isobar-spectator threshold. If the isobar invariant masses are below the decay threshold, then the RPE branch points move in the complex plane below the isobar-spectator threshold. For convenience, however, we will always call this the RPE cut, and emphasize that a real particle exchange occurs only if it is kinematically accessible. Note that although the value of the isobar mass dictates the physics of the OPE, the OPE is blind to the decay products of the isobar and the physical threshold in $s$ is $\max\{(\sqrt{\sigma_j}+m_j)^2,(\sqrt{\sigma_k'}+m_k)^2\}$. 

We can understand the analytic structure of the OPE by writing a dispersive representation in $s$. Eq.~\eqref{eq:OPE_Swave_imag} is nonzero in two regions, leading to the relation
\begin{equation}\label{eq:OPE_disp}
\begin{split}
\wt{\E}^S_{kj}(\sigma_k',s,\sigma_j) & = \int_{\Gamma_V} ds' \frac{1}{s' - s - i\epsilon} \frac{1}{4 \lvert \p_k'^{\star} \rvert \lvert \p_j^{\star} \rvert} \\
&  + \int_{\Gamma_R} ds'  \frac{1}{s' - s- i\epsilon} \frac{1}{4\lvert \p_k'^{\star} \rvert \lvert \p_j^{\star} \rvert},
\end{split}
\end{equation}
where $\Gamma_V$ is the contour over the VPE cut and $\Gamma_R$ is the contour over the RPE cut. The integrand has four branch points associated with the thresholds and pseudo-thresholds of the initial and final momenta. We choose to orient the branch cuts such that the lowest branch point ($\min\{(\sqrt{\sigma_k'}-m_k)^2, (\sqrt{\sigma_j}-m_j)^2\}$) has a cut running to $-\infty$, the highest branch point ($\max\{(\sqrt{\sigma_k'}+m_k)^2, (\sqrt{\sigma_j}+m_j)^2\}$) has a cut running to $+\infty$, and the other two branch points have a branch cut joining them.
The contour $\Gamma_V$ is always taken above the real axis, whereas the contour $\Gamma_R$ depends on the external masses.  The physical amplitude is defined as the boundary value when $s \to s + i\epsilon$
, below the RPE cut.

\begin{figure*}[t]
\centering
\includegraphics[trim={0 10cm 0 0}, clip, width=1.\textwidth]{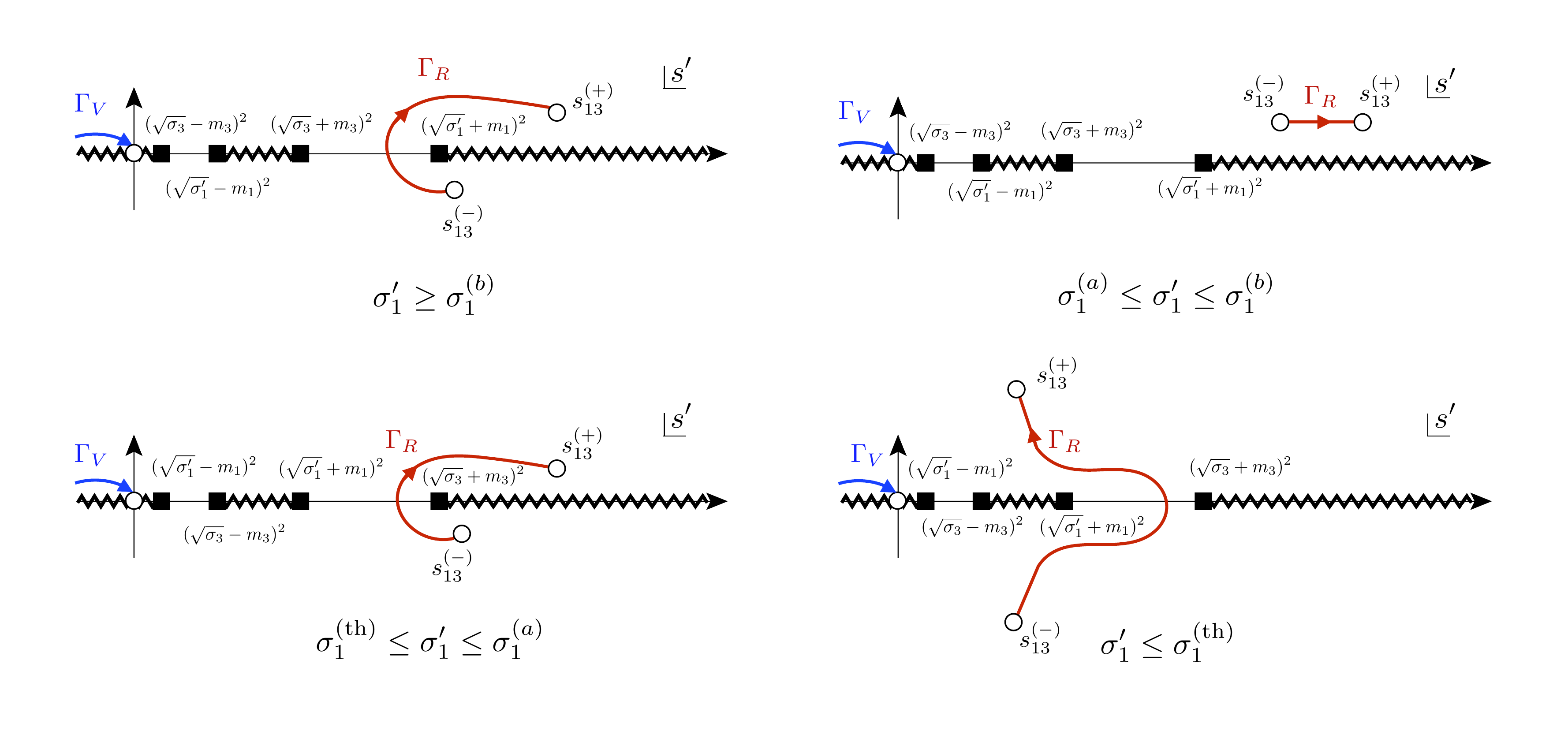}
\put(-470,150){{(a)}}
\put(-220,150){{(b)}}
\put(-470,30){{(c)}}
\put(-220,30){{(d)}}
\label{fig:OPE_cuts}\par
\caption{Cut structure of the OPE integrand Eq.~\eqref{eq:OPE_Swave_imag} in the $s'$-plane, and the OPE integration paths for the RPE contour ($\Gamma_R$ in red) and the VPE contour ($\Gamma_V$ in blue) for the dispersive integral Eq.~\eqref{eq:OPE_disp}. The four cases as a function of $\sigma_1'$ are: (a) $\sigma_1' \ge \sigma_1^{(b)}$, (b) $\sigma_1^{(a)}\le \sigma_1' \le \sigma_1^{(b)}$, (c) $\sigma_1^{(\mathrm{th})} \le \sigma_1' \le \sigma_1^{(a)}$, and (d) $\sigma_1' \le \sigma_1^{(\mathrm{th})}$. Real particle exchange cannot occur in case (d). In the logarithmic representation Eq.~\eqref{eq:OPE_Swave}, the RPE cut is circular.
}
\label{fig:OPE_integrand_cuts}
\end{figure*}

\begin{figure*}[t]
\centering
\includegraphics[width=1.\textwidth]{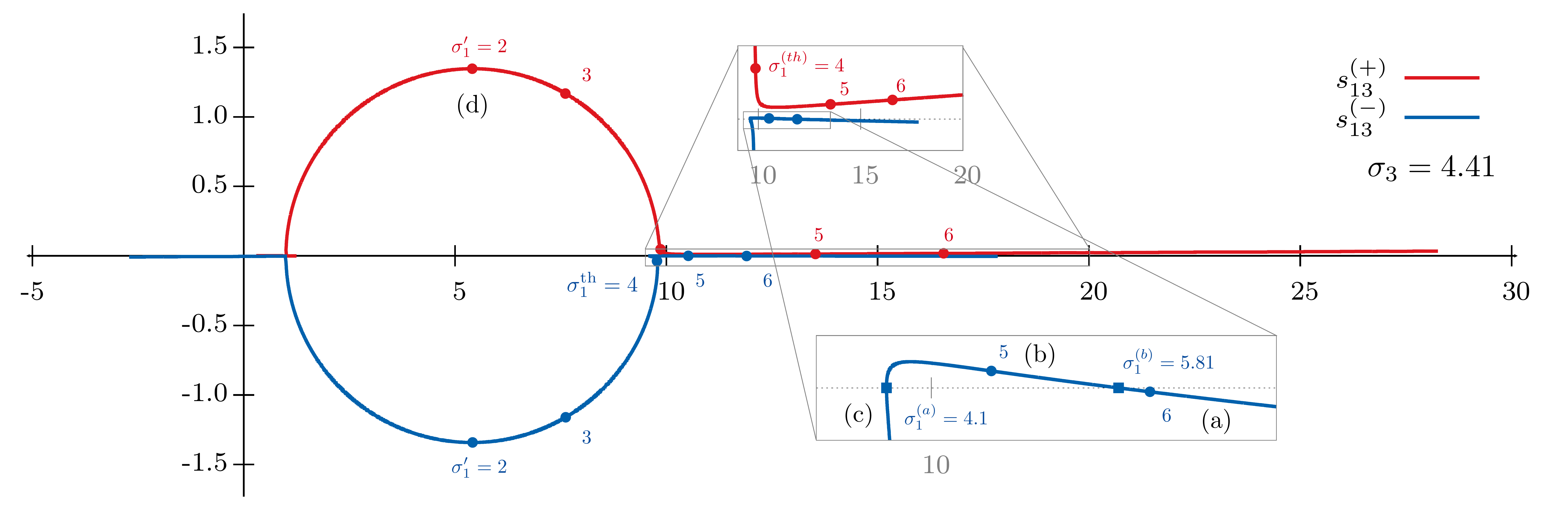}
\caption{Motion of $s_{13}^{(\pm)}$ in the $s$-plane as a function of $\sigma_1'$ for fixed $\sigma_3 = 4.41$, and unit external masses, $m_1 = m_2 = m_3 = 1$. Shown in red is $s_{13}^{(+)}$ and in blue $s_{13}^{(-)}$. The points indicate various $\sigma_1'$ values. Note that the physical region begins at $\sigma_1' = 4$. The inset shows the infinitesimal region where $s_{13}^{(-)}$ curve crosses the real axis at the points $\sigma_1^{(a)} = 4.1$ and $\sigma_1^{(b)} = 5.81$. The labels (a)-(d) indicate the regions described in the text and depicted in Fig.~\ref{fig:OPE_integrand_cuts}.
}
\label{fig:spm}
\end{figure*}

For fixed $\sigma_j > \sigma_j^{(\mathrm{th})}$, the RPE cut can be categorized by different regions in $\sigma_k'$. In Fig.~\ref{fig:OPE_integrand_cuts}, we illustrate how the analytic structure of the integrand and the integration contours change in these regions. Assuming a small imaginary part $\mu_{jk}^2 \to \mu_{jk}^2 - i\epsilon$, the RPE branch points have a finite imaginary part for $\sigma_k' < \sigma_k^{(\mathrm{th})}$,  with opposite signs. When $\sigma_k' > \sigma_k^{(\mathrm{th})}$, the branch points are infinitesimally close to the real axis. In the physical region, $s_{kj}^{(-)}$ has inflection points
at two locations of $\sigma_k'$:
\begin{subequations}\label{eq:inflect}
\begin{align}
\sigma_k^{(a)} & = m_j^2 + \mu_{jk}^2 + \frac{m_j (\sigma_j + \mu_{jk}^2 - m_k^2)}{\sqrt{ \sigma_j }}, \label{eq:inflectA}\\
\
\sigma_k^{(b)} & = -\frac{1}{2 m_k^2} \bigg[ 2 m_k^2 ( \sigma_j - m_j^2 ) - m_k^4  \label{eq:inflectB}\\
&  - (\sigma_j - \mu_{jk}^2)^2  + (m_k^2 + \mu_{jk}^2 - \sigma_j) \nn \\
&  \times \sqrt{4 m_j^2 m_k^2 + \lambda(\sigma_j,\mu_{jk}^2,m_k^2) } \bigg], \nn
\end{align}
\end{subequations}
which follow from $ds_{kj}^{(-)}/d\sigma_k' = 0$ corresponding to $\im s_{kj}^{(-)} = 0$.
Both $\sigma_k^{(a)}$ and $\sigma_k^{(b)}$ correspond to when $s_{kj}^{(-)}$ crosses the real $s$-axis. The $s_{kj}^{(+)}$ branch point always lies in the upper-half plane.

We can therefore classify the regions according to when the RPE branch points are both in the upper-half plane or when they approach the real axis.
\begin{enumerate}[(a)]
	\item $\sigma_k' \ge \sigma_k^{(b)}$, see region (a) in Fig.~\ref{fig:spm}. Here $s_{kj}^{(-)}$ is below the real axis, and $s_{kj}^{(+)}$ is above the real axis. The RPE cut connects these two points by crossing the real axis below the threshold $(\sqrt{\sigma_k'}+m_k)^2$. Real particle exchange in this case has consequences when considering the OPE processes embedded in the triangle diagram, which is discussed in the next section. For $k=1$ and $j = 3$, Fig.~\ref{fig:OPE_integrand_cuts}(a) shows the RPE and VPE contours, $\Gamma_R$ and $\Gamma_V$, respectively. Note that in the logarithmic representation, the RPE cut is circular, whereas in the dispersive representation, one can define the cut in any chosen manner as long as singularities are not crossed.
	\item $ \sigma_k^{(a)} \le \sigma_k' \le \sigma_k^{(b)}$, see region (b) in Fig.~\ref{fig:spm}. When $\sigma'_k$ decreases below the inversion point $\sigma^{(b)}_k$, $s_{kj}^{(-)}$ wanders above the real axis.  
    The RPE cut directly connects $s_{kj}^{(-)}$ to $s_{kj}^{(+)}$ without crossing the real axis. This is the typical case when considering the exchange of a real particle, illustrated in Fig.~\ref{fig:OPE_integrand_cuts}(b). Note that when $\sigma_k' = \sigma_j$ for equal masses $m_j = m_k$, then the integrand branch points merge into pole singularities.
	\item $\sigma_k^{(\mathrm{th})} \le \sigma_k' \le \sigma_k^{(a)}$, see region (c) in Fig.~\ref{fig:spm}. The RPE branch points again wrap around the real axis, \cf Fig.~\ref{fig:OPE_integrand_cuts}(c), with the cut crossing the real axis below the threshold $(\sqrt{\sigma_j}+m_j)^2$.
	\item $\sigma_k' \le \sigma_k^{(\mathrm{th})}$, see region (d) in Fig.~\ref{fig:spm}. The branch points $s_{kj}^{(\pm)}$ move deep into the complex plane, as shown in Fig.~\ref{fig:OPE_integrand_cuts}(d). In this region, the isobar cannot decay, and therefore it is unphysical for the $\3\to\3$ elastic scattering. Real particle exchange cannot occur, leaving only the virtual contributions.
\end{enumerate}

If we evaluate the OPE along the real $s$-axis in regions (a) or (c), we find that the real part of the OPE has a jump due to crossing the RPE cut.
In the logarithmic representation, this crossing occurs
when $z_{kj} = 0$, that is, when
\begin{equation}
\begin{split}
s_{kj}^{(0)} & = \frac{1}{2} \bigg[ m_j^2 + m_k^2 - 2 \mu_{jk}^2 + \sigma_j + \sigma_k' \\
& + \bigg(
   4 (m_j^2 - \sigma_j) (m_k^2 - \sigma_k') \\
&+ (m_j^2 + m_k^2 -
   2 \mu_{jk}^2 + \sigma_j + \sigma_k')^2 \bigg)^{1/2} \bigg].
\end{split}
\end{equation}
When choosing a different contour for $\Gamma_R$ in the dispersive representation, the location of this crossing depends on where real axis crosses the chosen contour.

\begin{figure*}
\centering
\subfigure{
\includegraphics[width=.33\textwidth]{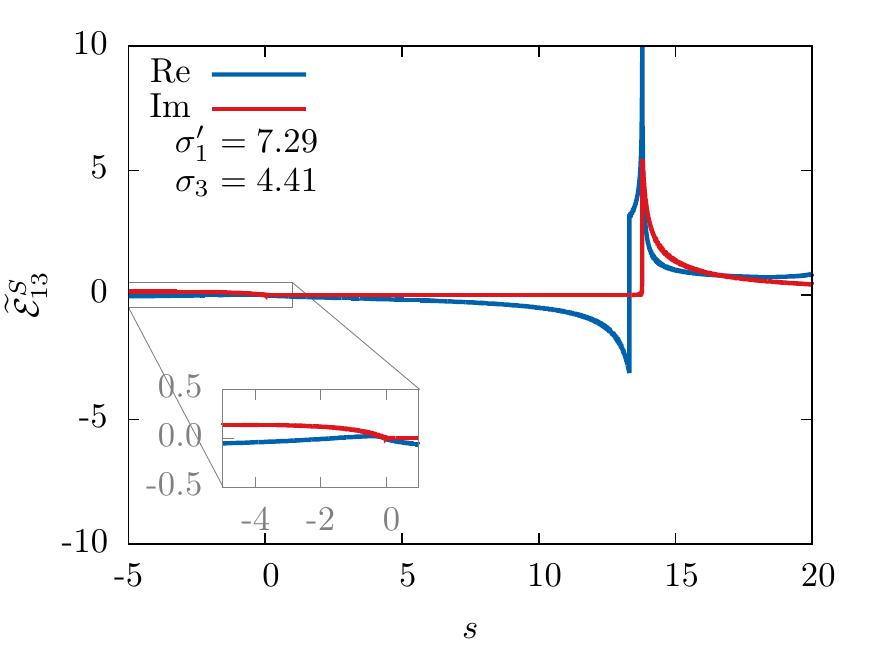} \label{fig:OPE1}
}
\put(-35,30){{(a)}}
\subfigure{
\includegraphics[width=.33\textwidth]{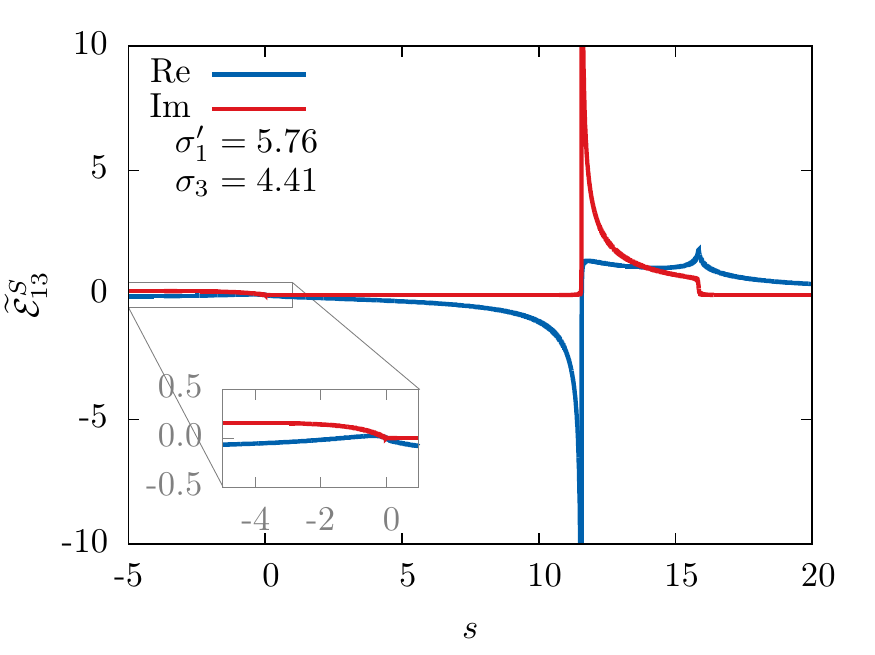}\label{fig:OPE2}
}
\put(-35,30){{(b)}}
\subfigure{
\includegraphics[width=.33\textwidth]{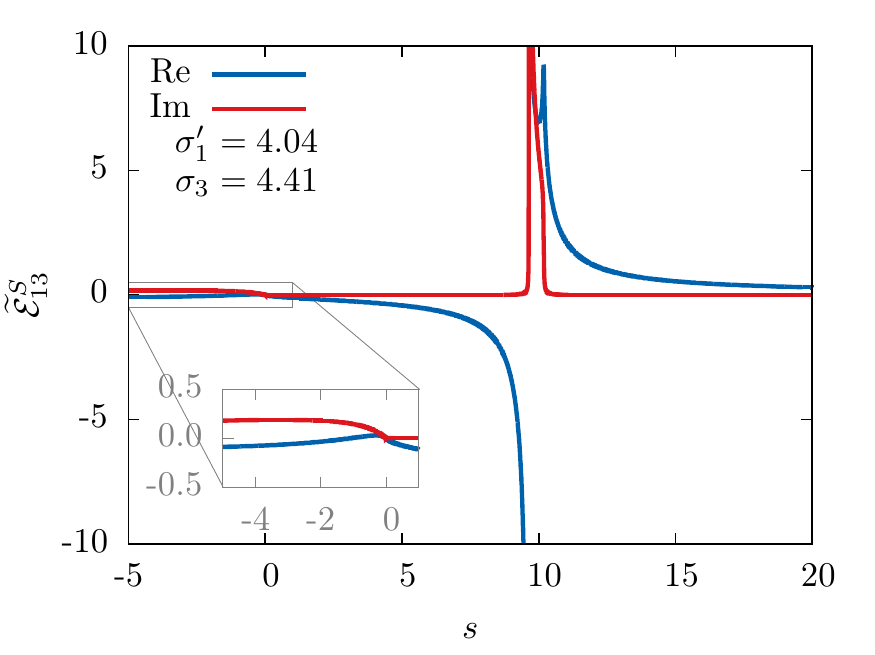}\label{fig:OPE3}
}
\put(-35,30){{(c)}}

\subfigure{
\includegraphics[width=.33\textwidth]{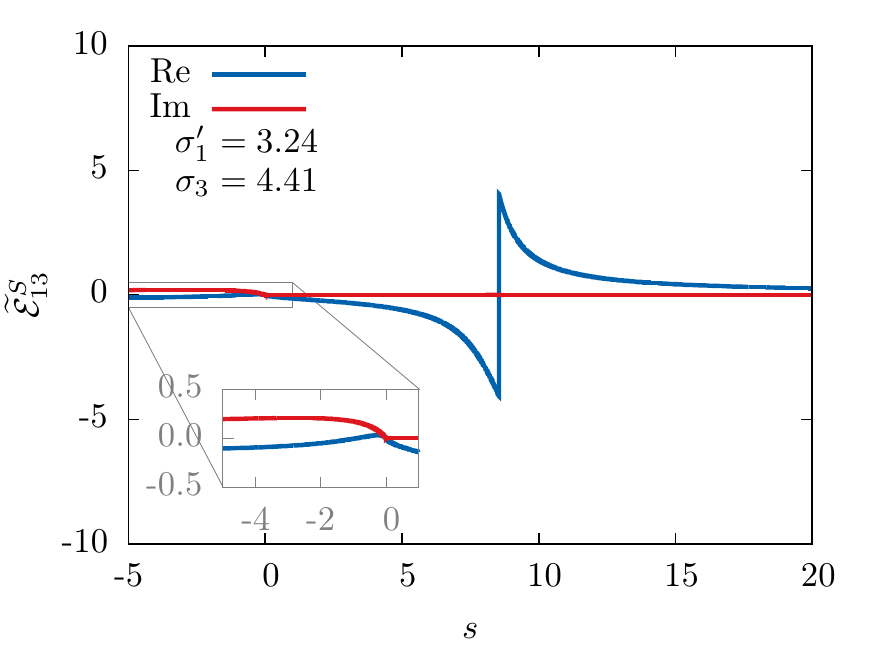}\label{fig:OPE4}
}
\put(-35,30){{(d)}}
\subfigure{
\includegraphics[width=.33\textwidth]{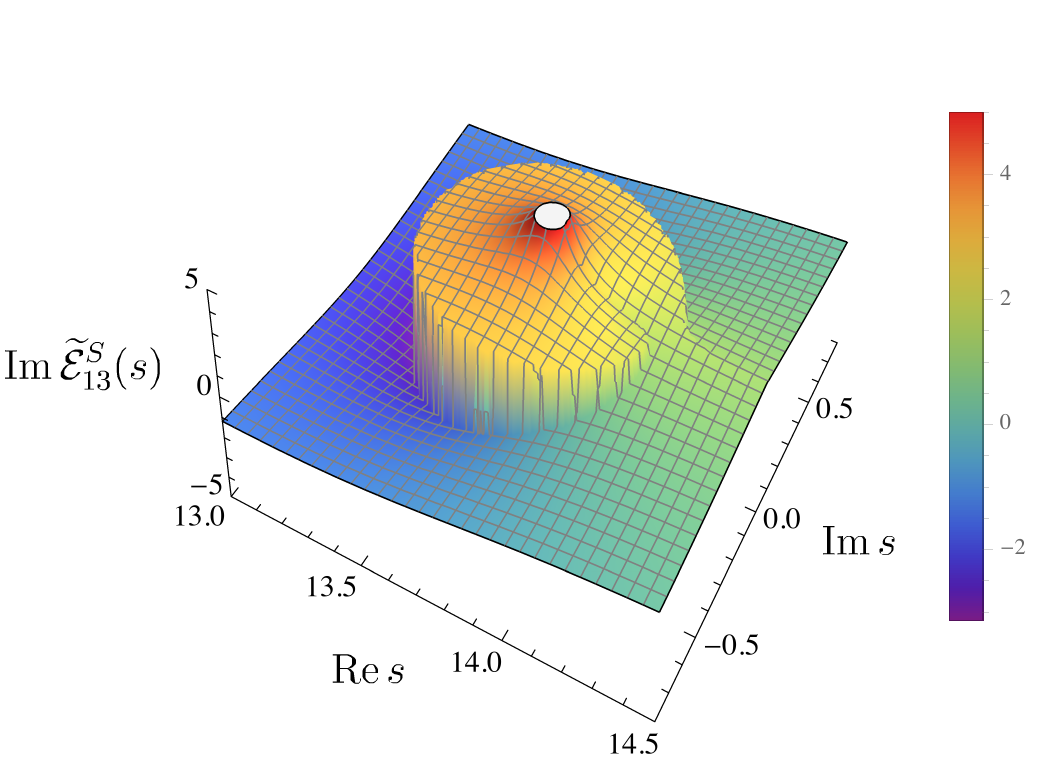}\label{fig:OPE5}
}
\put(-35,10){{(e)}}
\subfigure{
\includegraphics[width=.33\textwidth]{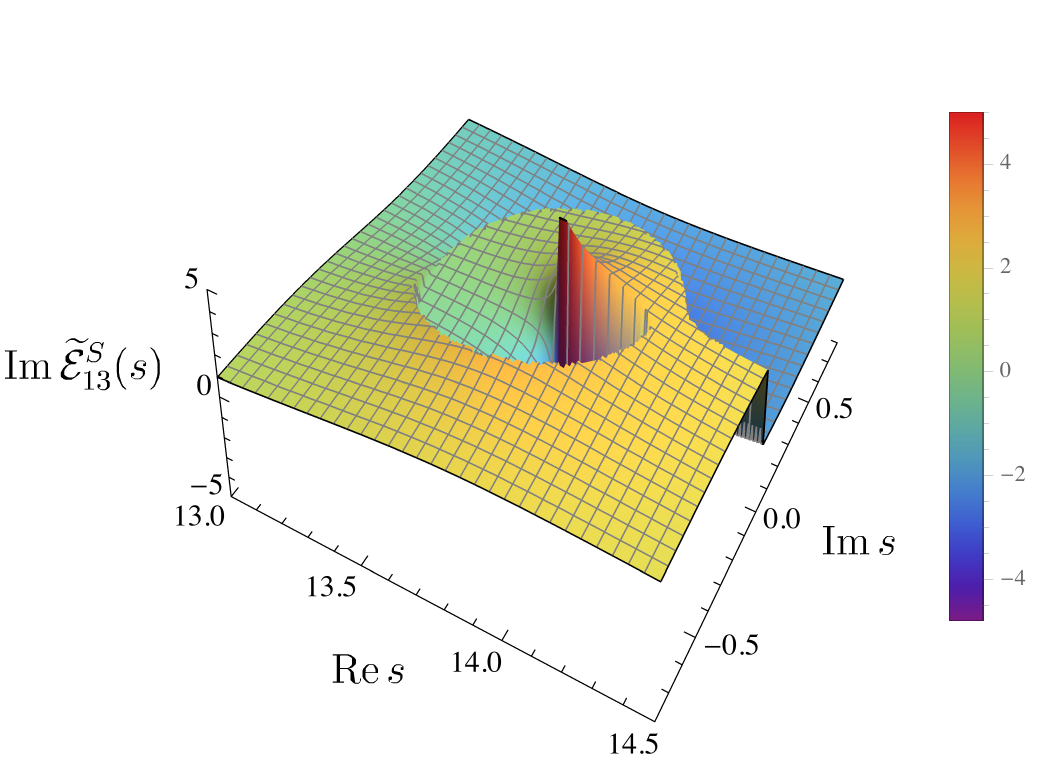}\label{fig:OPE6}
}
\put(-35,10){{(f)}}
\caption{OPE amplitudes Eq.~\eqref{eq:OPE_Swave} for external masses $m_1 = m_2 = m_3 =1 $ at fixed $\sigma_3 = 4.41$
for the four regions depicted in Fig.~\ref{fig:OPE_integrand_cuts}:
(a) $\sigma_1' = 3.24$ representing the unphysical region, (b) $4.04$, where the RPE cut wraps around the real axis, (c) $4.41$, where the RPE branch points are above the real axis, 
and (d) $7.29$ where the RPE cut wraps around the real axis. The insets show the contribution from the VPE cut. For these values, the inflection points are $\sigma_1^{(a)} = 4.1$ and $\sigma_1^{(b)} = 5.81$. The real and imaginary parts of the OPE Eq.~\eqref{eq:OPE_Swave} at $\sigma_1' = 7.29$ in the complex $s$-plane are shown in (e) and (f), respectively. The discontinuity at $s = s_{13}^{(0)} = 13.31$ occurs when evaluating the OPE across the RPE cut.
}
\label{fig:OPE_figs}
\end{figure*}

These cases are illustrated in Fig.~\ref{fig:OPE_integrand_cuts} for spectators $k=1$ and $j=3$. We plot the OPE amplitudes, Eq.~\eqref{eq:OPE_Swave}, as a function of $s$ for fixed $\sigma_3$ and $\sigma_1'$ in Fig.~\ref{fig:OPE_figs}. 
Figure~\ref{fig:OPE1} shows the OPE computed at $\sigma_1'$ in region (a). At this energy, the $s_{kj}^{(-)}$ is below the real axis, and the RPE cut wraps around the real axis, passing below the threshold $(\sqrt{\sigma_1'} + m_1)^2$. The jump in the real part at $s_{13}^{(0)}$ is due to crossing the RPE cut.
Figure~\ref{fig:OPE2}, is evaluated at $\sigma_1'$ in region (b), where both branch points are above the real axis. Here, we illustrate that as $\sigma_1'$ decreases, the width of the imaginary part decreases and the peak increases. The narrowing imaginary region physically represents that less phase space is available for real particle propagation in the intermediate state. 
Figure~\ref{fig:OPE3} is computed for $\sigma_1'$ in region (c), right above the two-particle threshold. 
There is a jump in the real part at $s_{13}^{(0)}$ from crossing the RPE cut. 
The final case is illustrated in Fig.~\ref{fig:OPE4}, where the OPE computed in the unphysical region (d). 
There is an imaginary part due to the VPE cut only, as it is kinematically inaccessible for the exchange of a real particle. The jump in the real part at 
$s_{13}^{(0)}$ comes from crossing the RPE cut.
Figures~\ref{fig:OPE5} and \ref{fig:OPE6} shows a 3-dimensional plot of the real and imaginary part of the logarithmic representation of the OPE, Eq.~\eqref{eq:OPE_Swave}. The circular cut is clearly visible connecting the RPE branch points. The physical region is taken as the region approaching the real axis, below the RPE cut.

To summarize, the analytic structure of the OPE is given by two branch cuts, the VPE and RPE cuts. The VPE cut is present for $-\infty < s \le 0$, and is associated with the exchange of an off-shell particle. For physical isobars, the RPE cut is in the physical region. We have shown different scenarios, identified by the isobar masses, in which the RPE branch points can approach the physical region, which impact the structure of the $B$-matrix kernels.

\subsection{Triangle Diagrams}\label{sec:triangle}
\begin{figure}[t!]
\centering
\includegraphics[width=0.8\columnwidth]{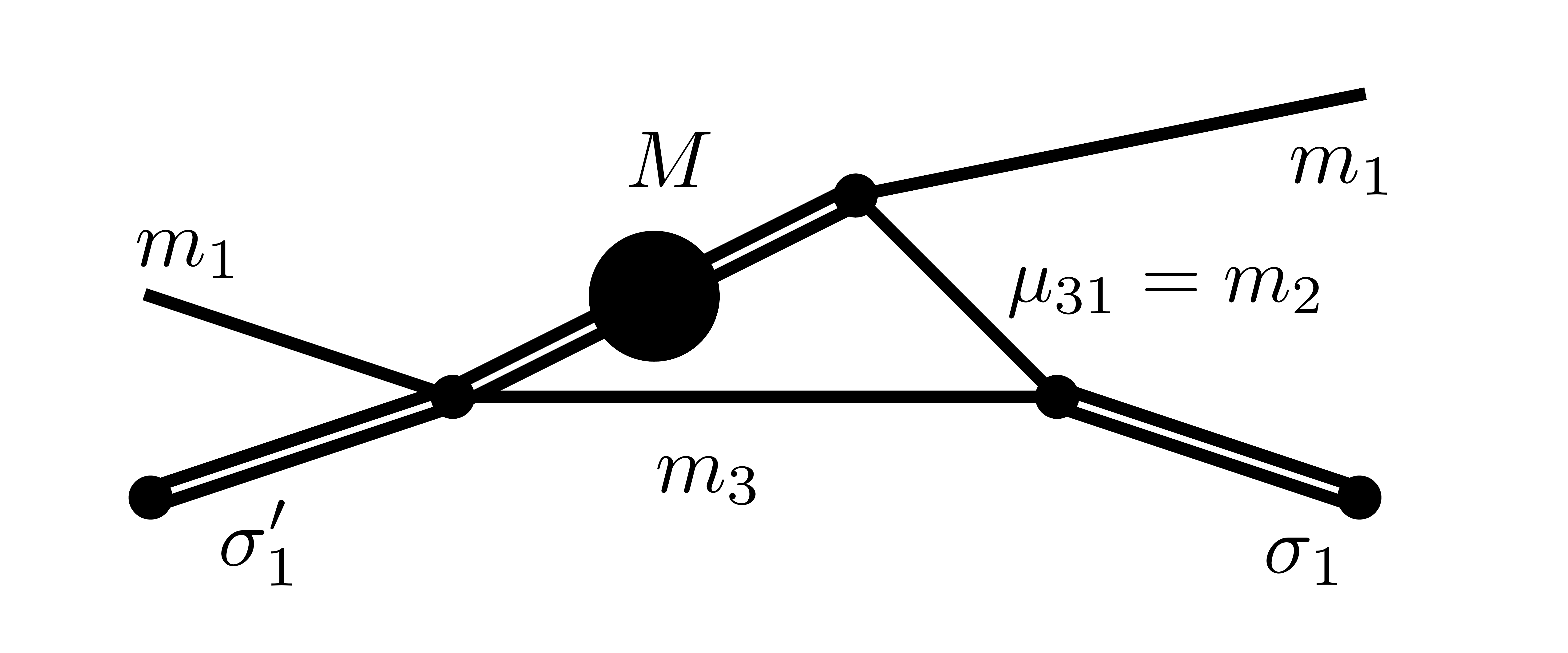}
\caption{The triangle diagram $\mathcal{T}_B$ contribution to the kernel $\Kc_{11}$. We take the isobar to have a narrow width with mass $M$. For numerical evaluations, $m_1 = m_3 = \mu_{31} = 1$. 
}
\label{fig:tri_label}
\end{figure}

\begin{figure}[t!]
\centering
\includegraphics[width=1.\columnwidth]{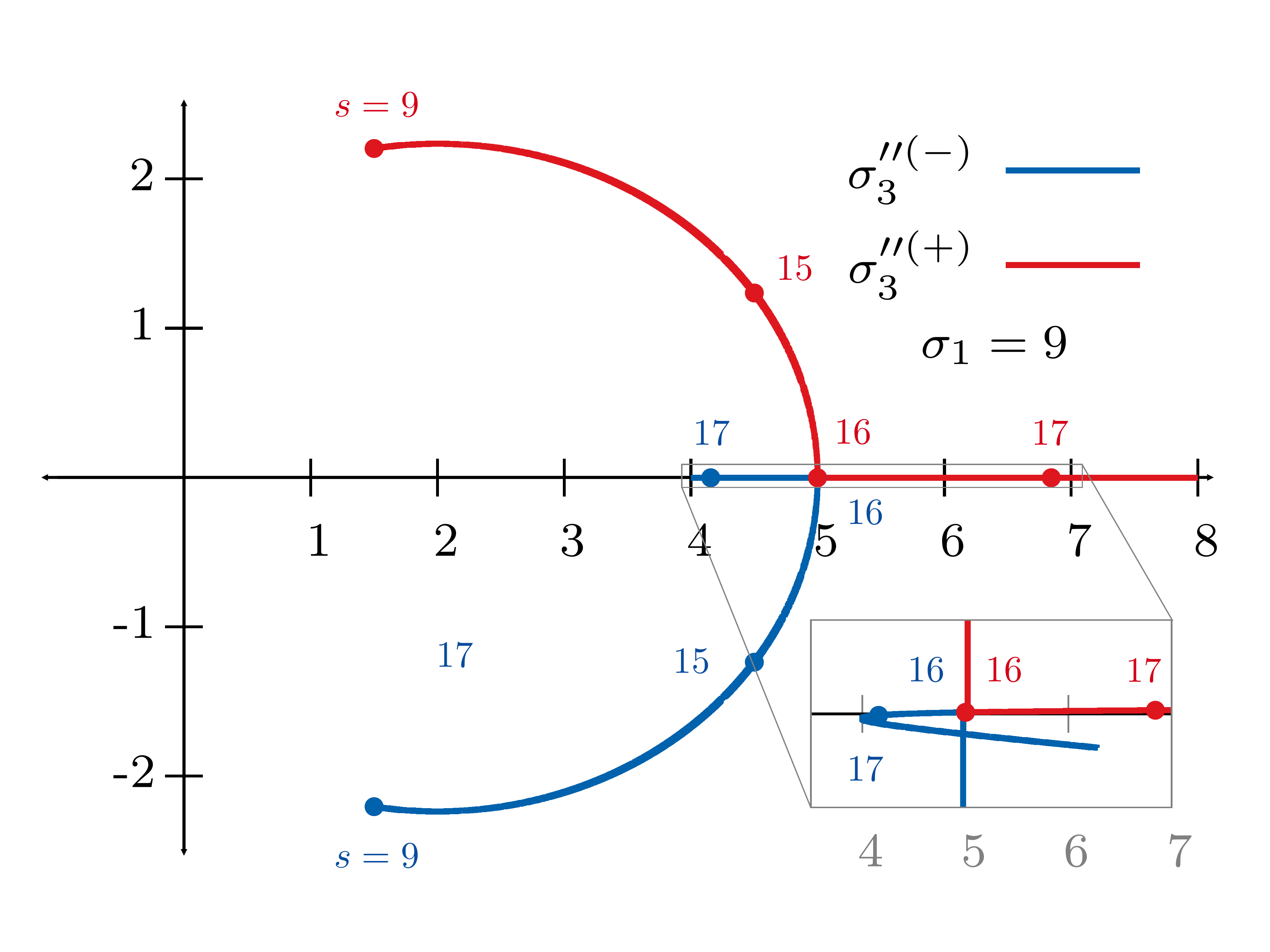}
\caption{Motion of the $\sigma_3''^{(\pm)}$ in the $\sigma_3''$-plane as a function of $s$ for fixed $\sigma_1$. Shown in red is $\sigma_3''^{(+)}$ and in blue $\sigma_3''^{(-)}$. The points indicate various $s$ values starting from the three particle threshold, $s=(m_1+m_2+m_3)^2 = 9$. The inset shows that the branch points pinch the real $\sigma_1'$ axis at $s = (\sqrt{\sigma_1} + m_1)^2 = 16$, which is responsible for a pinch singularity in $\mathcal{T}_B$. }
\label{fig:sigpm}
\end{figure}

To understand resonance poles of $\3\to\3$ systems, the analytic structure of the $B$-matrix parameterization, Eq.~\eqref{eq:Bmat1234} must be understood in the complex $s$-plane. This means understanding the properties of the $B$-matrix kernels. Here, we investigate the triangle diagram, and leave the box diagram for future studies.
Let us work with the triangle $\mathcal{T}_B \equiv \mathcal{T}_{11}^{(2)}$ introduced in Eq.~\eqref{eq:kernel_tri2}, where all angular momenta are in $S$-wave. For convenience, let $\wt{\R} = 1$, thus the amplitudes are independent of $\sigma_1'$, and given by
\begin{equation}
\begin{split}
\mathcal{T}_B(s) = \int_{\sigma_3^{(\mathrm{th})}}^{(\sqrt{s} - m_3)^2} d\sigma_3'' \, \tau_{3}(s,\sigma_3'') \wt{\E}^S_{31}(\sigma_3'',s,\sigma_1) ,
\end{split}
\end{equation}
where 
$\tau_3(s,\sigma_3'') = \rho_3(s,\sigma_3'')D_{3}^{-1}(\sigma_3'')$
, and the dependence of $\mathcal{T}_B$ on $\sigma_1$ has been understood. To ensure the correct analytic properties of the isobar amplitude, we introduce its dispersive representation
\begin{equation}
\begin{split}
 f_3(\sigma_3'') = \frac{1}{\pi}\int_{\sigma_3^{(\mathrm{th})}}^{\infty} d\wh{\sigma} \, \frac{\im{f_3(\wh{\sigma})}}{\wh{\sigma} - \sigma_3'' - i\epsilon},
\end{split}
\end{equation}
giving the form for $\mathcal{T}_B$
\begin{equation}\label{eq:TB_1}
\begin{split}
 \mathcal{T}_B(s) & = \frac{1}{\pi} \int_{\sigma_{3}^{(\mathrm{th})}}^{\infty} d\wh{\sigma}\, \im{f_3(\wh{\sigma})}\, \\
 &  \times \int_{\sigma_3^{(\mathrm{th})}}^{(\sqrt{s} - m_3)^2} d\sigma_3'' \frac{\rho_{3}(s,\sigma_3'') \wt{\E}_{31}^{S}(\sigma_3'',s,\sigma_1) }{\wh{\sigma} - \sigma_3'' - i\epsilon}.
\end{split}
\end{equation}
We see the $\sigma_3''$-integral does not depend on 
$f_3(\wh{\sigma})$, so for simplicity we take the narrow width limit 
$\im{f_3(\wh{\sigma})} = \pi \delta(\wh{\sigma} - M^2)$
, where $M$ is the mass of the isobar. The narrow width limit shifts the unitarity cut in the triangle diagram to begin at the threshold \mbox{$s = (M + m_3)^2$}. However, for a general isobar shape, Eq.~\eqref{eq:TB_1} can be used to sum over its distribution, recovering the correct unitarity branch cut starting at $s_{\textrm{th}}$.
Therefore, the triangle diagram has the form
\begin{equation}\label{eq:TB}
\begin{split}
\mathcal{T}_B(s) = \int_{\sigma_3^{(\mathrm{th})}}^{(\sqrt{s} - m_3)^2} d\sigma_3'' \frac{\rho_{3}(s,\sigma_3'') \wt{\E}_{31}^{S}(\sigma_3'',s,\sigma_1) }{M^2 - \sigma_3'' - i\epsilon}.
\end{split}
\end{equation}
Figure~\ref{fig:tri_label} shows the triangle diagram in consideration. The $B$-matrix triangle contains singularities on the physical sheet. These are due to the $s$-singularities  in $\rho_3$ and $\wt{\E}_{13}^S$, and to endpoint singularities when the integration limits hit the $\sigma_3''$-singularities of the integrand. The upper integration limit gives a branch cut
for $s<0$.
Since $\rho_3(s,\sigma_3'') \propto \{[\sigma_3'' - (\sqrt{s} + m_3)^2][\sigma_3'' - (\sqrt{s} - m_3)^2]\}^{1/2} / s$, there is a pole at $s = 0$.
When the integration variable hits the lower limit $\sigma_3'' = \sigma_3^{(\mathrm{th})}$,  there are two branch point singularities at $s = (\sqrt{\sigma_3^{(\mathrm{th})}} \pm m_3)^2 = (m_1 + m_2 \pm m_3)^2$. In the narrow width limit, the unitarity cut opens when the upper limit of the integral hits the pole in the isobar propagator, for $s = (M + m_3)^2$. On the real $s$-axis, the OPE has a discontinuity in the real part when $z_{31} = 0$, \ie at $\sigma_3'' = \sigma_3^{(\mathrm{th})}$. This discontinuity from crossing the RPE cut is present in $\mathcal{T}_B$. Two more singularities occur when $\sigma_3''$ hits the two inflection points $\sigma_3^{(a)}$ and $\sigma_3^{(b)}$, which are defined in Eqs.~\eqref{eq:inflect}.
The OPE pinches the real axis at $\sigma_3'' = \sigma_3^{(a)}$, and generates a singularity in $\mathcal{T}_B$ at the initial state threshold $s = (\sqrt{\sigma_1} + m_1)^2$. This can be understood by realizing that the OPE branch points, Eq.~\eqref{eq:spm}, can alternatively be written in terms of $\sigma_3''$ as a function of $s$ for fixed $\sigma_1$. The branch points are then $\sigma_3''^{(\pm)}$, where
\begin{equation}
\begin{split}
\sigma_3''^{(\pm)} & = \frac{1}{2\sigma_1} \bigg[  \sigma_1(s + m_1^2 + m_3^2 + \mu_{13}^2)  \\
& + (m_1^2 - s)(m_3^2 - \mu_{13}^2) - \sigma_1^2  \\
& \pm \lambda^{1/2}(s,\sigma_1,m_1^2)\lambda^{1/2}(\sigma_1,\mu_{13}^2,m_3^2) \bigg],
\end{split}
\end{equation}
and $\sigma_3^{\prime\prime(-)}$ lies infinitesimally below the real axis in the physical region.
Figure~\ref{fig:sigpm} shows the motion of $\sigma_3''^{(\pm)}$ in the complex $\sigma_3''$-plane as a function of $s$ for fixed $\sigma_1$.
At the three particle threshold, $s = (m_1 + m_2 + m_3)^2$, the branch points have finite imaginary part and are on opposite sides in the $\sigma_3''$-plane. As $s$ approaches the initial state threshold $s = (\sqrt{\sigma_1} + m_1)^2 $, the 
$\sigma_3''^{(\pm)}$ branch points pinch
the real axis at $\sigma_3'' = \sigma_3^{(a)}$. Since the $\mathcal{T}_B$ integration is on the real axis starting from $\sigma_3^{(\mathrm{th})}$, the integration path is pinched, causing a singularity in $\mathcal{T}_B$ at $s = (\sqrt{\sigma_1} + m_1)$. At $\sigma_3'' = \sigma_3^{(b)}$, the branch point migrates back below the real axis at a value greater than the threshold $\sigma_3^{(\textrm{th})}$ close to the real axis. When $M^2 > \sigma_3^{(b)}$, this effect generates the triangle singularity~\cite{Peierls:1961zz,Aitchison:1966,Eden:1966dnq}. The triangle singularity has been studied as a possible mechanism to explain anomalous structures observed in heavy flavor experiments~\cite{Guo:2015umn,Guo:2016bkl,Ketzer:2015tqa,Szczepaniak:2015eza}. The peak of the triangle singularity coincides with the $s_{31}^{(-)}$ branch point, \ie $s_{\textrm{tr}} = s_{31}^{(-)}$.

Aside for the unitarity branch cut starting at $s = (M + m_3)^2$ and the triangle singularity, these additional singularities in the physical $s$-plane are not allowed by analyticity. The extra singularities are moved to the second sheet when we consider the integration over the isobar shape, \cf Eq.~\eqref{eq:TB_1}, leaving only the unitarity cut starting at $s = (m_1 + m_2 + m_3)^2$ and the triangle singularity.

\begin{figure*}[t!]
\centering
\includegraphics[width=1.\columnwidth]{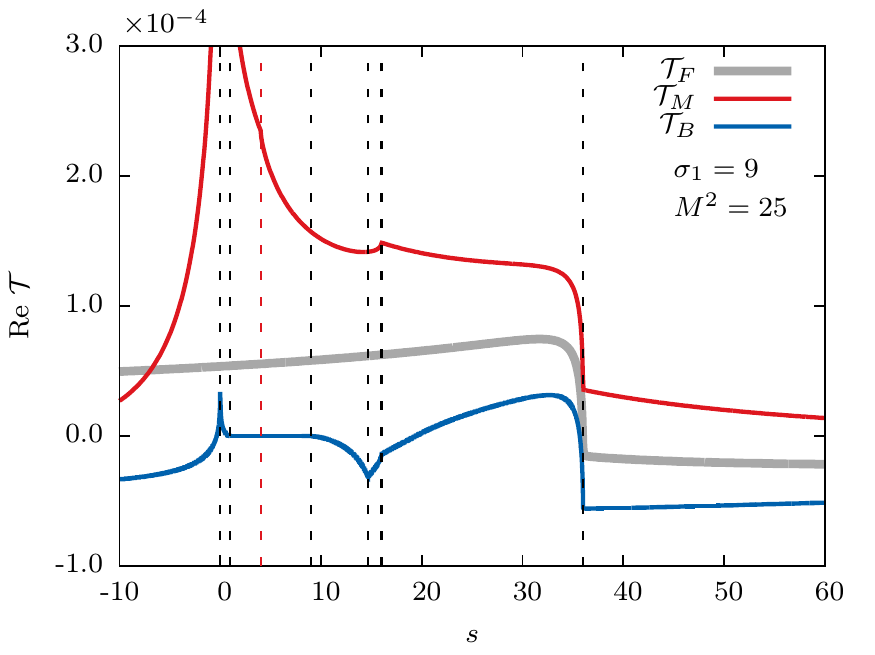}
\includegraphics[width=1.\columnwidth]{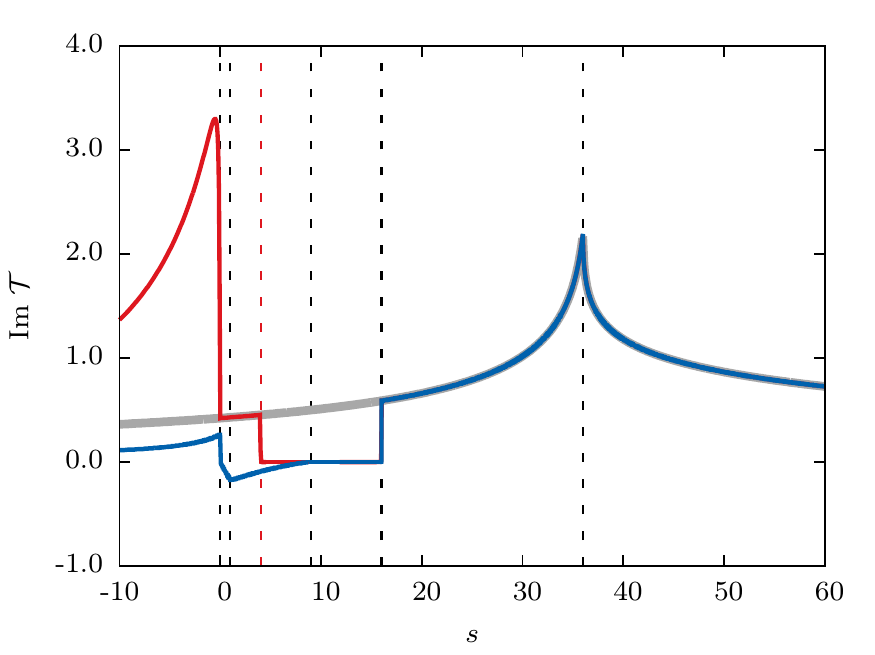}
\caption{Real (left) and imaginary (right) parts of the Feynman triangle diagram, Eq.~\eqref{eq:TF} (gray), the $B$-matrix triangle, Eq.~\eqref{eq:TB} (blue), and the Mai et al. triangle, Eq.~\eqref{eq:TM} (red). The external masses are set to unity, and the external isobar masses are $\sigma_1' = 9$ and $M^2 = 25$. The dashed vertical lines indicate the locations of singularities in the $B$-matrix as described in the text: (from left to right) the $s=0$ singularity (where the explicit $1/s$ pole in $\rho_3(s,\sigma_3'')$ makes $\mathcal{T}_B$ and $\mathcal{T}_M$ diverge), $s = (\sqrt{\sigma_3^{(\textrm{th})}} - m_3)^2 = 1$ ($\mathcal{T}_B$ is singular and $\mathcal{T}_M$  is regular), $s = (\sqrt{\sigma_3^{(\textrm{th})}} + m_3)^2 = 9$, the crossing of the RPE cut at $s = s_{31}^{(0)} = 14.64$, the initial state threshold $s = (\sqrt{\sigma_1} + m_1) = 16$, and the normal threshold singularity at $s= (M + m_3)^2 = 36$. The red dashed line indicates the pinch singularity at $s = (\sqrt{\sigma_1} - m_1)^2 = 4$ that occurs only in $\mathcal{T}_M$. 
}
\label{fig:triReal}
\end{figure*}

\begin{figure}[t!]
\centering
\includegraphics[width=1.\columnwidth]{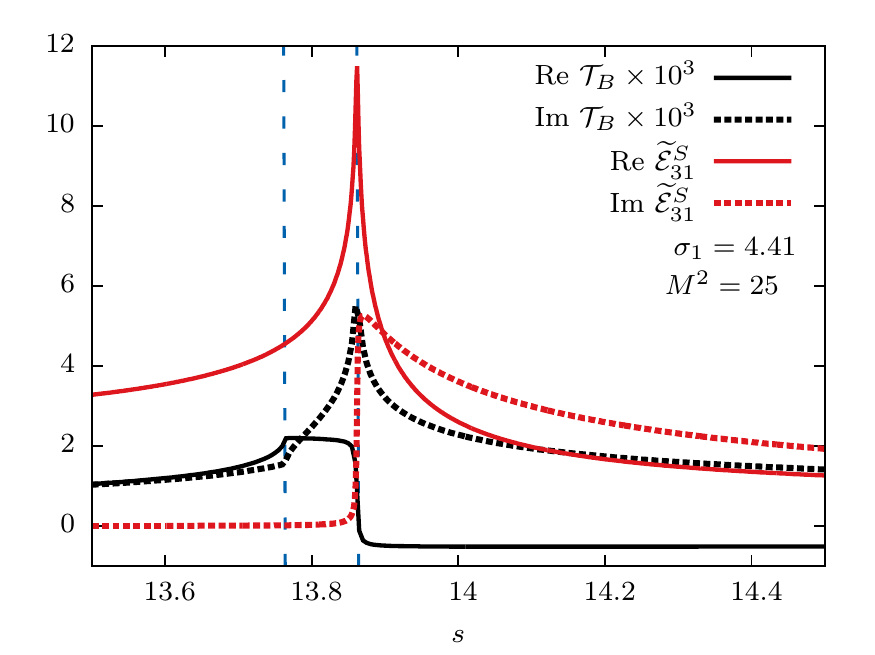}
\caption{The $B$-matrix triangle Eq.~\eqref{eq:TB} in with the triangle singularity. Shown in black are the real (solid) and imaginary (dashed) parts evaluated at $\sigma_1 = 4.41$ and $M^2 = 4.41$. Shown in red are the real (solid) and imaginary (dashed) parts of the OPE piece of the triangle, Eq.~\eqref{eq:OPE_Swave}, where $\sigma_3'' = M^2$. The blue dashed lines indicate the threshold $(M+m_3)^2 = 13.7641$ and the lower RPE branch point at $s_{31}^{(-)} = s_{\textrm{tr}} = 13.8619$. The normal threshold accounts for the first peak in the triangle diagram, while the second peak is caused by the triangle singularity. Note we scaled the triangle diagram to account for the phase space normalization of the triangle.}
\label{fig:triOPE}
\end{figure}

\begin{figure*}
\centering
\begin{minipage}{0.64\textwidth}
\subfigure{
\includegraphics[width=\linewidth]{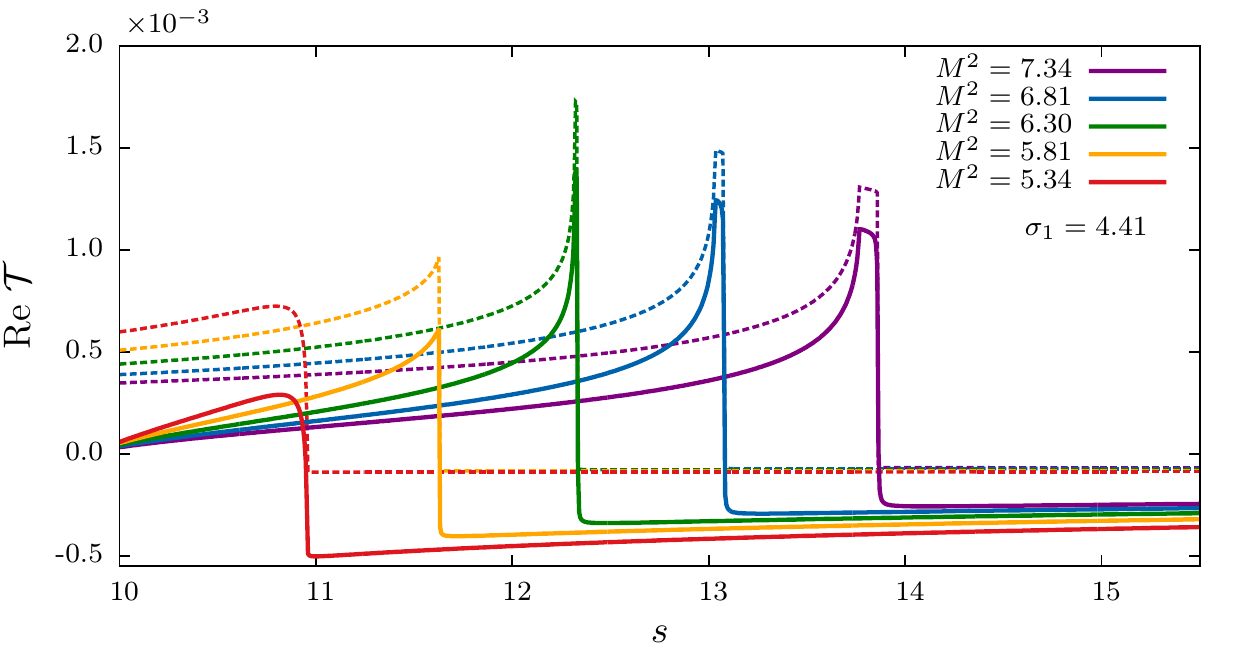} \label{fig:TriAllReal}
}
\put(-35,0){{(a)}}

\subfigure{
\includegraphics[width=\linewidth]{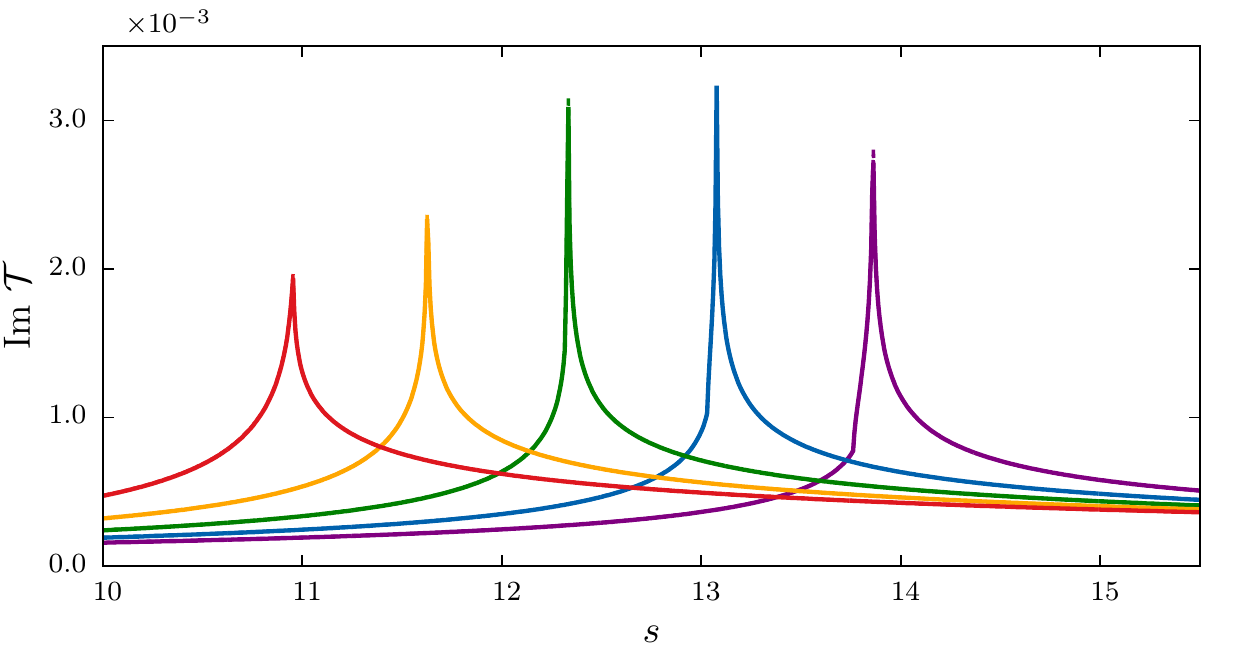}\label{fig:TriAllImag}
}
\put(-35,0){{(b)}}
\end{minipage}
\hspace*{\fill}
\begin{minipage}{0.33\linewidth}
\subfigure{
\includegraphics[width=\linewidth,height=4in]{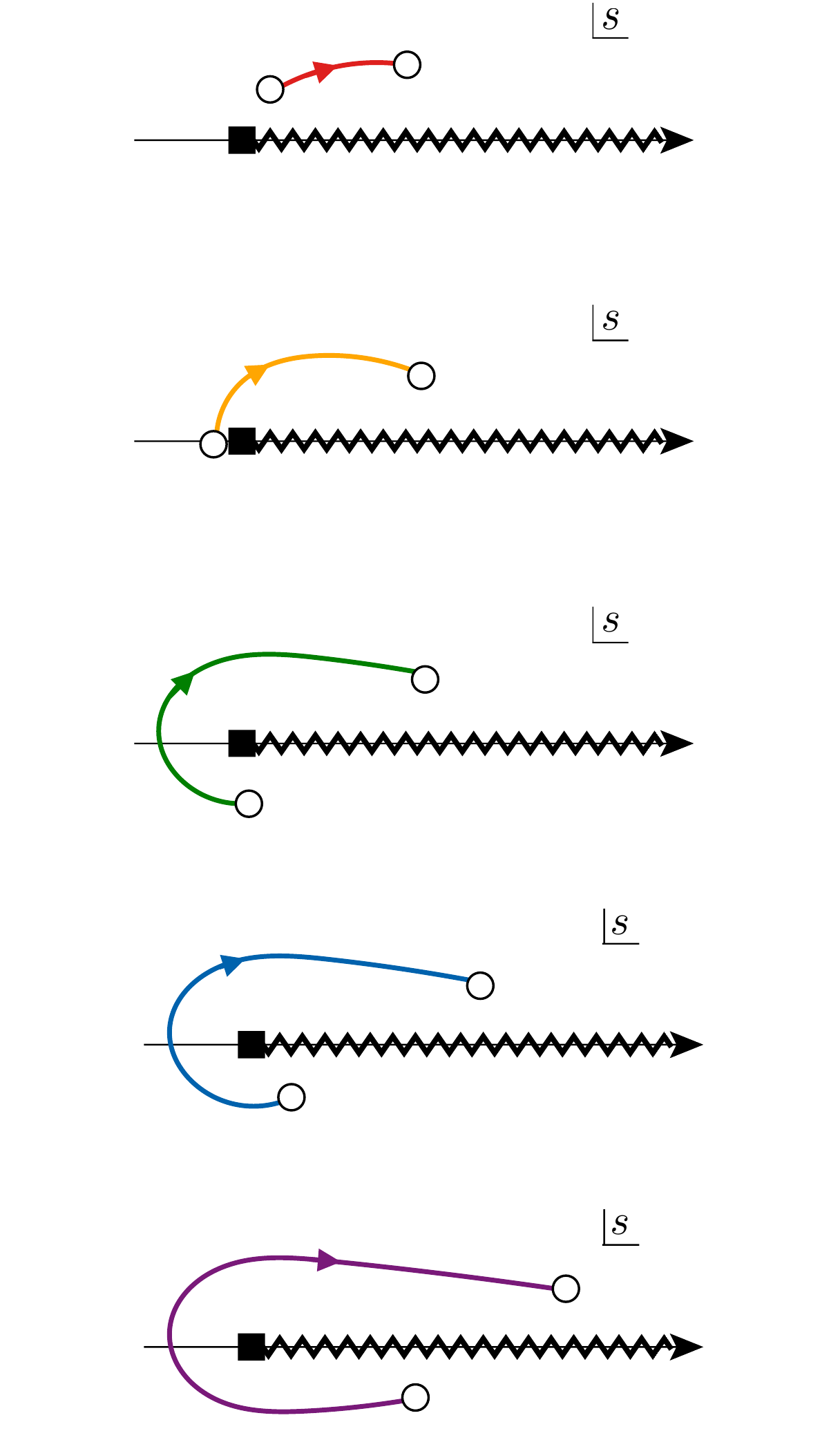}\label{fig:Tricontours}
}
\put(-80,-20){{(c)}}
\end{minipage}
\caption{The Feynman triangle Eq.~\eqref{eq:TF} and the $B$-matrix triangle Eq.~\eqref{eq:TB} in the vicinity of the triangle singularity for fixed $\sigma_1 = 4.41$ and varying $M^2$ in the range $5.33 \le M^2 \le 7.34$. 
$\mathcal{T}_B$ is shown with solid lines, $\mathcal{T}_F$ is shown with dashed lines. Real parts are shown in (a) and imaginary parts in (b). The imaginary parts for $\mathcal{T}_F$ and $\mathcal{T}_B$ coincide in this region. For $\sigma_1 = 4.41$, the triangle singularity region begins at $M^2 = \sigma_3^{(b)} = 5.81$, which manifest as a second threshold in the line shape beginning at $s = s_{31}^{(-)} = s_{\textrm{tr}} = 11.63$. The corresponding orientation of the OPE branch points are illustrated in (c). }
\label{fig:tri_compare_figsB}
\end{figure*}

We compare the structure of Eq.~\eqref{eq:TB} with that of a Feynman diagram triangle in a QFT (see Appendix \ref{sec:app_F} for a review of the Feynman triangle), which can be written as
\begin{equation}\label{eq:TF}
\mathcal{T}_{F}(s) = \int_{\Gamma_{T}} ds' \frac{\rho_3(s',M^2)\wt{\E}_{13}^{S}(M^2,s',\sigma_1)}{s' - s - i\epsilon},
\end{equation}
where $\Gamma_T$ is the path from the threshold $(M + m_3)^2$ to $\infty$, and
the $S$-wave amplitudes are normalized according to Eq.~\eqref{eq:PWIS_expansion}.
Figure \ref{fig:triReal} 
 shows the real and imaginary parts, respectively, of the two triangle diagrams $\mathcal{T}_F$ and $\mathcal{T}_B$, below the region of the triangle singularity. Notice that the Feynman triangle has only a normal threshold singularity at $s = (M + m_3)^2$, and is smooth everywhere else. The imaginary parts of both triangles are identical above threshold, as required by unitarity. The $B$-matrix triangle has noticeable kinks in both the real and imaginary parts below threshold, corresponding to the singularities discussed above. The black dashed lines indicate the location of the singularities. Starting from low energy, the first additional singularity is the $s=0$ singularity from the phase space. The next two singularities occur at $s = (\sqrt{\sigma_3^{(\mathrm{th})}} - m_3)^2 $ and $(\sqrt{\sigma_3^{(\mathrm{th})}} + m_3)^2 $, which are from the phase space evaluated at the lower integration limit. The real part contains a singularity from evaluating the OPE across the RPE cut. 
Note that the imaginary part does not contain this jump, consistent with the OPE description in the previous section. The next singularity occurs at the initial state threshold $s = (\sqrt{\sigma_1} + m_1)^2 $, which is due to the pinching of the $\sigma_3''$ contours by the OPE branch points. Finally the normal threshold at $s = (M + m_3)^2$.
Figure~\ref{fig:triOPE} shows $\mathcal{T}_B$ and the OPE in the region where the triangle singularity develops. The line shape shows the production threshold at $(M+m_3)^2$, and the peak at $s = s_{\textrm{tr}}$. The OPE branch point $s_{31}^{(-)}$ clearly coincides with the triangle peak. 

Figure~\ref{fig:tri_compare_figsB} shows the  $\mathcal{T}_B$ and $\mathcal{T}_F$ as a function of $s$ at fixed $\sigma_1$ and varying $M^2$ in the region below and above $M^2 = \sigma_3^{(b)}$. Figure~\ref{fig:TriAllReal} shows the real parts, and Fig.~\ref{fig:TriAllImag} shows the imaginary parts. 
At $M^2 = \sigma_3^{(b)}$, the triangle singularity develops, corresponding to when $s = s_{31}^{(-)} = s_{\textrm{tr}}$ 
. One can see a second threshold in the line shape above threshold $(M + m_3)^2$. Figure~\ref{fig:Tricontours} shows the RPE cut of the OPE in the $s$-plane at the corresponding values for the triangle amplitude.

We also compare the $B$-matrix triangle with the analogous one from Mai\etal~\cite{Mai:2017vot}, that we denote as $\mathcal{T}_M$, 
\begin{equation}\label{eq:TM}
\mathcal{T}_M(s) = \int_{-\infty}^{(\sqrt{s} - m_3)^2} d\sigma_3'' \frac{\rho_{3}(s,\sigma_3'') \wt{\E}_{31}^{S}(\sigma_3'',s,\sigma_1) }{M^2 - \sigma_3'' - i\epsilon},
\end{equation}
where we take their contact term equal to unity, and
the lower integration limit in their model accounts for the physics in the unphysical region. As $\sigma_3'' \to -\infty$, the OPE amplitude goes like $1/\sigma_3''$, while the phase space grows as $\sigma_3''$, thus the integrand goes like $1/\sigma_3''$ and the function is logarithmically divergent. Numerically, we choose to cut off the integral at some large value, \eg $-200$ to investigate the behavior. For $\mathcal{T}_M$, all lower limit endpoint singularities in $s$ from the phase space and OPE are moved toward $-\infty$. The $s=0$ pole from the phase space persist, and the normal threshold singularity at $s = (M+m_3)^2 $ is present since it is from the upper limit. The pinch singularity at $s = (\sqrt{\sigma_1} + m_1)^2 $ is also present, as well as the pinch singularity at $s = (\sqrt{\sigma_1} - m_1)^2$, \cf Fig.~\ref{fig:sigpm}. The second pinch singularity occurs when the integration over $\sigma_3''$ hits $\sigma_3'' = \sigma_3^{(c)}$, where
\begin{equation}
\sigma_3^{(c)} = m_1^2 + \mu_{31}^2 - \frac{m_1 (\sigma_1 + \mu_{31}^2 - m_3^2)}{\sqrt{\sigma_1}},
\end{equation}
is a third inflection point in the unphysical region, occurring at $s$ is at the threshold $s = (\sqrt{\sigma_1} - m_1)^2$ (when $\im{s_{kj}^{(-)}} = 0$). This pinch singularity is absent in the $B$-matrix triangle, as the integral is only over the physical region. Figure~\ref{fig:triReal} compares the line shapes of all three triangles, $\mathcal{T}_B$, $\mathcal{T}_F$, and $\mathcal{T}_M$. Although $\mathcal{T}_M$ has a logarithmic divergence, we fix the lower integration limit to $-200$. We see how the line shape below threshold smooths out except at the remaining singularities, shown with the black dashed lines. The red dashed line indicates the second pinch singularity in $\mathcal{T}_M$.

The Feynman triangle can be recovered from $\mathcal{T}_M$ with the method discussed by Aitchison and Pasquier~\cite{Aitchison:1966lpz}, where the isobar approximation for $\mathbf{1} \to \3$ decays was studied. Using their inversion technique, it was found that 
the Feynman triangle can be written as a dispersive integral over the isobar invariant mass as in Ref.~\cite{Mai:2017vot}, plus additional terms. The latter are real in the physical region, but cure the below threshold singularities shown in the $B$-matrix. The additional terms also cancel the logarithmic divergence, leaving a finite amplitude.

\subsection{Removal of Unphysical Singularities}
As shown, the $B$-matrix parameterization contains additional singularities which do not match the expected analytic behavior of the amplitudes. This happens in both our formulation and Mai\etal~\cite{Mai:2017vot}.  One possible venue for improving that is to substitute the $B$-matrix kernels, 
Eq.~\eqref{eq:kernels_all},
with the Feynman one. This is in the same spirit of the Chew-Mandelstam phase space in the $\2\to\2$ parameterizations, which removes the unphysical singularities of the phase space. 
Although the kernels will now have the proper analytic structure (no physical sheet singularities except for the unitarity cut), the resulting amplitude will still contain singularities from iterating the kernel. Consider the solution for $\wt{\A}_{33}$ in Eq.~\eqref{eq:Bmat2}, where the kernel is replaced by the Feynman one, $\Kc_{33} \to \Kc_{33}^{F}$, where
\begin{equation}
\Kc_{33}^{F}(s) = \int_{{\textrm{th}}}^{\infty} ds' \frac{\wt{\B}_{31}(s') \rho_3(s') \wt{\B}_{13}(s')}{s' - s - i\epsilon}.
\end{equation}
Now expand the solution Eq.~\eqref{eq:Bmat2} in an infinite series,
\begin{equation}\label{eq:FeynmanSwitch}
\wt{\A}_{33} = \Kc_{33}^{F}(s) + \Kc_{33}^{F}(s)\tau_{3}(s)\Kc_{33}^{F}(s) + \cdots.
\end{equation}
The first term is the kernel, composed of Feynman diagrams which have the correct analytic properties. Let the kernel consist only of the triangle diagram, then the second term is two Feynman triangles joined with a $\tau$-function. The equivalent Feynman diagram would have two exchanges integrated over the four-momenta, which is not equivalent to what is shown in Eq.~\eqref{eq:FeynmanSwitch} due to the $\tau$-function.
This diagram, as well as the higher-order ones, contain non-analyticities in a similar manner to what was shown for the triangle diagram. The unintegrated singularities from the phase space are always present. Therefore the simple kernel substitution does not produce the correct analytic behavior in the $B$-matrix solution. However, it can still be advantageous, as it corrects some of the unphysical singularities in the present $B$-matrix solution.

The remaining singularities should disappear if one was to solve the proper Bethe-Salpeter equations of the underlying QFT.
The $B$-matrix parameterization is indeed reminiscent of that for $\2\to\2$ scattering. We examine some differences between these formalisms. The $B$-matrix parameterization is a covariant 
integral equation for the on-shell isobar-spectator amplitudes.
It satisfies unitarity relations and does not have additional imposed constraints from analyticity. Thus, for complex energies on the physical Riemann sheet, the $B$-matrix parameterization contains the unitarity cut, and has additional $s$-singularities from the $\tau$ and OPE.

The Bethe-Salpeter equation is a covariant integral equations that incorporate an infinite number of exchanges for any given QFT~\cite{Itzykson:1980rh}. 
Solving it amounts to summation of exchange diagrams, similarly to the $B$-matrix. The resulting amplitudes are analytic functions in the complex $s$-plane, as the QFT amplitudes inherently obey analyticity constraints. The physical sheet thus has only the allowed singularities, such as the unitarity cut and possible bound state poles. 
Lippmann-Scwhinger equations are nonrelativistic equations for the scattering amplitude in a given potential model. 
The $B$-matrix has similarities to the Lippmann-Schwinger equations in that both involve in a three-dimensional integral over the momenta~\cite{Mai:2017vot}. In this work however, we focus on the physical region, and truncate the isobar mass integration appropriately. Conversely, the Bethe-Salpeter equation contains integrations over four-momenta, which results in integrating over the off-shell behavior of the amplitude. Introducing dispersive integrations in the $B$-matrix amounts to the same procedure, and would remove the unphysical singularities.

\section{Conclusions}\label{sec:Conclusion}
In summary, we have discussed the phenomenological description of $\3\to\3$ elastic scattering of spinless particles. The $\3\to\3$ amplitude was described in the isobar representation.
We constructed the unitarity relations for the isobar-spectator amplitudes for general partial wave quantum numbers. 
For a practical use, the infinite sums are truncated, leading to the standard isobar approximation.
We parameterize the isobar-spectator partial wave amplitudes with the $B$-matrix formalism, which automatically satisfies the unitarity.  
The $B$-matrix parameterization explicitly includes the one pion exchange as a long-range contribution required by unitarity. The short-range part is not constrained by unitarity, and it can be incorporated by a specific (model-dependent) choice of the parameterization. This gives to the framework enough freedom to incorporate QCD resonances.
The approach here differs from Mai\etal~\cite{Mai:2017vot} in that the $\2\to\2$ amplitudes required as input are only needed to be known in the physical energy regions. 
The singularities of the OPE directly impact the analytic structure of the $B$-matrix kernels, and are discussed explicitly for the triangle-like diagram.
The singularities in the unphysical region of our solution differ from the Mai\etal ones, and from the Feynman diagram triangle. This results in a different value for the real part of the amplitudes in the physical region.
Further studies are needed to understand how to remove unexpected singularities from the $B$-matrix. 
We also compare our formalism to the most recent ones discussed in the literature to extract three-body scattering amplitudes from lattice QCD. In particular, the main difference with Refs.~\cite{Hansen:2014eka,Hansen:2015zga,Hansen:2016ync,Briceno:2017tce,Briceno:2018mlh} consists in the order of how the partial wave expansion and the one particle exchange ladder summation is performed. It remains to be seen whether the two operations commute, and whether the resulting amplitudes coincide. 

Future studies will investigate the continuation to the unphysical energy sheets. This venue will allow us to constrain the role of one particle exchange in generating resonant structures, as it is assumed in some molecular models for the $XYZ$ states.

\begin{acknowledgments}
We thank Ian Aitchison, Ra\'ul Brice\~no, Michael D\"oring, and Maxim Mai for many useful discussions, and Arkaitz Rodas for useful comments on the manuscript. We thank the Institute for Nuclear Theory at the University of Washington for its hospitality during the completion of this work. KS was supported by the U.S.~National Science Foundation REU grant PHY-1757646. This work was supported by the U.S.~Department of Energy under grants No.~DE-AC05-06OR23177 and No.~DE-FG02-87ER40365,
PAPIIT-DGAPA (UNAM, Mexico) grants
No.~IA101717 and No.~IA101819,
CONACYT (Mexico) grant No.~251817,
Research Foundation -- Flanders (FWO),
U.S.~National Science Foundation under award 
No.~PHY-1415459,
and German Bundesministerium f\"{u}r Bildung und Forschung.
\end{acknowledgments}

\appendix
\section{Kinematics for $\3\to\3$ Reactions}\label{sec:app_A}
In this Appendix, we discuss some of the technical details of the kinematics for $\3\to\3$ processes. We first consider the system in the CMF, $\P^{\star} = \P'^{\star} = \0$. The momenta in terms of invariants are
\begin{equation}
\lvert \p_{j}^{\star} \rvert  = \frac{\lambda^{1/2}(s,m_{j}^2,\sigma_{j})}{2\sqrt{s}}, \quad \lvert \p_{k}'^{\star} \rvert  = \frac{\lambda^{1/2}(s,m_{k}^2,\sigma_{k}')}{2\sqrt{s}}.
\end{equation}
Considering the particles $j$ and $k$ as spectators, then the recoiling two particles has a total momentum $\P_{j}^{\star} = -\p_{j}^{\star}$ and $\P_{k}^{\star\,\prime} = -\p_k'^{\star}$, for the initial and final system, respectively. The invariants $t_{jk}$ and $u_{jk}$ are related to the CMF scattering angle between spectators via
\begin{subequations}
\begin{align}
t_{jk} & = (p_j - p_k')^2 \\
& = m_{j}^2 + m_{k}^2 -\frac{1}{2s} (s + m_j^2 - \sigma_{j} ) (s + m_k^2 - \sigma_{k}' ) \nonumber \\ & 
+ \frac{1}{2s} \lambda^{1/2}(s,\sigma_{j},m_j^2)\lambda^{1/2}(s,\sigma_{k}',m_k^2)z_{jk}^{\star}, \nonumber \\
u_{jk} & = ((P - p_j)- p_k')^2 \\
& = \sigma_{j} + m_{k}^2 -\frac{1}{2s} (s + \sigma_{j} - m_j^2) (s + m_k^2 - \sigma_{k}' ) \nonumber  \\ & 
- \frac{1}{2s} \lambda^{1/2}(s,\sigma_{j},m_j^2)\lambda^{1/2}(s,\sigma_{k}',m_k^2)z_{jk}^{\star},\nonumber
\end{align}
\end{subequations}
where $z_{jk}^{\star} = \cos\Theta_{jk}^{\star}$.
The cosine of the CMF angle between particles $j$ and $k$ is
\begin{equation}\label{eq:angle_appA}
\begin{split}
\cos\theta_{kj}^{\star}  = \frac{2s(\sigma_{j} + m_{k}^2 - \mu_{jk}^2) - (s + \sigma_{j} - m_j^2) (s + m_k^2 - \sigma_{k} )  }{ \lambda^{1/2}(s,\sigma_{j},m_j^2)\lambda^{1/2}(s,\sigma_{k},m_k^2) },
\end{split}
\end{equation}
where $\mu_{jk}$ is the mass of the particle that is neither $j$ nor $k$. 

The remaining variables needed to completely describe the $\3\to\3$ process are found by examining the IRFs. The initial IRF$_j$ and final IRF$'_k$ are defined when $\P_j = \0$ and $\P_k' = \0$, respectively. We use the convention that initial and final state variables are evaluated in their own respective IRF. The momentum of the first particle in the initial pair is denoted as $\q_j$ in the IRF$_j$. Similarly, the first particle in the final pair is $\q_k'$ in the IRF$'_k$. For example, for the final spectator $3$ in the IRF$'_3$, $\q_3'$ is the final momentum of particle 1, and in the IRF$_1$ of spectator 1, $\q_1$ is the initial momentum of particle 3. In terms of invariants, these momenta are
\begin{equation}
\lvert \q_3' \rvert = \frac{\lambda^{1/2}(\sigma_3',m_1^2,m_2^2)}{2\sqrt{\sigma_3'}}, \,\, \lvert \q_1 \rvert = \frac{\lambda^{1/2}(\sigma_1,m_3^2,m_2^2)}{2\sqrt{\sigma_1}}.
\end{equation}
The spectator momenta in these frames are
\begin{equation}
\lvert \p_3' \rvert = \frac{\lambda^{1/2}(s,\sigma_3',m_3^2)}{2\sqrt{\sigma_3'}},
\end{equation}
for the final state and
\begin{equation}
\lvert \p_1 \rvert = \frac{\lambda^{1/2}(s,\sigma_1,m_1^2)}{2\sqrt{\sigma_1}},
\end{equation}
for the initial state. The helicity angles of the first particle in the IRFs are given by $\chi_j$ and $\chi_k'$, for the initial and final states, respectively. The helicity angles are defined w.r.t.
the opposite line-of-flight of the spectator. The azimuthal angles for the initial and final state are $\gamma_j$ and $\gamma_k'$, respectively. The azimuthal angles are defined as the angle between the plane of the two particles in the CMF, and the IRFs, \cf Fig.~\ref{fig:3bodykins}. Note that the azimuthal angles $\gamma_j$ and $\gamma_k'$ are invariant with respect to the Lorentz boost between the IRFs and CMF, so $\gamma_j = \gamma_j^{\star}$ and $\gamma_k' = \gamma_k'^{\star}$. 

The invariant masses of the other two pairs in their respective frames are related to the helicity angles. For example, in the IRF$'_3$,
\begin{equation}
\begin{split}
\sigma_1' & = (P - p_1')^2 \\
& = s + m_1^2 - \frac{1}{2\sigma_3'}(s+\sigma_3' - m_3^2)(\sigma_3' + m_1^2 - m_2^2)  \\ &
+ \frac{1}{2 \sigma_3'} \lambda^{1/2}(s,\sigma_3',m_3^2)\lambda^{1/2}(\sigma_3',m_1^2,m_2^2) \cos\chi_3' , \\
\sigma_2' & = (P - p_2')^2 \\
& = s + m_2^2 - \frac{1}{2\sigma_3'} (s+\sigma_3' - m_3^2)(\sigma_3' + m_2^2 - m_1^2)  \\ &
- \frac{1}{2\sigma_3'} \lambda^{1/2}(s,\sigma_3',m_3^2) \lambda^{1/2}(\sigma_3',m_1^2,m_2^2) \cos\chi_3'.
\end{split}
\end{equation}

\section{Derivation of $\3\to\3$ Unitarity Relations}\label{sec:app_B}
In this appendix we derive the general elastic unitarity relations for the $\3\to\3$ elastic scattering of distinguishable spinless particles \cite{Fleming:1964zz,Holman:1965,Eden:1966dnq}. For convenience, in this section we adopt the notation that the normalization of a single particle state is $\braket{\p_k'|\p_j} = (2\pi)^{3}\,2\omega_j \delta^{(3)}(\p'_k - \p_j)\delta_{jk} \equiv \wt{\delta}(p'_k - p_j)\delta_{jk}$, and the invariant measure is $\wt{d}p_j \equiv d^{3}\p_j / (2\pi)^{3}\,2\omega_j $. The $S$-matrix is a unitary operator, $S^{\dag}S = \1$, which implies that $T - T^{\dag} = iT^{\dag}T$, where $S = \1 + iT$. We consider the system in an energy range above the three particle threshold, but below the first inelastic threshold, $s_{\mathrm{th}} \le s < s_{\mathrm{inel}}$
. Taking matrix elements of this operator between initial and final states $\ket{\p}$ and $\ket{\p'}$, and inserting the completeness relation $\1= \int \wt{d}p_1''\wt{d}p_2'' \wt{d}p_3''\,\ket{\p''}\bra{\p''}$, gives the unitarity relation
\begin{equation}\label{eq:unitarity_app}
\begin{split}
& \bra{\p'}T\ket{\p} -\bra{\p'} T^{\dag}\ket{\p} \\
& \qquad = i  \int \wt{d}p_1'' \wt{d}p_2'' \wt{d}p_3'' \bra{\p'}T^{\dag}\ket{\p''}\bra{\p''}T\ket{\p}.
\end{split}
\end{equation}
Since $T = T_d + T_c$, where $T_d = \sum_{j} \1_j \otimes T^{(j)}$, then the matrix element is
\begin{equation} \label{eq:Tc+Td_app}
\begin{split}
\bra{\p'}T\ket{\p} & = \bra{\p'}T_{c}\ket{\p} \\
& \, + \sum_{j} \wt{\delta}(p_j' - p_j) \bra{\p'}T^{(j)} \ket{\p} .
\end{split}
\end{equation}
The matrix elements $\bra{\p'} T^{\dag}\ket{\p}$ are equal to $\bra{\p'}T\ket{\p}^{*}$ by the property of Hermitian analyticity \cite{Olive:1962,Eden:1966dnq}. Thus the left hand side of Eq.~\eqref{eq:unitarity_app}  gives the imaginary part of the matrix element,
\begin{equation}
\begin{split}
\text{LHS} & = 2i\,\im\bra{\p'}T_c\ket{\p}  \\
& + \sum_{j}\wt{\delta}(p_j'-p_j) 2i\,\im \bra{\p'}T^{(j)}\ket{\p}.
\end{split}
\end{equation}
The right hand side of Eq.~\eqref{eq:unitarity_app} is evaluated by substituting
Eq.~\eqref{eq:Tc+Td_app} and expanding the product into four terms,

\begin{widetext}

\begin{equation}
\begin{split}
\text{RHS}  & =  i \int \widetilde{d}p_1'' \wt{d}p_2'' \wt{d}p_3''  \bigg[  \bra{\p'}T_c^{\dag}\ket{\p''} \bra{\p''}T_c\ket{\p} \\
& \qquad \qquad \qquad \quad + \sum_{k} \wt{\delta}(p_k''-p_k')\bra{\p'}T^{(k)\,\dag}\ket{\p''}\bra{\p''}T_c\ket{\p}  \\
& \qquad \qquad \qquad \quad + \sum_{j} \wt{\delta}(p_j''-p_j)\bra{\p'}T_c^{\dag}\ket{\p''}\bra{\p''}T^{(j)}\ket{\p} \\
& \qquad \qquad \qquad \quad + \sum_{j,k} \wt{\delta}(p_k''-p_k')\wt{\delta}(p_j''-p_j) \bra{\p'}T^{(k)\,\dag}\ket{\p''} \bra{\p''}T^{(j)}\ket{\p}  \bigg].
\end{split}
\end{equation}
The fourth term contains two cases, one where $j=k$, and one where $j\ne k$, so we split the sum into the two distinct terms
\begin{equation}\label{eq:DiscSplit_app}
\begin{split}
 \sum_{j,k} \widetilde{\delta}(p_k''-p_k')\wt{\delta}(p_j''-p_j) & \bra{\p'}T^{(k)\,\dag}\ket{\p''} \bra{\p''}T^{(j)}\ket{\p} \\
&  = \sum_{j} \wt{\delta}(p_j''-p_j')\wt{\delta}(p_j''-p_j) \bra{\p'}T^{(j)\,\dag}\ket{\p''} \bra{\p''}T^{(j)}\ket{\p} \\
& \quad + \sum_{\substack{ j,k \\ j\ne k}} \wt{\delta}(p_k''-p_k')\wt{\delta}(p_j''-p_j) \bra{\p'}T^{(k)\,\dag}\ket{\p''} \bra{\p''}T^{(j)}\ket{\p}.
\end{split}
\end{equation}
We can write $\wt{\delta}(p_j''-p_j')\wt{\delta}(p_j''-p_j) = \wt{\delta}(p_j' - p_j)\wt{\delta}(p_j'' - p_j)$ in the first term in Eq.~\eqref{eq:DiscSplit_app}, thus we can identify the disconnected unitarity relation as being proportional to the spectator singularity $\wt{\delta}(p_j' - p_j)$,
\begin{equation}\label{eq:2to2DiscUnit_app}
2\im \bra{\p'}T^{(j)}\ket{\p} =  \int \wt{d}p_{j_1}'' \wt{d}p_{j_2}''  \bra{\p'}T^{(j)\,\dag}\ket{\p''} \bra{\p''}T^{(j)}\ket{\p},
\end{equation}
and the connected unitarity relation 
\begin{equation}\label{eq:3to3ConnUnit_app}
\begin{split}
2\im \bra{\p'}T_c\ket{\p} & =   \int \widetilde{d}p_1'' \widetilde{d}p_2'' \widetilde{d}p_3'' \bigg[ \bra{\p'}T_c^{\dag}\ket{\p''} \bra{\p''}T_c\ket{\p} \\
& + \sum_{k} \wt{\delta}(p_k''-p_k')\bra{\p'}T^{(k)\,\dag}\ket{\p''}\bra{\p''}T_c\ket{\p}  \\
& + \sum_{j} \wt{\delta}(p_j''-p_j)\bra{\p'}T_c^{\dag}\ket{\p''}\bra{\p''}T^{(j)}\ket{\p} \\
& + \sum_{\substack{ j,k \\ j\ne k}} \wt{\delta}(p_k''-p_k')\wt{\delta}(p_j''-p_j) \bra{\p'}T^{(k)\,\dag}\ket{\p''} \bra{\p''}T^{(j)}\ket{\p} \bigg].
\end{split}
\end{equation}
The momenta with $j_1$ and $j_2$ in Eq.~\eqref{eq:2to2DiscUnit_app} identify the first and second particle in the pair. Substituting Eqs.~\eqref{eq:3to3c_Amp} and \eqref{eq:2to2d_Amp} into Eqs.~\eqref{eq:2to2DiscUnit_app} and \eqref{eq:3to3ConnUnit_app}, and evaluating the phase space integrals yield the unitarity relations Eqs. \eqref{eq:2to2discUnit} and \eqref{eq:3to3connUnit}.

\section{Derivation of PWIS Unitarity Relations}\label{sec:app_C}
Using the assumption of the isobar model Eq.~\eqref{eq:Isobar_Expand}, we derive a set of unitarity relations for the amputated PWIS amplitudes. Inserting Eq.~\eqref{eq:Isobar_Expand} into the unitarity relations Eq.~\eqref{eq:3to3connUnit} leads to a unitarity relation for the $\A_{kj}$ isobar-spectator amplitude,
\begin{equation}\label{eq:app_C_3to3UnitIsobar}
\begin{split}
 \im{\A_{kj}(\p';\p)} & = \frac{1}{\pi(32\pi^2)^2} \sum_{n} \int_{\sigma_{n}^{(\mathrm{th})}}^{(\sqrt{s} - m_n)^2} d\sigma_n'' \, \frac{\lvert \q_n'' \rvert \lvert \p_n''^{\star} \rvert }{ \sqrt{\sigma_n''} \sqrt{s}} \int d\wh{\P}_n''^{\star}  \int d\wh{\q}_n'' \, \A_{nk}^{*}(\p'';\p') \A_{nj}(\p'';\p) \Theta(s - s_{\mathrm{th}})  \\
\
& + \frac{1}{\pi(32\pi^2)^2} \sum_{\substack{n,r \\ n \ne r}}  \int_{\sigma_{n}^{(\mathrm{th})}}^{(\sqrt{s} - m_n)^2} d\sigma_n'' \, \frac{\lvert \q_n''\rvert \lvert \p_n''^{\star} \rvert }{ \sqrt{\sigma_n''} \sqrt{s}} \int d\wh{\P}_n''^{\star}  \int d\wh{\q}_{n}'' \,\A_{rk}^{*}(\p'';\p') \A_{nj}(\p'';\p) \Theta(s - s_{\mathrm{th}}) \\
\
&  + 
\rho_2(\sigma_k')\int d\wh{\q}_k'' \,\F_k^{*}(\p'';\p') \A_{kj}(\p'';\p)\rvert_{\p''_k = \p_k'} \Theta(\sigma_k' - \sigma_{k}^{(\mathrm{th})}) \\
\
& + 
\rho_2(\sigma_j)\int d\wh{\q}_j'' \,\A_{kj}^{*}(\p'';\p') \rvert_{\p''_j = \p_j} \F_{j}(\p'';\p)\Theta(\sigma_j - \sigma_{j}^{(\mathrm{th})})  \\
\
& + 
\rho_2(\sigma_k')\sum_{\substack{r \\ k \ne r}}\, \int d\wh{\q}_k''\,\F_k^{*}(\p'';\p') \A_{rj}(\p'';\p)\rvert_{\p''_k = \p_k'} \Theta(\sigma_k' - \sigma_{k}^{(\mathrm{th})}) \\
\
& + 
\rho_2(\sigma_j)\sum_{\substack{n \\ n \ne j}}\, \int d\wh{\q}_j'' \,\A_{kn}^{*}(\p'';\p') \rvert_{\p''_j = \p_j} \F_{j}(\p'';\p)\Theta(\sigma_j - \sigma_{j}^{(\mathrm{th})})  \\
\
& + \pi \, \delta(u_{jk} - \mu_{jk}^2)\,\F_{k}^{*}(\p'';\p')\rvert_{\p''_j = \p_j} \F_{j}(\p'';\p) \rvert_{\p''_k = \p_k'} (1 - \delta_{jk}) ,
\end{split}
\end{equation}
\end{widetext}
where we wrote the three-body phase space factor in the first two terms,
\begin{equation}\label{eq:app_C_3bodyPhaseSpace1}
\begin{split}
& \frac{1}{2(2\pi)^5}\int \frac{d^{3}\p_1''}{2\omega_1''}\frac{d^{3}\p_2''}{2\omega_2''}\frac{d^{3}\p_3''}{2\omega_3''} \delta^{(4)}(P'' - P)  \\
\
& = \frac{1}{\pi (32\pi^2)^2} \int_{\sigma_n^{(\mathrm{th})}}^{(\sqrt{s} - m_n)^2} d\sigma_n'' \, \frac{\lvert \q_n''\rvert \lvert \p_n^{\prime\prime\,\star} \rvert }{ \sqrt{\sigma_n''} \sqrt{s}} \int d\wh{\P}_n''^{\star}  \int d\wh{\q}_n''  ,
\end{split}
\end{equation}
\begin{figure*}[t!]
    \centering
    \includegraphics[ width=0.7\textwidth]{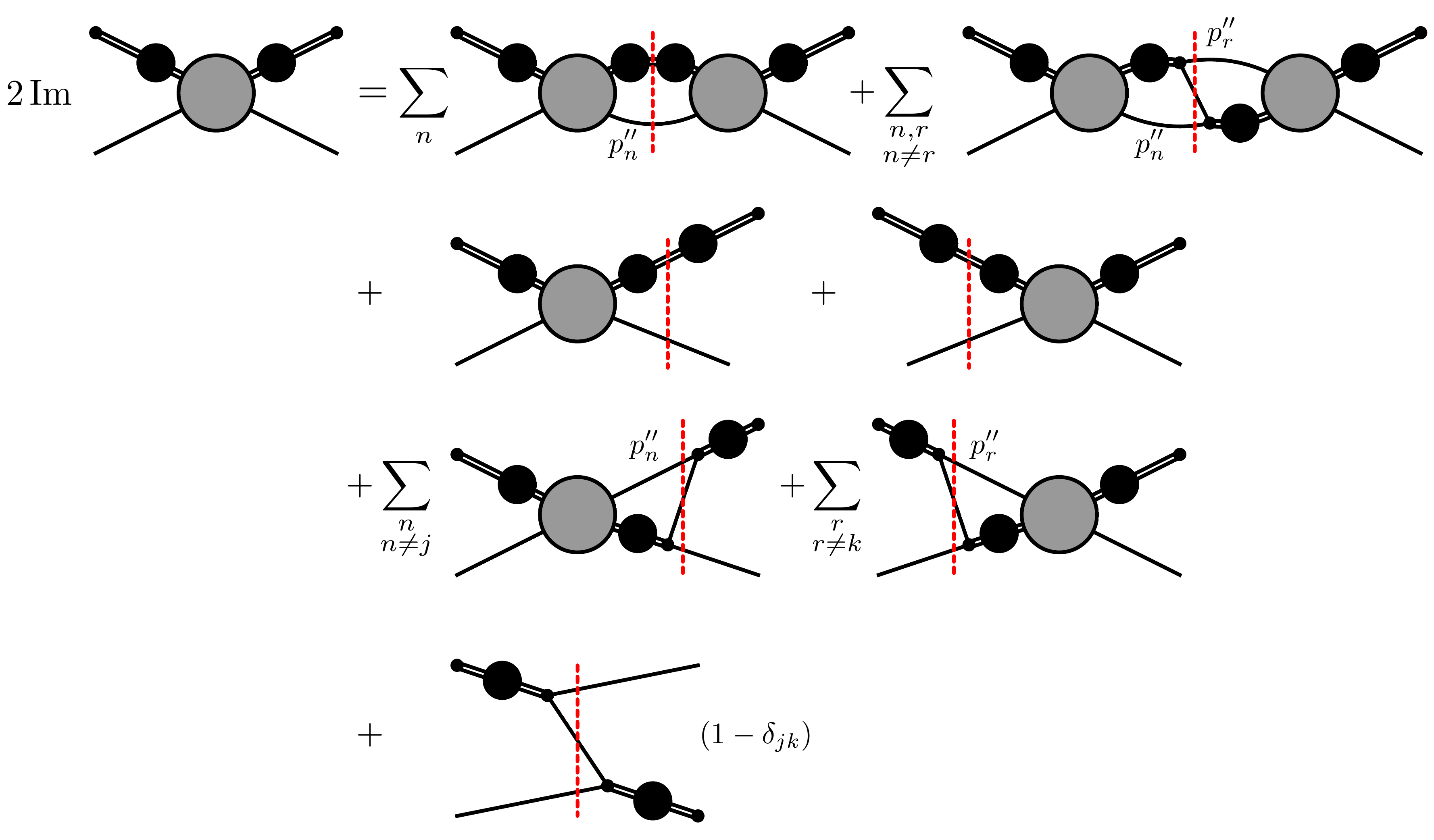}
    \caption{Diagrammatic representation for the isobar-spectator unitarity relation in Eq.~\eqref{eq:app_C_3to3UnitIsobar}.
    }
    \label{fig:isobar_spectator_unitarity}
\end{figure*}
where $\wh{\P}_n''^{\star} $ is the orientation of the isobar associated with spectator $n$ in the intermediate state, and $\wh{\q}_n'' $ is the orientation of the first particle in the intermediate isobar in its rest frame. The terms in the intermediate state have been split up into two groups, diagonal and off-diagonal. The diagonal terms in Eq.~\eqref{eq:app_C_3to3UnitIsobar} (first, third, and fourth line) are terms such that the isobar propagates in the intermediate state without breaking up. The off-diagonal terms (second, fifth, and sixth line in Eq.~\eqref{eq:app_C_3to3UnitIsobar}) are ones where the isobar breaks up in the intermediate state, and combines with the spectator to form a new isobar. Figure~\ref{fig:isobar_spectator_unitarity} shows the diagonal and off-diagonal topologies in the intermediate state.

The partial wave expansion Eq.~\eqref{eq:PWIS_expansion} can be derived by considering the expansion in three steps. The first step is to perform the expansion of the isobars into definite spin and helicity,
\begin{equation} \label{eq:app_C_isobar_strip}
\begin{split}
\A_{kj} = \sum_{s_k',\lambda_k'}\sum_{s_j,\lambda_j} Y_{\lambda_k'}^{s_k' }(\wh{\q}_k') \A_{s_k' \lambda_k' ; s_j \lambda_j} Y_{\lambda_j}^{s_j \, *}(\wh{\q}_j),
\end{split}
\end{equation}
where $\lambda_j$, $\lambda_k'$ are defined along the direction of the isobar in the CMF. The expansion removes the $\wh{\q}_k'$ and $\wh{\q}_j$ dependence in the helicity amplitude $\A_{s_k' \lambda_k' ; s_j \lambda_j}$. Second, the helicity amplitude can be expanded in partial waves,
\begin{equation}\label{eq:app_C_hel_expansion}
\begin{split}
\A_{s_k' \lambda_k' ; s_j \lambda_j} & = \sum_{J} \N_{J}^2 \, \A_{s_k' \lambda_k' ; s_j \lambda_j}^{J}(\sigma_k',s,\sigma_j) \\
& \times \sum_{M} \D_{M \lambda_k'}^{(J)\,*}(\wh{\P}_{k}'^{\star}) \D_{M \lambda_j}^{(J)}(\wh{\P}_j^{\star}),
\end{split}
\end{equation}
where $\N_J^2 = (2J + 1) / 4\pi$.
Finally, since parity is not a good quantum number in the helicity basis, we convert the helicity partial wave amplitudes into $LS$ partial wave amplitudes,
\begin{equation}\label{eq:app_C_LS_hel}
\A_{\ell_k' s_k' ; \ell_j s_j}^{J} = \sum_{\lambda_k',\lambda_j} C^{J}_{\ell_k' s_k' \lambda_k'} C^{J}_{\ell_j s_j \lambda_j}  \A_{s_k' \lambda_k' ; s_j \lambda_j}^{J} ,
\end{equation}
where $C_{\ell s \lambda}^{J} = \sqrt{ (2\ell + 1) / (2 J + 1)} \braket{J \lambda | \ell 0 s \lambda}$, and has the completeness relation
\begin{equation}\label{eq:app_C_rec_ortho}
\sum_{\lambda} C_{\ell s \lambda}^{J} C_{\ell' s' \lambda}^{J} = \delta_{JJ'}\delta_{\ell\ell'} \delta_{ss'}.
\end{equation}
Combining Eqs.~\eqref{eq:app_C_isobar_strip}, \eqref{eq:app_C_hel_expansion}, and \eqref{eq:app_C_LS_hel} yields the partial wave expansion Eq.~\eqref{eq:PWIS_expansion}. We apply the expansion to Eq.~\eqref{eq:app_C_3to3UnitIsobar} to obtain the PWIS unitarity relations.
The diagonal terms are most directly evaluated using the orthonormality condition, Eq.~\eqref{eq:ZfcnOrtho}. Since $\p_k'' = \p_k'$ in the third term, and $\p_j'' = \p_j$ in the fourth term, then the intermediate isobar orientation is identical to that of the final and initial state isobar, respectively. The integrations over $\wh{\q}_k'$ and $\wh{\q}_j$ can be performed by writing Eq.~\eqref{eq:2to2PW} using the spherical harmonic addition theorem.

The off-diagonal terms are more challenging, as they involve two different angles in the intermediate state, thus the rotational functions will not directly integrate out. We can make use of the group properties of rotations to simplify the intermediate rotational functions to a recoupling coefficient. The off-diagonal terms on the right-hand side of Eq.~\eqref{eq:app_C_3to3UnitIsobar} under the expansion Eq.~\eqref{eq:PWIS_expansion} will contain terms of the form
\begin{equation}
\begin{split}
& \D_{\lambda_n 0}^{(s_n)\,*} (\wh{\q}_n)  \D_{M\lambda_n}^{(J)\,*}(\wh{\P}_n^{\star}) \D_{M\lambda_r}^{(J)}(\wh{\P}_r^{\star}) \D_{\lambda_r 0}^{(s_r)}(\wh{\q}_r) \\
& \quad  = d_{\lambda_n 0}^{(s_n)} (\cos\chi_n) \D_{M \lambda_n}^{(J)\,*}(R_n^{\star})  \D_{M \lambda_r}^{(J)}(R_r^{\star}) d_{\lambda_r 0 }^{(s_r)}(\cos\chi_r),
\end{split}
\end{equation}
where $n\ne r$, and we combined the terms with $\gamma_n$ and the orientation of the isobar to the set of angles $R_n^{\star} = (\alpha_n^{\star},\beta_n^{\star},\gamma_n^{\star})$, where $\alpha_n^{\star}$ is the azimuthal angle of the isobar and $\beta_n^{\star}$ is the polar angle, w.r.t. some fixed coordinate system. Note that since we boost along the direction of the isobar to go between CMF and IRFs, $\gamma_n = \gamma_n^{\star}$. The angles $R^{\star}$ are the Euler angles describing the orientation of the three particles in their CMF. Since these two sets of angles describe the same configuration of three particles, with the only difference being which particle is the spectator, the angles $R_n^{\star}$ and $R_r^{\star}$ must be related by a rotation.

\begin{figure}[b!]
    \centering
    \includegraphics[trim={2cm 4cm 1cm 5cm},clip, width=1.0\columnwidth]{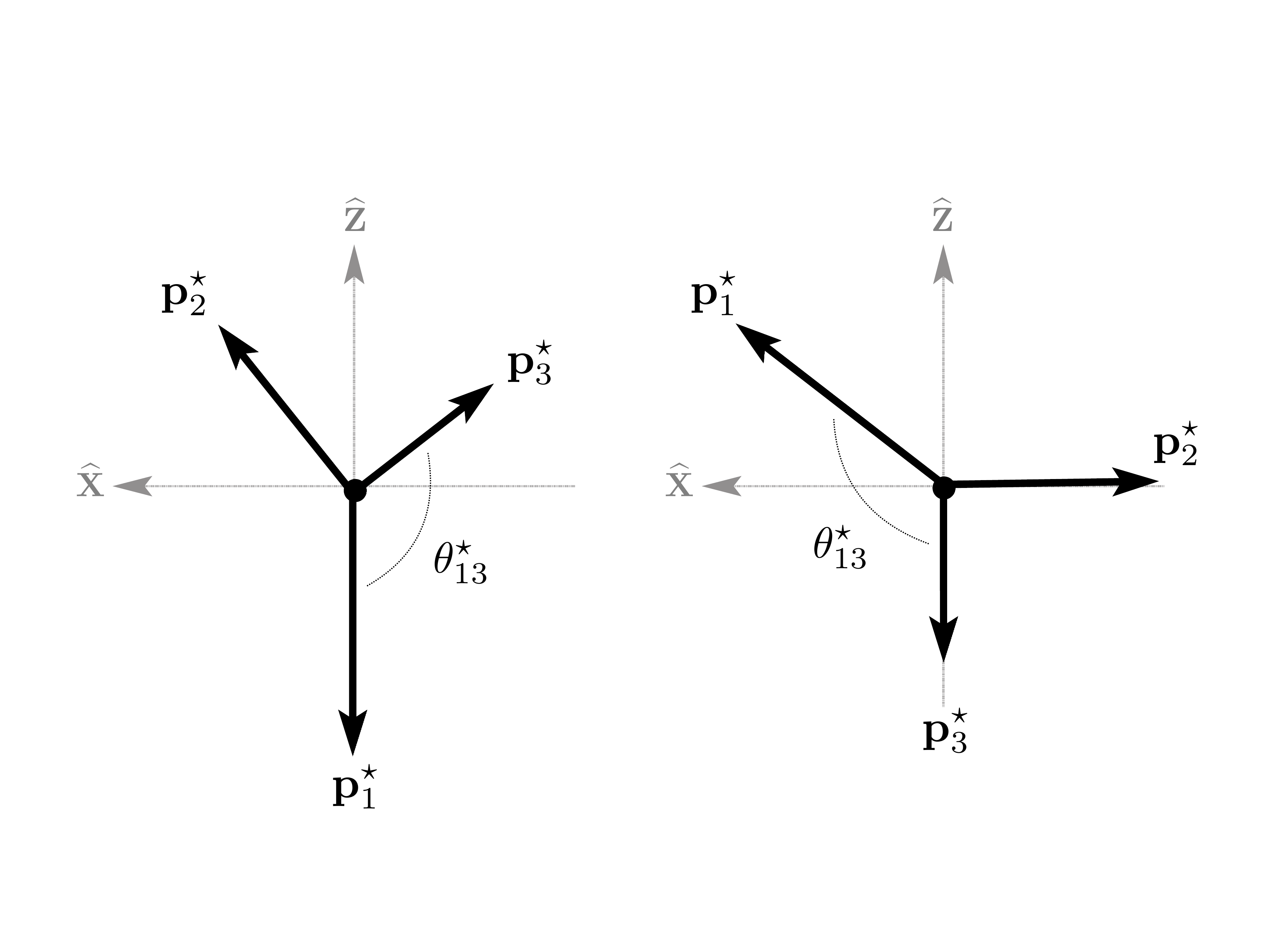}
   \put(-145,10){\colorbox{white}{(a)}}
   \put(-30,10){\colorbox{white}{(b)}}
    \caption{The standard configurations considering (a) particle 1 as the spectator and (b) particle $3$ as the spectator. 
    }
    \label{fig:SC}
\end{figure}

Each set of angles can be found by rotating from some initial standard configuration. We define the standard configuration such that the three particle system lies in the $xz$-plane, where the spectator momenta is along the negative $z$-axis, \cf Fig.~\ref{fig:SC}. Then, the difference in the Euler angles is a rotation about the $y$-axis,
\begin{equation}
R_r^{\star} = R_n^{\star} r_{nr}^{\star},
\end{equation}
where $r_{nr}^{\star}$ is the rotation relating the two standard configurations \cite{Ascoli:1975mn,Giebink:1985zz}. Here, the rotation is
about the $y$-axis, $r_{nr}^{\star} = R_{y}(\theta_{nr}^{\star})$, where $\theta_{nr}^{\star}$ is given in Eq.~\eqref{eq:angle_appA}. Thus, the rotation is a function of the invariant masses of the isobars, $\theta_{nr}^{\star} = \theta_{nr}^{\star}(\sigma_n,s,\sigma_r)$. Note that the inverse rotation is given by $r_{rn }^{\star} = r_{nr}^{\star\,-1}$.

Therefore, we can relate the two Wigner $\D$-matrices using the group addition property
\begin{equation}
\begin{split}
\D_{M\lambda_r}^{(J)}(R_{r}^{\star})   =  \sum_{\lambda_n} \D_{M \lambda_n}^{(J)}(R_{n}^{\star})  \D_{\lambda_n \lambda_r}^{(J)}(r_{nr}^{\star}).
\end{split}
\end{equation}
The integration over the Euler angles in the intermediate state can be performed, leaving one rotation that recouples the isobars,
\begin{equation}\label{eq:recouple_property}
\begin{split}
\int d R_{n}^{\star} \,& \D_{M\lambda_n}^{(J)\,*}(R_{n}^{\star}) \D_{M \lambda_r}^{(J)}(R_{r}^{\star}) \\
& =  \sum_{\lambda}  \int d R_n^{\star}\, \D_{M\lambda_n}^{(J)\,*}(R_{n}^{\star})\D_{M \lambda}^{(J)}(R_{n}^{\star})\D_{\lambda \lambda_r}^{(J)}(r_{nr}^{\star})  \\
& = \frac{8\pi^2}{2J+1} 
d_{\lambda_n \lambda_r}^{(J)}(\cos{\theta_{nr}^{\star}}).
\end{split}
\end{equation}
where $dR_n^{\star} = d\alpha_n^{\star} d\cos\beta_n^{\star} d\gamma_n^{\star}$.

Since the angle $\chi$ can be written in terms of the invariant masses of the isobars, it is advantages to write the phase space in terms of the Dalitz variables,
\begin{equation}\label{eq:app_C_3bodyPhaseSpace2}
\begin{split}
& \frac{1}{(2\pi)^5}\int \frac{d^{3}\p_1''}{2\omega_1''}\frac{d^{3}\p_2''}{2\omega_2''}\frac{d^{3}\p_3''}{2\omega_3''} \delta^{(4)}(P'' - P) \\
\
& = \frac{1}{\pi s \, (32\pi^2)^{2}} \int_{\sigma_n^{(\mathrm{th})}}^{(\sqrt{s} - m_n^2)^2} d\sigma_n'' \int_{\sigma_r^{(-)}}^{\sigma_r^{(+)}} d\sigma_r''\, \int d R_{n}''^{\star} ,
\end{split}
\end{equation}
where we used
\begin{equation}\label{eq:app_C_3bodyPhaseSpace3}
d\sigma_r'' = 2 \frac{\sqrt{s}}{\sqrt{\sigma_n''}} \lvert \q_n'' \rvert  \lvert \p_n''^{\star} \rvert \,d\cos\chi_n'',
\end{equation}
to rewrite Eq.~\eqref{eq:app_C_3bodyPhaseSpace1}.  The Dalitz region is bounded by $\sigma_n^{(\mathrm{th})} \le \sigma_n'' \le (\sqrt{s} - m_n)^2$ and $\sigma_r^{(-)} \le \sigma_r'' \le \sigma_r^{(+)}$, where $\sigma_r^{(\pm)} = \sigma_r^{(\pm)}(\sigma_n'')$ is found by the physical boundary $\cos\chi_n'' = \pm 1$, \eg for $n = 1$ and $r=3$, then
\begin{equation}
\begin{split}
\sigma_3^{(\pm)} & = m_1^2 + m_2^2 - \frac{1}{2\sigma_1''} (\sigma_1'' - s + m_1^2)(\sigma_1'' + m_2^2 - m_3^2) \\
\
& \pm \frac{1}{2\sigma_1''} \lambda^{1/2}(s,\sigma_1'',m_1^2)\lambda^{1/2}(\sigma_1'',m_2^2,m_3^2).
\end{split}
\end{equation}

The last piece needed is the partial wave projection of the OPE term. To evaluate the partial wave projection, we write the delta-function as
\begin{equation}\label{eq:app_C_ope_term1}
\delta( u_{jk} - \mu_{jk}^2) = \frac{1}{2\lvert \p_k'^{\star} \rvert \lvert \p_j^{\star} \rvert} \delta(z_{kj}^{\star} - z_{kj}),
\end{equation}
where $z_{kj}$ is defined in Eq.~\eqref{eq:z_mu}. Then, the completeness relation for the delta-function allows us to write Eq.~\eqref{eq:app_C_ope_term1} as
\begin{equation}
\delta(z_{kj}^{\star} - z_{kj}) = \sum_{J}\left( \frac{2J + 1}{2}\right) d_{\lambda' \lambda}^{(J)}(z_{kj}^{\star}) d_{\lambda'\lambda}^{(J)}(z_{kj}),
\end{equation}
where $\lambda$ and $\lambda'$ are arbitary, and thus we may choose them to align with $\lambda_j$ and $\lambda_k'$, respectively. Then, $d_{\lambda_j \lambda_k'}^{(J)}(z_{kj}^{\star})$ is written in terms of the angles $\wh{\P}_j^{\star}$ and $\wh{\P}_k'^{\star}$, via the group addition property,
\begin{equation}
d_{\lambda'\lambda}^{(J)}(z_{kj}^{\star}) = \sum_{M} \D_{M \lambda'}^{(J)\,*}(\wh{\P}_k'^{\star})\D_{M\lambda}^{(J)}(\wh{\P}_j^{\star}).
\end{equation}

\begin{figure}[b!]
    \centering
    \includegraphics[trim={4cm 5cm 8cm 3.5cm},clip,  width=0.9\columnwidth]{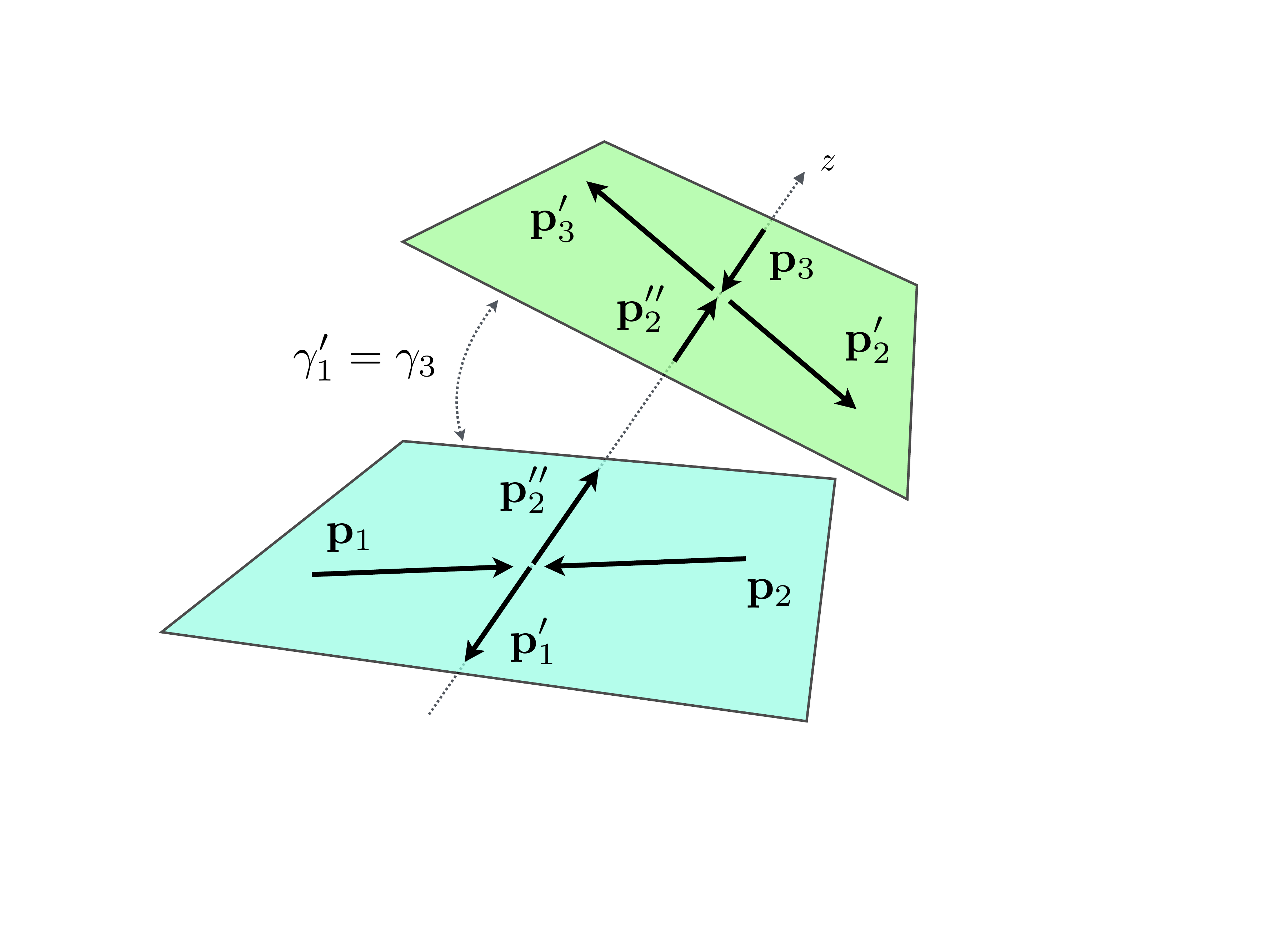}
    \caption{Relation between the initial and final IRF planes for the OPE amplitude.}
    \label{fig:ope_planes}
\end{figure}

Finally, the $\2\to\2$ amplitudes are written using Eq.~\eqref{eq:2to2PW}
\begin{equation}
\begin{split}
F_{k}^{*}F_j & = f^{*}_{s_k'}(\sigma_k') f_{s_j}(\sigma_j) P_{s_k'}(\wh{\ov{\q}}_j \cdot \wh{\q}_k') P_{s_j}(\wh{\q}_j \cdot \wh{\ov{\q}}_k') \\
\
&   =f^{*}_{s_k'}(\sigma_k') f_{s_j}(\sigma_j)  \sum_{\lambda_k'} \D_{\lambda_k'0}^{(s_k')\,*}(\wh{\q}_k')\D_{\lambda_k'0}^{(s_k')}(\wh{\ov{\q}}_{j}) \\
& \qquad \times \sum_{\lambda_j} \D_{\lambda_j 0}^{(s_j)\,*}(\wh{\ov{\q}}_k') \D_{\lambda_j 0}^{(s_j)}(\wh{\q}_j),
\end{split}
\end{equation}
where $\bar{\q}_j$ is the momentum of the first particle in the final state in the IRF$_j$, and $\bar{\q}_k'$ is the momentum of the first particle in the initial state in the IRF$'_k$.
Since $\p_k'' = \p_k'$ and $\p_j'' = \p_j$, then the azimuthal angles are identical, $\gamma_k' = \gamma_j$. The helicity angles of the first particle in the opposite frames are defined as $\ov{\chi}_j$ and $\ov{\chi}_k'$, \cf Fig.~\ref{fig:ope_planes}. However, we can easily identify that $\chi_k' = \ov{\chi}_k'$ and $\chi_j = \ov{\chi_k'}$ since the intermediate spectator is aligned for the OPE and the IRFs merely differ by the rotation about $z$.

Combining all of this together yields the PWIS unitarity relations,
\begin{widetext}
\begin{equation}\label{eq:app_C_PWIS_unitarity}
\begin{split}
 \im\, & \A^{J}_{ \ell_k' s_k'  ; \ell_j s_j }(\sigma_k',s,\sigma_j) \\
 & = \frac{1}{\pi(32\pi^2)^2} \sum_{n} \sum_{\ell_n'' , s_n''} \int_{\sigma_{n}^{(\mathrm{th})}}^{(\sqrt{s} - m_n)^2} d\sigma_n'' \, \frac{\lvert \q_n'' \rvert \lvert \p_n''^{\star} \rvert }{ \sqrt{\sigma_n''} \sqrt{s}}  \, \A^{J\,*}_{\ell_n'' s_n'' ; \ell_k' s_k' }(\sigma_n'',s,\sigma_k') \A^{J}_{\ell_n'' s_n'' ; \ell_j s_j }(\sigma_n'',s,\sigma_j) \Theta(s - s_{\mathrm{th}})  \\
& + \frac{1}{2\pi s(32\pi^2)^2} \sum_{\substack{n,r \\ n \ne r}} \sum_{\ell_n'' , s_n''}\sum_{\ell_r'' , s_r''} \int_{\sigma_{n}^{(\mathrm{th})}}^{(\sqrt{s} - m_n)^2} d\sigma_n'' \,\int_{\sigma_r^{(-)}(\sigma_n'')}^{\sigma_r^{(+)}(\sigma_n'')} d\sigma_r'' \,    \\
& \qquad\qquad \times \C_{\ell_n'' s_n'' ; \ell_r'' s_r''}^{J}(\sigma_n'',s,\sigma_r'')\,  \A^{J\,*}_{\ell_r'' s_r''; \ell_k' s_k' }(\sigma_r'',s,\sigma_k') \A^{J}_{\ell_n'' s_n'' ;  \ell_j s_j }(\sigma_n'',s,\sigma_j) \Theta(s - s_{\mathrm{th}}) \\
& + 
\rho_2(\sigma_k')\, f_{s_k'}^{*}(\sigma_k') \A_{\ell_k' s_k' ; \ell_j s_j}^{J}(\sigma_k',s,\sigma_j) \\
&  +
\rho_2(\sigma_j)\, \A_{\ell_k' s_k' ; \ell_j s_j}^{J \, *}(\sigma_k',s,\sigma_j) f_{s_j}(\sigma_j) \\
& +  \frac{1}{64 \pi^2 \sqrt{s}}  \frac{1}{\lvert \p_k^{\prime\,\star} \rvert}  \sum_{\substack{r \\ r \ne k}} \sum_{\ell_r'', s_r''}\, \int_{\sigma_r^{(\mathrm{th})}}^{(\sqrt{s}-m_r)^2} d\sigma_r'' \,  \C_{\ell_k' s_k' ; \ell_r'' s_r''}^{J}(\sigma_k',s,\sigma_r'') \, f_{s_k'}^{*}(\sigma_k') \, \A_{\ell_r'' s_r''  ; \ell_j s_j }^{J}(\sigma_r'',s,\sigma_j) \Theta(\sigma_k' - \sigma_{k}^{(\mathrm{th})}) \\
&  + \frac{ 1 }{64 \pi^2 \sqrt{s}} \frac{1}{\lvert \p_j^{\star} \rvert}  \sum_{\substack{n \\ n \ne j}} \sum_{\ell_n'' , s_n''} \,  \int_{\sigma_n^{(\mathrm{th})}}^{(\sqrt{s}-m_n)^2} d\sigma_n'' \,   \C_{\ell_n'' s_n'' ; \ell_j s_j}^{J}(\sigma_n'',s,\sigma_j) \, f_{s_j}(\sigma_j) \, \A_{ \ell_k' s_k'; \ell_n'' s_n''}^{J\,*}(\sigma_k',s,\sigma_n'') \Theta(\sigma_j - \sigma_{j}^{(\mathrm{th})})  \\
&  + \frac{\pi}{2\lvert \p_j^{\star} \rvert \lvert \p_k'^{\star} \rvert} \, f_{s_k'}^{*}(\sigma_k') f_{s_j}(\sigma_j)\,\C_{\ell_k' s_k';\ell_j s_j}^{J}(\sigma_k',s,\sigma_j)
(1 - \delta_{jk}) \Theta(1 - \lvert z_{kj} \rvert^2)  ,
\end{split}
\end{equation}

\end{widetext}
where the recoupling coefficients are defined in Eq.~\eqref{eq:recoupling_coef}. Notice that the first, third, and fourth term involve direct channel exchanges in the intermediate state, while the others involve rescattering between cross channels.

Finally, we introduce the amputated amplitude $\wt{\A}_{\ell_k' s_k' ; \ell_j s_j}^{J}$, defined in Eq.~\eqref{eq:amputation}. The amputation eliminates the unitarity cut from the $\2\to\2$ amplitude in the two particle subsystem in the third and fourth term of Eq.~\eqref{eq:app_C_PWIS_unitarity}. Taking the imaginary part of Eq.~\eqref{eq:amputation},
\begin{equation}\label{eq:ImagFact}
\begin{split}
 &  \im{\big[ f_{s_{k}'}(\sigma_{k}')\wt{\A}_{\ell_k' s_k' ; \ell_j s_j}(\sigma_{k}',s,\sigma_{j}) f_{s_{j}}(\sigma_{j}) \big]}  \\
 & \, = \im{\big[f_{s_{k}'}(\sigma_{k}')\big]}\wt{\A}_{\ell_k' s_k' ; \ell_j s_j}(\sigma_{k}',s,\sigma_{j}) f_{s_{j}}(\sigma_{j}) \\
 &\quad + f_{s_{k}'}^{*}(\sigma_{k}') \im{\big[\wt{\A}_{\ell_k' s_k' ; \ell_j s_j}(\sigma_{k}',s,\sigma_{j}) \big]} f_{s_{j}}(\sigma_{j}) \\
 &\quad + f_{s_k'}^{*}(\sigma_k')\wt{\A}_{\ell_k' s_k' ; \ell_j s_j}^{*}(\sigma_k',s,\sigma_j) \im{\big[ f_{s_j}(\sigma_j)\big]}.
\end{split}
\end{equation}
The amputation removes the contribution from the isobar amplitude unitarity cut using Eq.~\eqref{eq:2to2UnitPW}, 
leaving only rescattering corrections to the isobar shape. We then arrive at the amputated PWIS unitarity relations Eq.~\eqref{eq:PWIS_unitarity}.

\section{The $B$-matrix and Unitarity}\label{sec:app_E}
In this appendix, we demonstrate that the $B$-matrix parameterization satisfies the unitarity relations Eq.~\eqref{eq:PWIS_unitarity}, specifically for $\wt{\A}_{13}$. Recall that the $B$-matrix parameterization for $\wt{\A}_{13}$ is
\begin{equation}\label{eq:app_Bmat1}
\begin{split}
\wt{\A}_{13}(s) & = \wt{\B}_{13}(s)  + \wt{\B}_{13}(s)\tau_{3}(s)\wt{\A}_{33}(s),
\end{split}
\end{equation}
which has an imaginary part
\begin{equation}\label{eq:app_E_derive1}
\begin{split}
\im{\wt{\A}_{13}(s)} & = \im{\wt{\B}_{13}}(s) \\
&  + \im{\wt{\B}_{13}(s)}\tau_{3}(s)\wt{\A}_{33}(s) \\
&  + \wt{\B}_{13}^{*}(s)\im{\tau_{3}(s)}\wt{\A}_{33}(s) \\
&  + \wt{\B}_{13}^{*}(s)\tau_{3}^{*}(s)\im{\wt{\A}_{33}(s)}.
\end{split}
\end{equation}
From Eqs.~\eqref{eq:tau} and \eqref{eq:2to2UnitPW}, \begin{equation}
\im{\tau_n(s,\sigma_n)} = \rho_3(s,\sigma_n)\rho_2(\sigma_n) \lvert f_{s_n}(\sigma_n) \rvert^{2},
\end{equation}
and since $\wt{\R}_{kj}$ is real, $\im\wt{\B}_{kj} = \im{\wt{\E}_{kj}}$, which is known from projecting Eq.~\eqref{eq:OPE} into partial waves.

The imaginary part of $\wt{\A}_{33}$ is found by using Eq.~\eqref{eq:Bmat2},
\begin{equation}\label{eq:app_E_derive2}
\begin{split}
\im{\wt{\A}_{33}(s)} & = \im{\left[\mathbbm{1} - \Kc_{33}(s)\tau_{3}(s) \right]^{-1}}\Kc_{33}(s) \\
& + \left[\mathbbm{1} - \Kc_{33}^{*}(s)\tau_{3}^{*}(s) \right]^{-1} \im{\Kc_{33}(s)},
\end{split}
\end{equation}
where the kernel $\Kc_{33}(s) = \wt{\B}_{31}(s)\tau_1(s) \wt{\B}_{13}(s)$
. The imaginary part of $\left[\mathbbm{1} - \Kc_{33}(s)\tau_3(s)\right]^{-1}$ is found by the identity $\im\left[A^{-1}A\right] = \im A^{-1} A + A^{*\,-1}\im A  = 0$, giving
\begin{equation}\label{eq:app_E_derive3}
\begin{split}
 \im{\left[ \mathbbm{1} - \Kc_{33}(s)\tau_3(s) \right]^{-1}} & =\left[ \mathbbm{1} - \Kc_{33}^{*}(s) \tau_{3}^{*}(s) \right]^{-1} \\
&  \times \im\left[ \Kc_{33}(s)  \tau_{3}(s)\right] \\
&  \times \left[ \mathbbm{1} - \Kc_{33}(s) \tau_{3}(s) \right]^{-1},
\end{split}
\end{equation}
with $\im\left[ \Kc_{33}(s)  \tau_{3}(s)\right] = \im\Kc_{33}(s) \tau_3(s) + \Kc_{33}^{*}(s) \im \tau_3(s)$. Combining Eqs.~\eqref{eq:app_E_derive1}, \eqref{eq:app_E_derive2}, and \eqref{eq:app_E_derive3} give
\begin{widetext}
\begin{equation}\label{eq:app_E_derive4}
\begin{split}
\im{\wt{\A}_{13}(s)} & = \im{\wt{\B}_{13}}(s) \\
& + \im{\wt{\B}_{13}(s)}\tau_{3}(s)\wt{\A}_{33}(s) \\
\
&  + \wt{\B}_{13}^{*}(s)\im{\tau_{3}(s)}\wt{\A}_{33}(s) \\
\
&  + \wt{\B}_{13}^{*}(s)\tau_{3}^{*}(s)\left[\mathbbm{1} - \Kc_{33}^{*}(s)\tau_{3}^{*}(s) \right]^{-1} \im{\Kc_{33}(s)}\tau_3(s)\left[\mathbbm{1}-\Kc_{33}(s)\tau_3(s) \right]^{-1}\Kc_{33}(s) \\
\
& + \wt{\B}_{13}^{*}(s)\tau_{3}^{*}(s)\left[\mathbbm{1} - \Kc_{33}^{*}(s)\tau_{3}^{*}(s) \right]^{-1} \Kc_{33}^{*}(s)\im{\tau_3}(s)\left[\mathbbm{1}-\Kc_{33}(s)\tau_3(s) \right]^{-1}\Kc_{33}(s) \\
\
&  + \wt{\B}_{13}^{*}(s)\tau_{3}^{*} \left[\mathbbm{1} - \Kc_{33}^{*}(s)\tau_{3}^{*}(s) \right]^{-1} \im{\Kc_{33}(s)}.
\end{split}
\end{equation}
The imaginary part of the kernel is
\begin{equation}
\begin{split}
\im{\Kc_{33}(s)} & = \im{\wt{\B}_{31}(s)} \tau_1(s) \wt{\B}_{13}(s) \\
& + \wt{\B}_{31}^{*}(s) \im{\tau_1(s)} \wt{\B}_{13}(s) \\
& + \wt{\B}_{31}^{*}(s) \tau_1^{*}(s) \im{\wt{\B}_{13}(s)}.
\end{split}
\end{equation}
We use Eq.~\eqref{eq:denom_shift} to shift the last three lines of Eq.~\eqref{eq:app_E_derive4} in terms $\Kc_{11}=\wt{\B}_{13}(s)\tau_3(s)\wt{\B}_{31}(s)$,
\begin{equation}\label{eq:app_E_derive5}
\begin{split}
\im{\wt{\A}_{13}(s)} & = \im{\wt{\B}_{13}}(s) \\
& + \im{\wt{\B}_{13}(s)}\tau_{3}(s)\wt{\A}_{33}(s) \\
& + \wt{\B}_{13}^{*}(s)\im{\tau_{3}(s)}\wt{\A}_{33}(s) \\
\
& + \left[\mathbbm{1} - \Kc_{11}^{*}(s)\tau_{1}^{*}(s) \right]^{-1}\wt{\B}_{13}^{*}(s)\tau_{3}^{*} \im{\wt{\B}_{31}(s)}\tau_1(s)\wt{\B}_{13}(s) \tau_3(s)\left[\mathbbm{1}-\Kc_{33}(s)\tau_3(s) \right]^{-1}\Kc_{33}(s) \\
\
& + \left[\mathbbm{1} - \Kc_{11}^{*}(s)\tau_{1}^{*}(s) \right]^{-1}\Kc_{11}^{*}(s)\im{\tau_1(s)}\wt{\B}_{13}(s)\tau_3(s)\left[\mathbbm{1}-\Kc_{33}(s)\tau_3(s) \right]^{-1}\Kc_{33}(s) \\
\
& + \left[\mathbbm{1} - \Kc_{11}^{*}(s)\tau_{1}^{*}(s) \right]^{-1}\Kc_{11}^{*}(s)\tau_1^{*}(s) \im{\wt{\B}_{13}(s) }\tau_3(s)\left[\mathbbm{1}-\Kc_{33}(s)\tau_3(s) \right]^{-1}\Kc_{33}(s) \\
\
& + \wt{\B}_{13}^{*}(s)\tau_{3}^{*}(s)  \left[\mathbbm{1} - \Kc_{33}^{*}(s)\tau_{3}^{*}(s) \right]^{-1}\Kc_{33}^{*}(s)\im{\tau_3}(s)\left[\mathbbm{1}-\Kc_{33}(s)\tau_3(s) \right]^{-1}\Kc_{33}(s) \\
\
& +  \left[\mathbbm{1} - \Kc_{11}^{*}(s)\tau_{1}^{*}(s) \right]^{-1}\wt{\B}_{13}^{*}(s)\tau_{3}^{*}(s) \im{\wt{\B}_{31}(s)}\tau_1(s)\wt{\B}_{13}(s) \\
\
&  +  \left[\mathbbm{1} - \Kc_{11}^{*}(s)\tau_{1}^{*}(s) \right]^{-1}\Kc_{11}^{*}(s)\im{\tau_1(s)}\wt{\B}_{13}(s) \\
\
& +  \left[\mathbbm{1} - \Kc_{11}^{*}(s)\tau_{1}^{*}(s) \right]^{-1}\Kc_{11}^{*}(s)\tau_1^{*}(s) \im{\wt{\B}_{13}(s) }.
\end{split}
\end{equation}
Grouping common terms in $\im\tau_n$ and $\im{\wt{\B}_{kj}}$, and identifying the forms of the amplitudes from Sec.~\ref{sec:B-Matrix}, yields
\begin{equation}\label{eq:app_E_derive6}
\begin{split}
\im{\wt{\A}_{13}(s)} & = \im{\wt{\B}_{13}}(s) \\
& + \im{\wt{\B}_{13}(s)}\tau_{3}(s)\wt{\A}_{33}(s) \\
\
& +  \wt{\A}_{11}^{*}(s)\tau_1^{*}(s) \im{\wt{\B}_{13}(s) }\\
\
& + \wt{\A}_{13}^{*}(s)\im{\tau_{3}(s)}\wt{\A}_{33}(s) \\
\
&  + \wt{\A}_{11}^{*}(s)\im{\tau_1(s)}\wt{\A}_{13}(s)\\
\
&  + \wt{\A}_{11}^{*}(s)\tau_1^{*}(s) \im{\wt{\B}_{13}(s) }\tau_3(s)\wt{\A}_{33}(s) \\
\
& +  \wt{\A}_{13}^{*}(s)\tau_{3}^{*}(s) \im{\wt{\B}_{31}(s)}\tau_1(s)\wt{\A}_{13}(s).
\end{split}
\end{equation}
Substituting for the imaginary parts of $\tau_n$ and $\wt{\B}_{kj}$ gives the PWIS unitarity relation for $\wt{\A}_{13}$. The unitarity relations for the other amplitudes can be found in a similar manner.

\end{widetext}

\section{The Feynman Triangle Diagram}\label{sec:app_F}
\begin{figure}[t!]
\centering
\includegraphics[width=0.8\columnwidth]{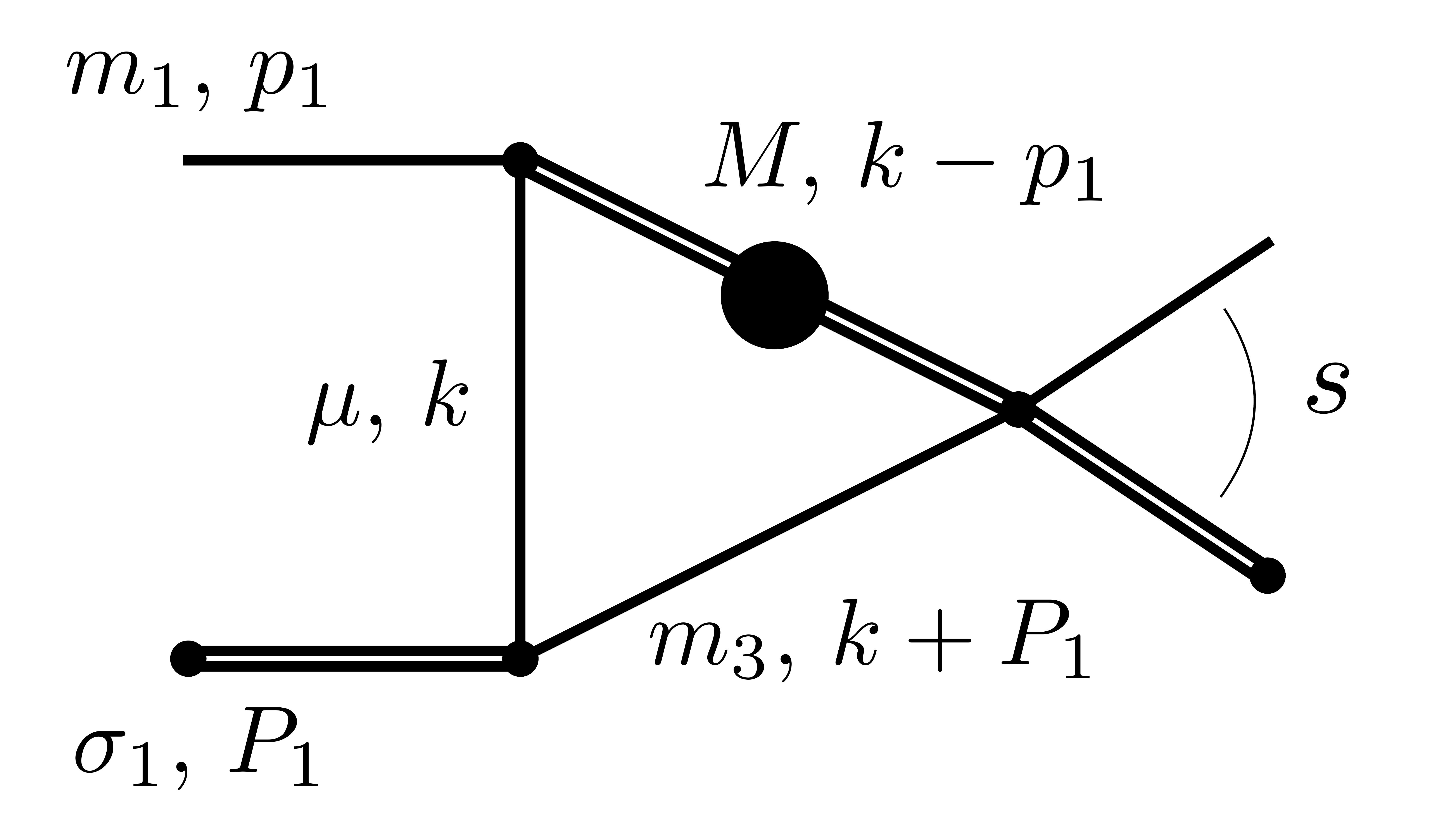}
\caption{The triangle diagram with loop momentum labels.}
\label{fig:tri_label_app}
\end{figure}

For reference, we state the basic formulae for computing the Feynman triangle diagram, \cf Ref.~\cite{Itzykson:1980rh}.
The perturbative Feynman diagram has the form
\begin{equation}
\mathcal{T}_{F}(s) = i\int \frac{d^{4}k}{(2\pi)^4}\frac{1}{D_1D_2D_3},
\end{equation}
shown in Fig.~\ref{fig:tri_label_app}, where the denominator is the product of internal propagators,
\begin{equation}
\begin{split}
D_1D_2D_3 & = \left[k^2 - \mu_{jk}^2 + i\epsilon\right] \\
&  \times \left[(k+P_1)^2 - m_3^2 + i\epsilon\right] \\
&  \times \left[(k-p_1)^2 - M^2 + i\epsilon\right].
\end{split}
\end{equation}

Using the Feynman parameterization and standard loop integration techniques, the Feynman diagram has the form
\begin{equation}
\mathcal{T}_{F}(s) = \frac{1}{16\pi^2} \int_{0}^{1} d\alpha_1 \int_{0}^{1-\alpha_1} d\alpha_2 \, F(s;\alpha_1,\alpha_2),
\end{equation}
where
\begin{equation}
\begin{split}
F^{-1}(s;\alpha_1,\alpha_2) & = M^2 \alpha_1 + m_3^2 \alpha_2 + \mu{}^2 (1 - \alpha_1 - \alpha_2) \\
& + m_1^2 \alpha_1( \alpha_1 - 1)  + \sigma_1 \alpha_2 (\alpha_2 - 1) \\
& - (s - \sigma_1 - m_1^2) \alpha_1 \alpha_2 - i\epsilon.
\end{split}
\end{equation}
The remaining integrals over the Feynman parameters can be computed either numerically, or by analytically performing the integral over $\alpha_2$, then numerically computing the remaining integral over $\alpha_1$.

\begin{figure}[t!]
\centering
\includegraphics[width=0.8\columnwidth]{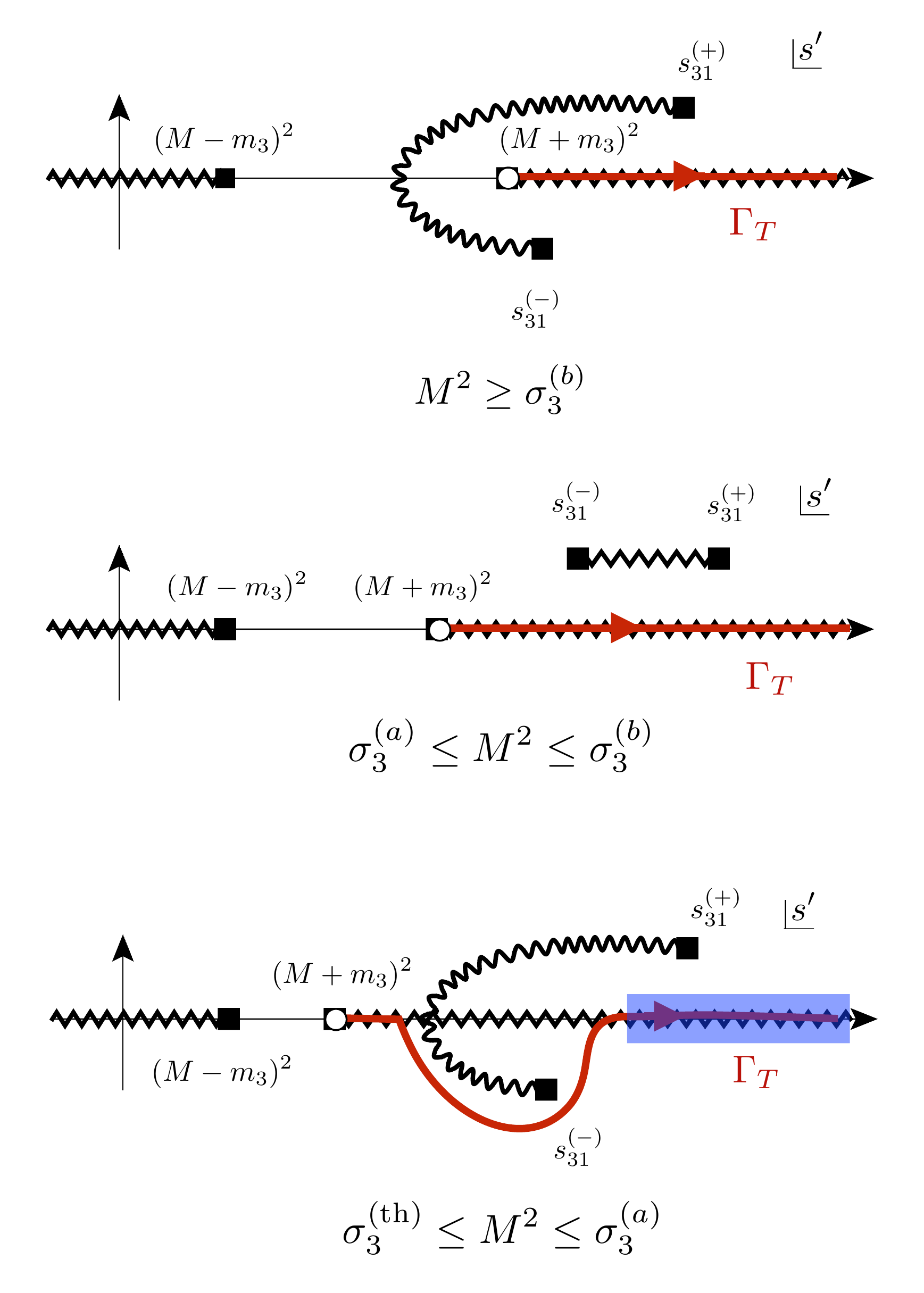}
\put(-150,10){{(c)}}
\put(-150,110){{(b)}}
\put(-150,190){{(a)}}
\caption{Contours for dispersive triangle Eq.~\eqref{eq:TF} shown in red, and the integrand cuts. The three cases are (a) $M^2 \ge \sigma_3^{(b)}$, (b) $\sigma_3^{(a)} \le M^2 \le \sigma_3^{(b)}$, and (c) $\sigma_3^{(\mathrm{th})} \le M^2 \le \sigma_3^{(a)}$. Case (a) corresponds to the usual triangle singularity, which occurs since the OPE branch points pinch the integration region. Case (c) happens when the initial state of the OPE has a higher threshold then the intermediate state. The blue region indicates the physical region from the initial threshold. Note that a triangle singularity does not occur in this case and the integration is not pinched.}
\label{fig:tri_contours}
\end{figure}
Alternatively, the Feynman triangle can be written with a dispersive representation in $s$ using the Cutkosky rules \cite{Itzykson:1980rh},
\begin{equation}\label{eq:TF_app}
\mathcal{T}_{F}(s) = \int_{\Gamma_{T}} ds' \frac{\rho_3(s',M^2)\wt{\E}_{13}(M^2,s',\sigma_1)}{s' - s - i\epsilon},
\end{equation}
where $\Gamma_T$ is the path from the threshold $(M + m_3)^2$ to $\infty$, $\rho_3(s,M^2)$ is given by Eq.~\eqref{eq:3bodyPhaseSpace}, $\wt{\E}_{13}$ is given by Eq.~\eqref{eq:OPE_Swave}, and
the $S$-wave amplitudes are normalized according to Eq.~\eqref{eq:PWIS_expansion}.
The phase space contributes branch point singularities from the threshold and pseudothreshold, $(M \pm m_3)^2$, and a pole at $s=0$. The OPE has branch points $s = s_{31}^{(\pm)}$ near the integration region. Following the discussion in Section~\ref{sec:ope}, the OPE branch points give us the following scenarios:
\begin{enumerate}[(a)]
	\item $M^2 \ge \sigma_3^{(b)}$. The RPE branch points pinch the integration region which starts at $s = (M+m_3)^2$. Figure~\ref{fig:tri_contours}(a) shows the integrand branch cuts and the dispersive contour.  The RPE branch point $s_{31}^{(-)}$ lie in the unphysical sheet close to threshold, causing the known as the triangle singularity \cite{Peierls:1961zz,Aitchison:1966,Eden:1966dnq, Szczepaniak:2015eza}. The triangle singularity produces an extra threshold in the physical region above the threshold $s = (M+m_3)^2$, and is associated with real particle exchange in the intermediate state.
The location of the triangle singularity occurs at
\begin{equation}
\begin{split}
s_{\textrm{tri}} & = \frac{1}{2 m_2^2} \bigg[ (m_3^2 - \sigma_1)(m_1^2 - M^2 ) - m_2^4 \\
& + m_2^2(m_3^2 + m_1^2 + \sigma_1 + M^2) \\
& \pm \lambda^{1/2}(m_2^2,m_3^2,\sigma_1)  \lambda^{1/2}(m_2^2,m_1^2,M^2) \bigg].
\end{split}
\end{equation}
    \item $\sigma_3^{(a)} \le M^2 \le \sigma_3'^{(b)}$. The RPE branch points are both above the real axis, and cause no additional singular behavior. Figure~\ref{fig:tri_contours}(b) illustrates this case.
    \item $\sigma_3^{(\mathrm{th})} \le M^2 \le \sigma_3^{(a)}$. The RPE branch points are again on opposite sides of the real axis. However, the integration region begins below the location where the RPE cut crosses the real axis. This is due to the fact that the initial state has a higher threshold then the intermediate state, so the physical region is above the RPE crossing location. This is illustrated in Fig.~\ref{fig:tri_contours} where the blue region indicates the physical region starting at the initial state threshold, and the integration contour is a path around the RPE branch point $s_{31}^{(-)}$. No singularity occurs in this region as the RPE branch points do not pinch the integration region.
\end{enumerate}
Notice that in contrast to the $B$-matrix triangle, Eq.~\eqref{eq:TB}, the dispersive triangle moves all the singularities from the phase space and OPE to the unphysical sheet. Thus, the only singularity present on the physical sheet is the unitarity cut starting at $s = (M + m_3)^2$.

\bibliographystyle{apsrev4-1}
\bibliography{bibliography.bib}
\end{document}